\documentclass[12pt]{iopart}

\usepackage[utf8]{inputenc} 
\usepackage[T1]{fontenc} 	
\usepackage[english]{babel} 

\usepackage{lmodern} 		
\usepackage{anyfontsize}
\usepackage{fix-cm}

\usepackage{amsmath} 		
\usepackage{amssymb} 		
\usepackage{mathtools} 		
\usepackage{float} 			
\usepackage{graphicx} 		
\usepackage{tabularx} 		
\usepackage{booktabs} 		
\usepackage{color, xcolor} 	
\usepackage{pdfpages} 		
\usepackage{extarrows} 		
\usepackage{multirow} 		
\usepackage{multicol} 		
\usepackage{enumitem} 		
\usepackage{xspace} 		
\usepackage{stackrel} 		
\usepackage{tikz} 			
\usepackage{braket} 		
\usepackage{bm} 			
\usepackage{tensor} 		
\usepackage{slashed} 		
\usepackage{siunitx} 		
\usepackage{lastpage} 		
\usepackage[square,sort&compress,comma,numbers]{natbib}
\usepackage[normalem]{ulem} 
\usepackage{fontawesome} 	
\usepackage{tocloft} 		
\usepackage{titlesec} 		
\usepackage{doi} 			
\usepackage[most]{tcolorbox} 					
\usepackage{hyperref} 		
\usepackage[nameinlink, capitalize]{cleveref} 	
\usepackage[nottoc, notlot, notlof]{tocbibind} 	
\usepackage[ruled, vlined]{algorithm2e} 		
\usepackage{makecell}
\usepackage[makeroom]{cancel}
\usepackage{feynmf}
\usepackage{cancel}
\usepackage[bottom]{footmisc}

\makeatletter
\def\BState{\State\hskip-\ALG@thistlm}
\makeatother

\makeatletter
\@ifundefined{pdfoutput}{}{\DeclareGraphicsRule{*}{mps}{*}{}}
\makeatother

\hypersetup{
	pdftitle={Universal ViT},
	pdfauthor={Luigi Favaro, Andrea Giammanco, and Claudius Krause},
	colorlinks=true, 			
	linkcolor={red!50!black}, 	
	citecolor={blue!50!black}, 	
	urlcolor={blue!80!black} 	
} 

\DeclareSymbolFont{usualmathcal}{OMS}{cmsy}{m}{n}
\DeclareSymbolFontAlphabet{\mathcal}{usualmathcal}

\SetArgSty{textnormal}
\SetKwComment{Comment}{{\small\#}~}{}
\SetCommentSty{mycommfont}

\setitemize{itemsep=0pt, parsep=0pt} 				
\setenumerate{itemsep=0pt, parsep=0pt} 				
\setitemize{itemsep=2pt,topsep=2pt,parsep=0pt,partopsep=0pt,leftmargin=*}
\setenumerate{itemsep=0pt,topsep=2pt,parsep=0pt,partopsep=0pt,labelindent=3pt,leftmargin=*}
\usepackage{amsmath}
 
\usepackage{amsthm} 		
\theoremstyle{definition}

\usetikzlibrary{arrows, shapes.geometric, arrows.meta, shapes, decorations.pathreplacing, fit, patterns, patterns.meta,positioning,calc}
\usepackage{tikz} 			

\definecolor{Rcolor}{HTML}{E99595}
\definecolor{Gcolor}{HTML}{C5E0B4}
\definecolor{Gcolor_light}{HTML}{F1F8ED}
\definecolor{dforestgreen}{HTML}{0B3B24}
\definecolor{Bcolor}{HTML}{9DC3E6}
\definecolor{Ycolor}{HTML}{FFE699}
\definecolor{Ycolor_light}{HTML}{FFF7DE}
\definecolor{blue_light}{HTML}{E3F2FD}
\definecolor{peach_light}{HTML}{ffe8c4}
\definecolor{deepterracotta}{HTML}{BF360C}
\definecolor{lavender}{HTML}{f4bbfc}
\definecolor{deeppurple}{HTML}{4A148C}
\definecolor{silver}{HTML}{F5F5F5}
\definecolor{rose}{HTML}{ffd6e4}
\definecolor{darkcherry}{HTML}{880E4F}
\newcommand{\tikznode}[2]{%
\ifmmode%

\tikz[remember picture,baseline=(#1.base),inner sep=0pt] \node (#1) {$#2$};%
\else
\tikz[remember picture,baseline=(#1.base),inner sep=0pt] \node (#1) {#2};%
\fi}

\tikzstyle{expr} = [rectangle, rounded corners=0.3ex, minimum width=1.5cm, minimum height=1cm, text centered, align=center, inner sep=0, fill=Ycolor, font=\large, draw]

\tikzstyle{small_cinn} = [double arrow, double arrow head extend=0cm, double arrow tip angle=130, inner sep=0, align=center, minimum width=1.1cm, minimum height=0.5cm, fill=Rcolor, draw]

\tikzstyle{small_cinn_black} = [small_cinn, minimum height=1.5cm, fill=black]

\tikzstyle{block_blue_light} = [rectangle, rounded corners, minimum width=6cm, minimum height=2.4cm, font=\large, fill=blue_light, draw]
\tikzstyle{smallblock_blue_light} = [rectangle, rounded corners, minimum width=3cm, minimum height=1.2cm, font=\normalsize, fill=blue_light, draw]

\tikzstyle{block_blue} = [rectangle, rounded corners=0.3ex, minimum width=5.5cm, minimum height=1.2cm, align=center, fill=Bcolor, draw, font=\large]

\tikzstyle{block_green_light} = [rectangle, rounded corners, minimum width=6cm, minimum height=2.4cm, font=\large, fill=Gcolor_light, draw]

\tikzstyle{block_green} = [rectangle, rounded corners=0.3ex, minimum width=5.5cm, minimum height=1.2cm, align=center, fill=Bcolor, draw, font=\large]

\tikzstyle{transformer_huge} = [rectangle, rounded corners, minimum width=8.5cm, minimum height=2.4cm, font=\large, fill=Gcolor, draw]

\tikzstyle{attention_huge} = [rectangle, rounded corners=0.3ex, minimum width=8cm, minimum height=1.2cm, align=center, fill=Gcolor, draw, font=\large]

\tikzstyle{txt_huge} = [align=center, font=\Huge, scale=2]
\tikzstyle{txt} = [align=center, font=\LARGE, minimum height=1cm]

\tikzstyle{arrow} = [thick,-{Latex[scale=1.0]}, line width=0.2mm, color=black]
\tikzstyle{line} = [thick, line width=0.2mm, color=black]
\tikzstyle{bkg} = [minimum width=22cm, minimum height=7.5cm, font=\large, fill=white]

\tikzstyle{encoder_black} = [trapezium, fill=black, rotate=270, minimum height=4cm, minimum width=3.2cm,align=center,draw]
\tikzstyle{encoder} = [trapezium, fill=Bcolor, rotate=270, minimum height=4cm, minimum width=3cm,align=center,draw]
\tikzset{trapezium stretches=true}
\tikzstyle{cinn} = [double arrow, double arrow head extend=0cm, double arrow tip angle=130, shape border rotate=90, inner sep=0, align=center, minimum width=2.1cm, minimum height=2.3cm, fill=Rcolor, draw,font=\LARGE]
\tikzstyle{cinn_black} = [cinn, minimum height=2.5cm, fill=black]
\tikzstyle{crc} = [circle, rounded corners=0.3ex, minimum width=1.5cm, minimum height=1cm, text centered, align=center, inner sep=0, fill=white, font=\LARGE, draw]
\newcommand{\orcid}[1]{\href{https://orcid.org/#1}{\includegraphics[width=8pt]{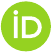}}}

\definecolor{red_cb}{HTML}{e41a1c}
\definecolor{blue_cb}{HTML}{377eb8}
\definecolor{green_cb}{HTML}{4daf4a}
\definecolor{purple_cb}{HTML}{984ea3}
\definecolor{orange_cb}{HTML}{ff7f00}

\definecolor{EmeraldGreen}{HTML}{1ea78d}
\definecolor{EnglishRed}{HTML}{b02427}
\hypersetup{colorlinks=true,urlcolor=EmeraldGreen,citecolor=EmeraldGreen,linkcolor=EnglishRed}

\newcommand{\pd}{p_\text{data}}

\newcommand{\XXLangle}{\biggl\langle}
\newcommand{\XXRangle}{\biggr\rangle}

\def\d{\text{d}}
\newcommand\one{\leavevmode\hbox{\small1\normalsize\kern-.33em1}}
\newcommand{\normal}{\mathcal{N}} 			

\newcommand{\loss}{\mathcal{L}} 	

\newcommand{\geant}{Geant4\xspace}

\newcommand{\arXiv}[2][]{%
	\ifthenelse{\equal{#1}{}}%
	{\href{http://arxiv.org/abs/#2}{arXiv:#2}}%
	{\href{http://arxiv.org/abs/#2}{arXiv:#2~[#1]}}}

\def\slashchar#1{\setbox0=\hbox{$#1$}           
   \dimen0=\wd0                                 
   \setbox1=\hbox{/} \dimen1=\wd1               
   \ifdim\dimen0>\dimen1                        
      \rlap{\hbox to \dimen0{\hfil/\hfil}}      
      #1                                        
   \else                                        
      \rlap{\hbox to \dimen1{\hfil$#1$\hfil}}   
      /                                         
   \fi}

\def\mathswitchr#1{\relax\ifmmode{\text{#1}}\else$\text{#1}$\xspace\fi}
\def\mathswitch#1{\relax\ifmmode#1\else$#1$\xspace\fi}

\begin{document}

\begin{flushright}
    \vspace*{-1cm}
    {\footnotesize IRMP-CP3-26-01,\;MBI-ML-26-01}
    \vspace*{0.5cm}
\end{flushright}
\title{A universal vision transformer for fast calorimeter simulations}

\author{Luigi Favaro$^1$\orcid{0000-0003-2421-7100}, Andrea Giammanco$^1$\orcid{0000-0001-9640-8294} and Claudius Krause$^{2}\footnote{corresponding author}$\orcid{0000-0003-0924-3036}}

\address{$^1$Centre for Cosmology, Particle Physics and Phenomenology (CP3), \\ Universit\'e catholique de Louvain, Louvain-la-Neuve, Belgium}

\address{$^2$Marietta Blau Institute for Particle Physics (MBI Vienna), \\ Austrian Academy of Sciences (ÖAW), Austria \\}

\ead{luigi.favaro@uclouvain.be, andrea.giammanco@cern.ch, claudius.krause@oeaw.ac.at}

\begin{abstract}
The high-dimensional complex nature of detectors makes fast calorimeter simulations a prime application for modern generative machine learning. Vision transformers (ViTs) can emulate the \geant response with unmatched accuracy and are not limited to regular geometries. Starting from the CaloDREAM architecture, we demonstrate the robustness and scalability of ViTs on regular and irregular geometries, and multiple detectors. Our results show that ViTs generate electromagnetic and hadronic showers with minimal deviations from \geant in multiple evaluation metrics, while maintaining the generation time in the $\mathcal{O}(10-100)$ ms on a single GPU. Furthermore, we show that pretraining on a large dataset and fine-tuning on the target geometry leads to reduced training costs and higher data efficiency, or altogether improves the fidelity of generated showers.
\end{abstract}

{\footnotesize \noindent{\it Keywords}: fast detector simulations, generative networks, conditional flow matching, transformers\\ }
{\footnotesize \it Submitted to Machine Learning: Science and Technology}

\section{Introduction}
Particle Physics is a numerically intensive, data-processing and simulation-heavy science. The large experiments ATLAS and CMS at the Large Hadron Collider (LHC) record data at a staggering rate of several GB/s~\cite{Albrecht:2024eio}, totaling now one Exabyte in recorded data~\cite{exabyte} at CERN. Simulation of collisions, the major backbone of statistical data analyses, needs to keep up with the amount of data as well. The upcoming runs of the LHC and the high-luminosity phase will increase the amount of required computing even further. To fully exploit the data and learn about the underlying laws of nature, it is therefore essential to develop efficient algorithms in every part of the analyses or simulation chains. Modern machine learning (ML) has the potential to contribute substantially to this endeavor~\cite{Butter:2022rso,Shanahan:2022ifi}, for example, by accelerating computationally intensive bottlenecks and opening up new  avenues for efficient analyses~\cite{Cranmer:2019eaq}.   
The rise of generative ML in computer science has introduced many new ideas for more efficient simulation in high-energy physics (HEP) in recent years, in particular for amplitude evaluation~\cite{Badger:2020uow,Bahl:2024gyt,Beccatini:2025tpk}, phase space generation~\cite{Gao:2020zvv,Bothmann:2020ywa,Heimel:2022wyj,Heimel:2023ngj,Heimel:2024wph}, (end-to-end) event generation~\cite{DiSipio:2019imz,Verheyen:2022tov,Butter:2019cae,Butter:2021csz,Quetant:2024ftg,Leigh:2023toe}, detector tracking~\cite{Novak:2025jqv}, and detector calorimeter simulation (see \cite{Hashemi:2023rgo,Krause:2024avx} for recent reviews on the vast literature and~\cite{Paganini:2017hrr,Paganini:2017dwg,Krause:2021ilc,Krause:2021wez,FaucciGiannelli:2023fow,Kach:2023rqw,Kach:2024yxi,ATL-SOFT-PUB-2020-006,DeepTreeCHEP,DeepTreeNIPS,Diefenbacher:2023vsw,Buss:2024orz,Krause:2022jna,Buckley:2023daw,Ernst:2023qvn,Pang:2023wfx,Schnake:2024mip,Amram:2023onf,Buhmann:2023bwk,Buhmann:2023kdg,Mikuni:2022xry,Mikuni:2023tqg,Kobylianskii:2024ijw,CaloDIT_ACAT,Liu:2024kvv,Cresswell:2022tof,Salamani:2021zet,ATLAS:2022jhk,G4TransDalila,CaloLatent,Cresswell:2024fst,Favaro:2024rle,Raikwar:2025fky,Buss:2025bec,Buss:2025kiu} for recent works).  First-principled simulations of the latter are computationally expensive and comprise a significant portion of the overall computing budget~\cite{Collaboration:2802918,Software:2815292}, so any kind of improvement has a direct and strong impact on the global computing efficiency. Generative networks provide a faster emulator of the detailed simulations based on \geant~\cite{Agostinelli:2002hh,1610988,ALLISON2016186}, while having the capability to oversample,
meaning they can amplify the statistical properties of the training dataset~\cite{Matchev:2020tbw,Butter:2020qhk,Bieringer:2022cbs,Watts:2024caw,Bahl:2025ryd}. 
Alternatively, individual steps of the simulation chain~\cite{Butter:2022rso} can be combined into a single generative network~\cite{CERN-CMS-NOTE-2023-003,Krammer:2024rdl,Mazurek:2025isd}. 

Despite all these advantages, generative ML networks for fast simulation are still computationally expensive. First, the training data need to be generated by traditional simulations, and second, the generative networks need to be trained. When switching to a new detector layout or even just changing the voxelization of a given geometry, the generative networks need to be retrained completely. It is therefore highly beneficial to make the overall training of the networks more efficient. 
For example, the voxelization that is applied to the raw hits can be adopted to better reflect the types of showers under consideration. Areas of larger activity would have a finer read-out and areas with less activity would be coarse~\cite{ATL-SOFT-PUB-2025-003}. As a result, the number of voxels to be considered in the subsequent training can be reduced, and the networks can be smaller. 
However, state of the art ML networks often assume regular geometries, and mapping irregular voxelizations to regular spaces is only achieved through the addition of artificial voxels, which comes at increased computational costs.

As an alternative, one can keep the general setup the same, but investigate how the training of generative networks can be made more efficient. For this, we start with an observation regarding calorimeter showers from different incident particles, detector geometries, and detector materials: Even though the specific details are very different, the showers still have many things in common:
\begin{itemize}
    \item Sparsity: A single shower deposits energy only in a fraction of the voxels, and most of the voxels will not receive an energy deposition at all. 
    \item Dynamic range: The energy that is deposited in the voxels spans several orders of magnitude and, in general, scales with the incident energy. Since a common approach to calorimeter simulation with generative ML is to split off the scale of the energy that is deposited from the normalized shower shapes~\cite{Krause:2021ilc}, the latter becomes less sensitive to the incident energies.  
    \item Spatial correlations: Showers emerge as spatially connected clusters, in some cases (depending on incident particle and detector materials), even tracks of individual particles become visible, thereby strongly correlating the energy depositions of nearby voxels. 
    \item Central activity: The incoming particle, generating the shower, is assumed to hit the center of the voxelized space. Especially for electromagnetic showers, the main activity in the following layers will be at the center of the considered volume and will form a single cluster. 
\end{itemize}
Motivated by these observations, we investigate transfer learning for calorimeter showers.
Instead of training the generative network entirely from scratch, i.e. randomly initialized network weights, we initialize a large fraction of weights from a network originally trained for another dataset.
While this could still involve different incident particles, detector materials, detector layouts, or voxelizations, the differences in the pre-processed distributions between the datasets are still rather small. Fine-tuning is therefore more efficient because the network has to learn a smaller shift in how the data are mapped to the Gaussian latent space.

While similar in spirit to foundation models~\cite{Golling:2024abg,Birk:2024knn,Mikuni:2024qsr,Wildridge:2024yeg,Birk:2025wai,Mikuni:2025tar,Tani:2025osu,Hallin:2025ywf,Bhimji:2025isp}, which are pretrained once to a big dataset and then fine-tuned to the application at hand, we prefer the term transfer learning in this case, as we only consider generative tasks (on different datasets) and not other tasks like classification or regression as one usually would for a foundation model. Transfer learning was studied before in HEP~\cite{Dreyer:2022yom,Chappell:2022yxd,Beauchesne:2023vie,Mokhtar:2025zqs,Bonilla:2025rrn,vigl2024finetuningfoundationmodelsjoint}, but mainly in the context of classification tasks. In the context of detector simulation, transfer learning has recently also been studied for detectors with fixed voxelization in~\cite{Raikwar:2025fky} and point-cloud architectures in~\cite{Gaede:2025shc}. In this work, we propose a general fine-tuning strategy for vision transformers applicable to any detector geometry and across particle types. Additionally, we present complete benchmark results on the LEMURS dataset~\cite{McKeown:2025gtw}, used for pretraining, and on multiple irregular geometries intractable in~\cite{Favaro:2024rle}. Finally, we discuss the importance of evaluation metrics for a proper evaluation of efficiency gains from pretraining.

The paper is organized as follows. In~\cref{sec:ml}, we introduce conditional flow matching and the generative networks we study. These include the original CaloDREAM~\cite{Favaro:2024rle} and our improvements to it, which we call CaloDREAM++. In~\cref{sec:data}, we present the datasets at the core of our studies, and~\cref{sec:metrics} discusses the evaluation metrics we employ to study the performance of the generative architectures. \cref{sec:baseline} and \cref{sec:baseline_irreg} show the performance of our generative network on datasets of regular and irregular geometries, respectively. In \cref{sec:finetuning} we discuss transfer learning and show how the pre
trained CaloDREAM++ adapts to new datasets. We conclude in~\cref{sec:conclusions}. In the appendices, we show additional high-level feature distributions of the datasets as well as the training hyperparameters.

\section{Methods}
\label{sec:ml}

\subsection{Conditional Flow Matching}
Our approach uses continuous normalizing flows
trained with Conditional Flow Matching (CFM)~\cite{lipman2023flowmatchinggenerativemodeling} as the underlining generative network.
This class of generative networks parametrizes the transition from
the data to the latent space as an ordinary differential equation (ODE):
\begin{align}
  \frac{dx(t)}{dt} = v(x(t),t)
  \qquad \text{with} \qquad
  x\in\mathbb{R}^d \; ,
\label{eq:sample_ODE}
\end{align}
where the velocity field $v(x(t),t) \in \mathbb{R}^d$ matches the dimensionality of the data.
A differential equation of this form can be related to the 
underlying density through the continuity equation
\begin{align}
\frac{\partial p(x,t)}{\partial t} + \nabla_x \left[ p(x,t) v(x,t) \right] = 0 \;.
\label{eq:continuity}
\end{align} 
The continuous transformation of the density $p(x,t)$, parametrized by $t$,
should satisfy the boundary conditions
\begin{align}
 p(x,t) \to 
 \begin{cases}
  \normal(x|0,1) \quad & t \to 0 \\
  \pd(x)  \quad & t \to 1  \;.
\end{cases} 
\label{eq:cfm_limits}
\end{align}
CFM is a simple prescription to train continuous normalizing flows 
upon selecting a conditional target trajectory.
A standard choice is a linear trajectory of the form
\begin{align}
    &x(t|\epsilon, x_0) = (1-t)\epsilon + tx_0, \quad\text{with} \\ 
    &\epsilon\sim\normal(0,1), \quad x_0\sim\pd(x) \notag \;.
\end{align}
Given Gaussian distributed random numbers, a training dataset, and 
uniformly distributed times, we can approximate the true velocity field
with a neural network $v_\Theta(x(t),t)$. 
Using linear trajectories, the network is optimized using a simple
mean-squared error loss of the form
\begin{align}
    \loss_\text{CFM} &= \XXLangle \left[ v_\Theta(x(t|\epsilon, x_0),t) - (x - \epsilon)\right]^2 \XXRangle_{t\sim U(0,1),\,\epsilon\sim \normal,\,x \sim \pd} \; .
\label{eq:didi_loss}
\end{align}
Conditional probability distributions can be learned by allowing
$v_\Theta$ to depend on additional inputs.
Sampling from the trained network requires solving the ODE with the learned velocity field,
\begin{equation}
x(t=1) = x(t=0) + \int_0^1 \d t\; v_\Theta(x(t), t)
\label{eq:cfm_sampling}
\end{equation}
This is typically done numerically with standard ODE solvers such as
Runge-Kutta methods, or more advanced Bespoke samplers\cite{shaul2024bespokenonstationarysolversfast}.

\subsection{CaloDREAM and CaloDREAM++}
We briefly review the core concepts of CaloDREAM~\cite{Favaro:2024rle} before
discussing the extended setting, CaloDREAM++, needed for our studies.
CaloDREAM combines two neural networks to generate the final 
calorimeter showers. An ``energy network'' produces the layer energy
ratio variables $u$~\cite{Krause:2021ilc} conditioned
on the incident energy $E_\text{inc}$ of the incoming particle.
More generally, we refer to the set of global conditions of the incident particle as $C$. In our studies, it contains per-shower information such as the incident energy, and the direction of
the incident particle parametrized with the azimuthal, $\theta$, and polar angles, $\phi$, but also labels that specify the detector used in a specific dataset. We define the variables $u$ 
such that they are in the range $[0, 1]$ by construction and allow for an
analytic formula which maps back to the total energy deposited in a layer.
The parametrization is described in~\labelcref{app:implementation}.
The ``shape network'' learns the conditional distribution for 
the normalized voxels $x$, given all the other energy and conditional variables.
We train the two networks independently, and the generation process
follows the sequential steps:
\begin{align}
    u &\sim p_{\Theta_e}(u|C) \qquad\quad \text{energy variables}, \notag \\ 
    x &\sim p_{\Theta_s}(x|C, u) \qquad \text{normalized voxels},
\end{align}
where $\Theta_e$ and $\Theta_s$ are the learnable weights of the energy and shape network, respectively. Namely, we sample per-layer energy depositions from the first generative network, given a set of global initial conditions. Then, we pass the sampled $u$-vector to the second generative network, together with the same global conditions, to obtain a calorimeter shower in normalized space. Finally, we use the analytic expression for the $u$-variables to retrieve the layer energies and rescale the generated calorimeter shower.

\subsubsection*{Energy network.}
The original CaloDREAM energy network built an embedding vector for each
energy ratio from the incident energy and the energy ratios from 
the previous layers. Sampling from this network required the 
sequential generation of layer energies, hence solving an ODE
for each energy ratio.
We accelerate the generation, especially for small batch sizes,
by avoiding this autoregressive sampling in favor of a parallel sampling of the full velocity vector.
The construction of the conditional information is unchanged.
A standard transformer encoder-decoder network takes as inputs the energy $u$-vector
and the global shower information $C$.
We represent the input of the transformer as a sequence with fixed-length 
embedding vectors. The embedding vector is manually constructed from the input
quantity of interest to which we append a one-hot encoding, and a zero-padding vector. The one-hot encoding is necessary to discriminate between the different inputs, while the zero-padding ensures that the input embedding vector has the correct length.
We use this approach to construct the input vector for both the transformer encoder and decoder.
While using learnable embedding vectors is also 
possible, we did not observe any improvements during evaluation.
For the encoder, each term of the sequence corresponds to
a global condition of the shower. Setting the length to $l$, the encoder
embedding vector is constructed as
\begin{equation}
    p_i = [C_i, \mathrm{onehot}(C_i), \vec 0]\;, \qquad p_i\in \mathbb{R}^{l} \;,
\end{equation}
where the index $i$ runs over the number of conditions.
The full encoder is a series of multi-head self-attention blocks implemented
in the following way
\begin{align*}
    \mathrm{SelfAttention: \;\; SA}(x)&= (\mathrm{LayerNorm} \;\circ\; \mathrm{Residual} \;\circ\; \mathrm{MultiHead SA}) (x) \;, \\
    \mathrm{FeedForward: \;\; FF}(x) &= (\mathrm{Linear} \;\circ\; \mathrm{ReLU} \;\circ\; \mathrm{Linear}) (x) \;, \\
    \mathrm{MLP: \;\; MLP} (x) &= (\mathrm{LayerNorm} \;\circ\; \mathrm{Residual} \;\circ\; \mathrm{FF})(x) \;,\\
    \mathrm{TransformerEncLayer: \;\; TEL}(x) &= (\mathrm{MLP} \;\circ\; \mathrm{SA})(x) \;,\\
    \mathrm{TransformerEncoder: \;\; TE}(x) &= (\mathrm{TEL} \;\circ\; \ldots \;\circ\; \mathrm{TEL})(x) \;,
\end{align*}
where MultiHead SA is a multi-headed self-attention~\cite{vaswani2017attention}, and the Residual step adds the output of the previous function to the inputs~\cite{he2015deepresiduallearningimage}. Additionally, we use LayerNorm~\cite{ba2016layernormalization}, rectified linear units (ReLU)~\cite{Nair2010RectifiedLU}, and Linear is a standard linear transformation with learnable weights and biases.
For the decoder, we use the output of the transformer encoder and the 
sequence of embedded $u$-variables. Such embedding follows a similar logic
and is constructed as
\begin{equation}
    q_i = [u_i, \mathrm{onehot}(u_i), \vec 0]\;, \qquad q_i\in \mathbb{R}^{l} \;.
\end{equation}
The transformer decoder is a series of layers with a self-attention operation, 
a cross-attention between the embedding vector and the output of the transformer 
encoder, and an MLP block.
The main ingredients for the transformer encoder are
\begin{align*}
    \mathrm{CrossAttentionBlock: \;\; CA}(x, y)&= (\mathrm{LayerNorm} \;\circ\; \mathrm{Residual} \;\circ\; \mathrm{CrossAttention})(x,y) \;, \\
    \mathrm{TransformerDecLayer: \;\; TDL}(x,y) &= (\mathrm{MLP} \;\circ\; \mathrm{CA}(\cdot, y) \;\circ\; \mathrm{SA})(x) \;, \\
    \mathrm{TransformerDecoder: \;\; TD}(x,y) &= (\mathrm{TDL} \;\circ\; \ldots \;\circ\; \mathrm{TDL})(x,y) \;,
\end{align*}
Our first update to the network includes the encoding of the energy information
in a single forward pass. In formulas, we extract the condition $c$ as
\begin{equation}
    c(u_0(t), \ldots, u_L(t), C) = \mathrm{TD}(q, \mathrm{TE}(p)) \;,
\end{equation}
where $L$ is the number of detector layers. We finally pass this information to
a multi-layer perceptron (MLP) with two linear layers and a sigmoid linear unit non-linearity, which predicts the full velocity field. 
Altogether, at each time step, we require a single network prediction during both training and inference, which can be written as
\begin{equation}
    v_{\Theta_e}(u(t), C, t) = \text{MLP}\left( c(u_0(t), \ldots, u_L(t), C), \mathcal{F}(t) \right)  \; ,
\end{equation}
where $\mathcal{F}$ indicates a Gaussian Fourier projection, typically used to encode the time information~\cite{tancik2020fourierfeaturesletnetworks}.
The training hyperparameters of the energy network 
are given in~\labelcref{app:hyperparams}.

\subsubsection*{Shape network.}
\begin{figure}[t]
    \centering
    \scalebox{0.75}{\begin{tikzpicture}

\node (block_embed) [block_green_light, text width=4cm, yshift=5.1cm, 
text depth=4.8cm, align=center, minimum height=5.5cm, minimum width=7cm, font=\large, fill=silver] {Detector specific};

\node (block) [block_blue_light, text width=4cm,
text depth=0.5cm, align=center, minimum height=3cm,font=\large] {Universal ViT Block};

\node (pe) [expr, above of=block, yshift=2.3cm, xshift=-1.2cm, minimum width=4.0cm, font=\large, fill=lavender]{Pos. embedding};
\node (embed_x) [expr, above of=pe, yshift=0.65cm, xshift=0cm, minimum width=4.0cm, font=\large, fill=Gcolor]{Linear projection};
\node (embed_c) [expr, above of=block, yshift=2.75cm, xshift=2.2cm, minimum width=2.0cm, font=\large, fill=peach_light]{MLP};
\draw [arrow, color=black] (pe.south) -- ([xshift=-1.2cm]block.north);
\draw [arrow, color=black] (embed_c.south) -- ([xshift=2.2cm]block.north);
\draw [arrow, color=black] (embed_x.south) -- (pe.north);

\node (patch_x) [expr, above of=embed_x, xshift=0cm, yshift=0.65cm, inner sep=4pt, fill=Ycolor, minimum width=2.7cm, font=\large]{To patches};
\draw [arrow, color=black] (patch_x.south) -- (embed_x.north);

\node (x) [txt, above of=block_embed, yshift=2.6cm, font=\large]{$x(t)$};
\node (c) [txt, above of=embed_c, yshift=0.6cm, font=\large]{$t, C$};
\draw [arrow, color=black] (x.south) -- (block_embed.north);
\draw [arrow, color=black] (c.south) -- (embed_c.north);

\node (vector) [txt, below of=block, yshift=-3.5cm, font=\large]{$v_{\Theta_s}(x(t), t, C)$};
\node (unpatch_v) [expr, below of=block, inner sep=4pt, yshift=-2.0cm, fill=Bcolor, minimum width=2.7cm, font=\large, fill=rose]{Detector specfic head};
\draw [arrow, color=black] (block.south) -- (unpatch_v.north);
\draw [arrow, color=black] (unpatch_v.south) -- (vector.north);

\node (block_2) [block_blue_light, right of=block, text width=4cm, xshift=9cm, 
text depth=6.6cm, align=center, minimum height=7.5cm,font=\large] {Patch operations};
\node (block_aff) [block_blue, above of=block_2, yshift=0.6cm, font=\large] {Modulation};
\node (block_atn) [block_blue, above of=block_2, yshift=-0.75cm, font=\large] {Self-attention};
\node (block_aff_2) [block_blue, below of=block_2, yshift=-0.30cm, font=\large] {Modulation};
\node (block_atn) [block_blue, below of=block_2, yshift=-1.7cm, font=\large] {Non-linear \\ transformation};

\draw [dashed, color=black] ([xshift=3.0cm]block.north) -- ([xshift=-3.0cm]block_2.north);
\draw [dashed, color=black] ([xshift=3.0cm]block.south) -- ([xshift=-3.0cm]block_2.south);

\node (block_d1) [smallblock_blue_light, above of=block_2, xshift=-3cm, yshift=5.0cm, minimum height=1.0cm, minimum width=1.5cm, font=\large] {};
\node (block_d2) [smallblock_blue_light, xshift=1.5cm, right of=block_d1, minimum height=1.0cm, minimum width=1.5cm, font=\large] {};
\node (block_d4) [smallblock_blue_light, xshift=2.0cm, right of=block_d2, minimum height=1.0cm, minimum width=1.5cm, font=\large] {};

\node (block_ds1) [smallblock_blue_light, above of=block_d1, yshift=0.50cm, minimum height=1.0cm, minimum width=1.5cm, font=\large, fill=silver] {det. 1};
\node (block_ds2) [smallblock_blue_light, xshift=1.5cm, right of=block_ds1, minimum height=1.0cm, minimum width=1.5cm, font=\large, fill=silver] {det. 2};
\node (block_ds4) [smallblock_blue_light, xshift=2.0cm, right of=block_ds2, minimum height=1.0cm, minimum width=1.5cm, font=\large, fill=silver] {det. n};

\node (block_dh1) [smallblock_blue_light, below of=block_d1, yshift=-0.1cm, minimum height=0.5cm, minimum width=1.5cm, font=\large, fill=silver] {};
\node (block_dh2) [smallblock_blue_light, xshift=1.5cm, right of=block_dh1, minimum height=0.5cm, minimum width=1.5cm, font=\large, fill=silver] {};
\node (block_dh4) [smallblock_blue_light, xshift=2.0cm, right of=block_dh2, minimum height=0.5cm, minimum width=1.5cm, font=\large, fill=silver] {};

\draw [loosely dotted, arrow, color=black] (block_ds1.south) -- (block_d1.north);
\draw [loosely dotted, arrow, color=black] (block_ds2.south) -- (block_d2.north);
\draw [loosely dotted, arrow, color=black] (block_ds4.south) -- (block_d4.north);

\draw [loosely dotted, arrow, color=black] (block_d1.south) -- (block_dh1.north);
\draw [loosely dotted, arrow, color=black] (block_d2.south) -- (block_dh2.north);
\draw [loosely dotted, arrow, color=black] (block_d4.south) -- (block_dh4.north);

\draw [loosely dotted, line width=0.5mm, color=black] ([xshift=0.5cm]block_d2.east) -- ([xshift=-0.5cm]block_d4.west);
\end{tikzpicture}}
    \caption{Schematic diagram of the vision transformer~\cite{peebles2023scalablediffusionmodelstransformers}, which highlights the detector-specific and the universal part of the architecture. The color coded detector-specific steps (see text for more details) indicate the components which may be reinitialized during fine-tuning. The universal ViT block only contains learnable transformations at patch-level objects. These weights, trained on a large corpus of data, can learn general features of calorimeter showers which are used as initializations for other detectors.}
    \label{fig:vit-diagram}
\end{figure}
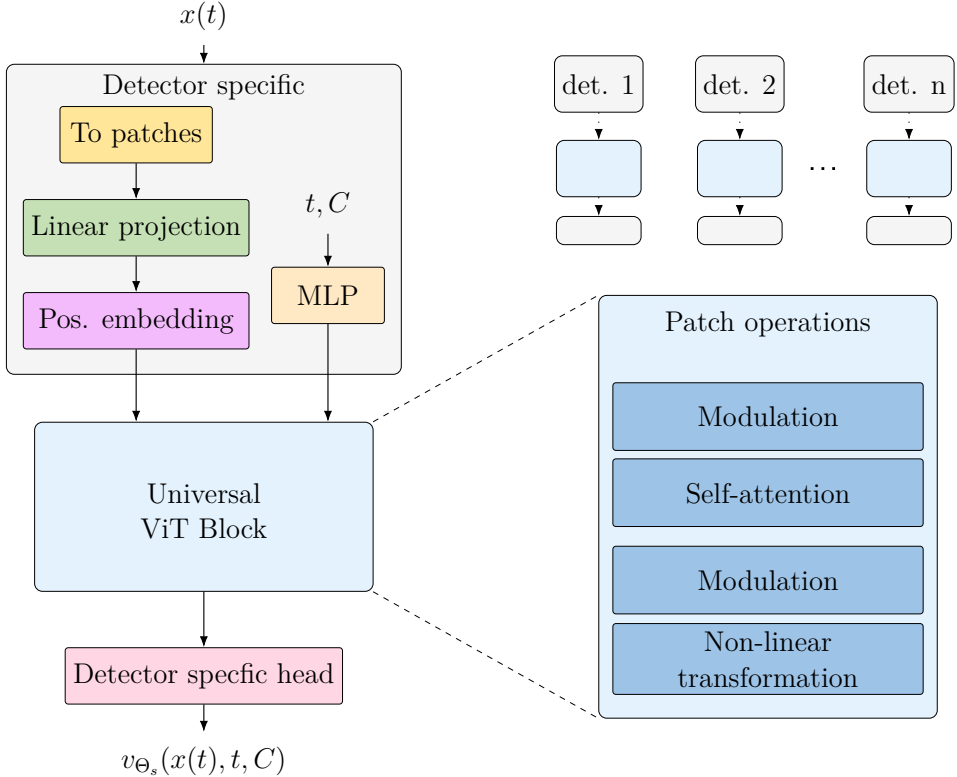

The shape network contains the majority of the learnable weights and
constitutes the most expensive component in terms of training data and resources.
We use a Vision Transformer (ViT), based on~\cite{peebles2023scalablediffusionmodelstransformers}, extended to operate
on three-dimensional inputs.
A ViT divides the calorimeter into patches, each of which contains an exclusive set of voxels.
The large-scale architecture consists of a series of transformer blocks which perform
a residual self-attention operation between embedded patches, followed by a dense network.
This part of the neural network can be considered ``universal'' as it handles a variable
number of patches, and it is common to any detector.
The remaining elements of the ViT serve to embed the patches and the conditions in a 
common latent space, and to map the final representation to the predicted velocity field 
$v_{\Theta_s}(x(t), t, C, u)$.
Before entering the universal block, we first create the patches and then embed them in
a latent vector with a simple linear projection. Two more embedding networks perform the same
operation on the physical conditions $(u, C)$ and time step $t$.
More details on the embedding steps of the input voxels are discussed in the following section.
These two latent vectors are added to construct a single conditional vector, 
then encoded in the transformer blocks through learnable affine transformations.
The modulations, self-attention, and the non-linear operations in the universal block closely follow the description presented in~\cite{Favaro:2024rle}. \Cref{fig:vit-diagram} shows a visualization of the main components of the shape network and its universal block.

\subsubsection*{Patching irregular geometries.}
\begin{figure}
    \centering
    \includegraphics[width=0.9\linewidth]{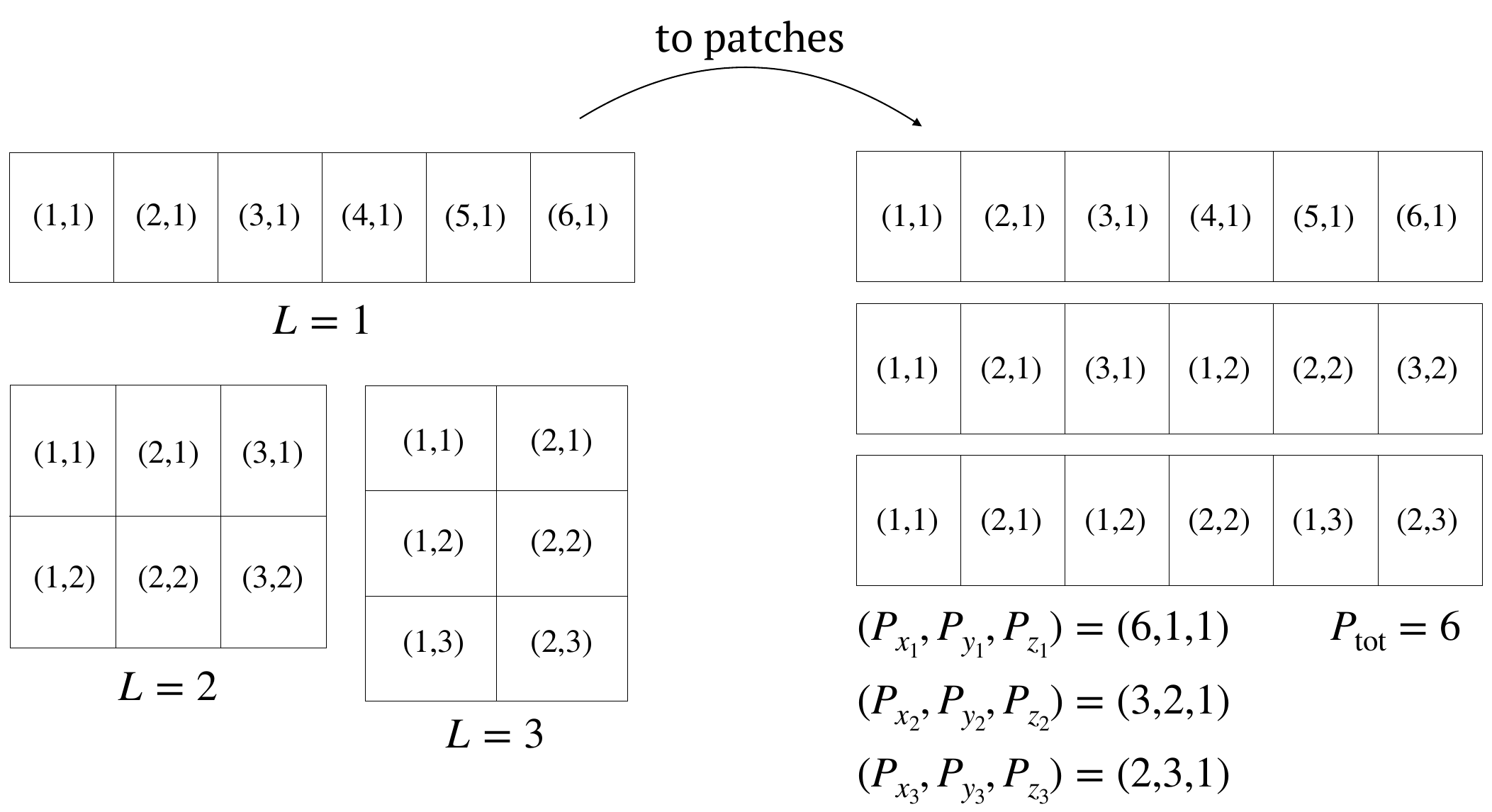}
    \caption{Example transformation to patches for a simple geometry with three layers and six voxels. Each layer uses a different grid $(P_{x_i}, P_{y_i}, P_{z_i})$, with the appropriate positional encoding indexing, as defined in~\cref{sec:ml}. The indices associated to each voxels are used to define the positional embedding.}
    \label{fig:patching}
\end{figure}
Representing a shower as a regular grid introduces inefficiencies when modeling 
calorimeters with high granularity. The required spatial resolution translates into
a large number of voxels, most of which have no energy deposition.
A more computationally effective approach optimizes the geometry in the detector
such that there are no biases in the downstream analysis while minimizing the number
of voxels~\cite{ATL-SOFT-PUB-2025-003}. Such optimization results in an irregular
grid of voxels in the direction of propagation of the shower.
The structure of a ViT is not limited to a regular grid. We show how a ViT is extended
to irregular geometries as long as a function which transforms voxels into patches 
can be defined.
The CaloChallenge-ds1~\cite{CaloChallenge_ds1_v3} and CaloHadronic~\cite{Buss:2025cyw} datasets are examples of geometries with varying
number of voxels per layer, or sub-detector.
We handle these cases by defining the length of a patch $P_\text{tot}$ and allowing for grouping
adjacent voxels according to the calorimeter grid. If there are $N$ different grids, we
define the patch sizes
\begin{equation}
    \{(P_{x_i}, P_{y_i}, P_{z_i})\}_{i=0}^{N}\;, \qquad\text{such that}\qquad P_{x_i}P_{y_i}P_{z_i} = P_\text{tot} \;,
    \label{eq:patches}
\end{equation}
for each set $i$. Here, the coordinates $(x,y,z)$ characterize the geometry and $P_k$ represents
the number of voxels selected for the $k$ coordinate. For a detector with $N_\text{tot}$ number of voxels, the total number of patches is $N_\text{patches} = \frac{N_\text{tot}}{P_\text{tot}}$.

Since transformers are permutation-invariant, the next embedding step consists of
applying a positional embedding to the created patches~\cite{vaswani2017attention}. This makes the transformer aware
of the position of the patches, and hence of the spatial location of the calorimeter cells that
are part of the patch.
We show an example of patching an irregular geometry in~\cref{fig:patching}. It shows a toy detector with three layers, each with six voxels but organized differently in the $(x,y)$-plane. We define the patch dimensions as defined in~\cref{eq:patches}, which leads to $N_\text{patches}=3$. In the more general case of different number of voxels per layer, the number of patches will vary with more patches dedicated to the largest part of the detector.
Such positional embedding should also respect the varying geometry and the corresponding
patching. Therefore, we introduce a three-dimensional sine embedding with learnable
frequencies.
Let $\left(X_{n_i}, Y_{n_i}, Z_{n_i} \right)$ be the number of patches along each direction 
for grid $i$, to incorporate spatial information into the transformer, 
we construct a three-dimensional positional encoding that respects the heterogeneous
detector layout.
For each grid, we define a cumulative coordinate in the depth direction
\begin{equation}
    z\in \left\{ 0, \frac{1}{L}, \ldots, \frac{L-1}{L}\right\}, \qquad L = \sum_{i=0}^{N} Z_{n_i} \; ,
\end{equation}
and two local coordinates in the transverse directions
\begin{equation}
    x\in \left\{ 0, \frac{1}{X_{n_i}}, \ldots, \frac{X_{n_i}-1}{X_{n_i}}\right\}, \qquad \
    y\in \left\{ 0, \frac{1}{Y_{n_i}}, \ldots, \frac{Y_{n_i}-1}{Y_{n_i}}\right\} \;.
\end{equation}
Finally, the 3D meshgrid $(x, y, z)$ is multiplied with a set of learnable frequencies $\omega_d$
initialized from a Gaussian distribution,
\begin{equation}
    \omega_d = 2\pi f_d \;, \quad f_d \sim \normal(0,1)\;,\quad f_d \in \mathbb{R}^{D/6} \;,
\end{equation}
where $D$ is the latent dimensionality of a single patch.
Therefore, the positional embedding vector, added to the inputs, is 
\begin{equation}
    \mathrm{P} = \left[ \sin(xw^T), \cos(xw^T), \sin(yw^T), \cos(yw^T), \sin(zw^T), \cos(zw^T)  \right] \;.
\end{equation}

\subsubsection*{Fine-tuning of pretrained networks.}
In the fine-tuning setting, our goal is to train a neural network on a large
dataset and ``finetune'', in a second training step, on a smaller dataset sampled from the 
target distribution.
A successful fine-tuning will show improved performance compared to a network trained from scratch at fixed resources, either the size of the training dataset or number of training iterations.
If the pretraining and target detectors are the same, the fine-tuning step is
straightforward: all pretrained weights can be preserved, and only the optimization
continues on the target dataset.
However, if the detector geometry differs between the pretraining and fine-tuning datasets,
several components of the architecture may no longer match in dimensionality.
This makes part of the pretrained parameters incompatible and requires careful
reinitialization.
In practice, we have to address the following:
\begin{itemize}
    \item \textcolor{dforestgreen}{Embedding layer}: the optimal patch size can be different between datasets and we have to realign the training. We find a simple interpolation between the original dimensionality and the final one to work well in practice;
    \item \textcolor{deepterracotta}{Embedding of the conditions}: the energy ratios and the global conditions may follow
    a different distribution depending on the incident particle, the detector specifics,
    and the incident energy. We devise a preprocessing which keeps the boundary
    of the distributions unchanged. This choice minimizes the distributional shift
    in the fine-tuning step; 
    \item \textcolor{darkcherry}{Final layer}: a change in the dimensionality of the final velocity vector requires a reinitialization of the final ``head'' layer. This step is similar to foundation models~\cite{Hallin:2025ywf} where the final head contains few learnable parameters and spatial information, and can be retrained quickly, while the pretrained transformer backbone provides a strong inductive bias;
    \item \textcolor{deeppurple}{Positional embedding}: the learnable position embedding is reinitialized if the total number of patches changes. In particular, we recompute the spatial grid and
    reinitialize the learnable frequencies.
\end{itemize}
The specific choices adopted in the various datasets are discussed in the corresponding 
result sections. In general, we train reinitialized layers with a learning rate five times larger than the base one.
We posit that general features, e.g the sparsity, of calorimeter showers are encoded
in the large ViT backbone during pretraining. The second fine-tuning step can leverage
these general aspects of calorimeter showers for a more efficient training
dynamic. Even if embedding and final layers may have to be reinitialized, 
the majority of the learnable weights ($>$99\%), which are still optimized during fine-tuning, are contained in the universal backbone.
The same network architectures can be used to train normalizing flows, for faster generation at the cost of accuracy~\cite{Favaro:2025ift}.

\section{Datasets}
\label{sec:data}
This section contains a description of the datasets used for our studies. 
We first benchmark the universal backbone architecture on well-known regular
datasets. Then, we move on to modeling multiple detectors with a single network, irregular geometries, and full detectors composed of an electromagnetic and hadronic calorimeter. We provide a visualization of the geometries in~\ref{app:datasets}.
We define a common architecture and train an energy and shape network for each
dataset.

\subsection{Regular geometries}
\subsubsection*{CaloChallenge datasets 2/3.}
The public datasets~\cite{CaloChallenge_ds1_v3,CaloChallenge_ds2,CaloChallenge_ds3}
used for the Fast Calorimeter Simulation Challenge~\cite{Krause:2024avx} are the ideal testbeds since they provide
the latest comparison to multiple generative networks.
Our starting points are the regular geometries developed for 
dataset-2 and 3. The simulation contains incoming electrons interacting
with layers of alternating active silicon (thickness
0.3~mm) and inactive tungsten absorber layers (thickness
1.4~mm) at $\eta=0$. The energy in the absorber layers is voxelized
in a space with 45 layers, with binning in the radial, $r$,
and angular, $\alpha$, directions.
Dataset-2 contains a total of 144 bins per layer, divided into $16 \times 9$ angular and radial voxels, while dataset-3 contains a total of 900
voxels per layer arranged into a $50 \times 18$ spatial grid.
For both datasets, the minimum readout energy, which is also used
as a threshold for calculating the sparsity, is $x_\text{th}=15.15$~keV.
The initial condition of the shower is characterized solely by its incident energy, which is log-uniformly distributed in $E_\text{inc} = 1~...~1000$~GeV.

\subsubsection*{LEMURS dataset.}
The LEMURS dataset~\cite{McKeown:2025gtw} is an
extended set built upon the CaloChallenge-ds2. It contains showers produced
by a single particle interacting with a voxelized representation of a 
calorimeter. Similarly to the CaloChallenge, the voxelization consists of a
cylinder with segmentation in the depth direction $z$ and in the transversal polar
coordinates $(r, \alpha)$. The LEMURS dataset contains a total of 5M showers evenly divided into
five different detectors: Par04SiW, Par04SciPb, ODD, FCCeeCLD, and FCCeeALLEGRO.
The Par04 geometry is the starting point for the CalChallenge-ds2/3 studies and,
in this dataset, is simulated with two possible compositions as active and
passive materials: silicon-tungsten (SiW) and scintillators-lead (SciPb).
The other three detectors are more realistic and are taken from the Open
Data Detector~\cite{corentin_allaire_2022_6445359} (ODD), and two detector proposals for the Future Circular Collider~\cite{FCC:2018byv}, namely
the CLIC-like detector~\cite{Bacchetta:2019fmz} (CLD) and the ALLEGRO~\cite{Mlynarikova:2025skz} designs. The full description of the detector
geometry can be found in~\cite{McKeown:2025gtw}.
The size and breadth of detectors contained in the LEMURS dataset make it
the prime candidate for studies on transfer learning and multi-purpose
training of generative networks.

Showers entering the detectors are simulated at different incident energies and
detector locations. The information of the incoming particle is described by
the incident energy $E_\text{inc}$ and the polar and azimuthal angles 
$(\phi, \theta)$ in the detector reference frame.
The global conditions are independently sampled from 
\begin{align}
E_\text{inc}&\sim \mathcal{U}(1, 10^3)\,\text{GeV}, \qquad \cos\theta \sim \mathcal{U}(\cos(0.87), \cos(2.27))\,, \notag \\ 
\quad \text{and} &\quad \phi\sim\mathcal{U}(-\pi, \pi)\, \text{rad}\;,
\end{align}
where $\mathcal{U}(a,b)$ is a uniform distribution with support $[a,b]$.
\subsection{Irregular geometries}
\subsubsection*{CaloChallenge dataset 1.}
\begin{table}[b!]
    \centering
    \begin{small} \begin{tabular}{l|c|ccccc}
      \toprule
      $E_\text{inc}$ & 256~MeV~...~131~GeV & 262~GeV   & 0.524~TeV  & 1.04~TeV  & 2.1~TeV   & 4.2~TeV  \\
      \midrule
      photons       & 10000 per energy & 10000  & 5000  & 3000  & 2000  & 1000 \\
      pions         & 10000 per energy & 9800   & 5000  & 3000  & 2000  & 1000 \\
      \bottomrule
    \end{tabular} \end{small}
    \caption{Sample sizes for different incident energies in
      dataset~1.}
    \label{tab:Einc.ds1}
\end{table}
The two remaining datasets of the CaloChallenge are of lower dimensionality but geometrically irregular. They provide calorimeter showers for central photons and
charged pions originally used in
AtlFast3~\cite{ATLAS:2021pzo}. The detector geometries for photon and pion showers have
five and seven layers, respectively,
totaling 368 voxels for photons and 533 voxels for pions.
The incident energies of both photons and pions are 15 discrete
values with increasing power of two, namely $E_\text{inc} = 256~\text{MeV}~...~4.2~\text{TeV}$. 
The energy $E_\text{inc}$ refers to the momentum of the particles; this
has implications for charged pions at low energy, which have non-negligible mass.
Due to the \geant generation time constraints, the sample sizes for larger energies are smaller.
Details are given in~\cref{tab:Einc.ds1}.

The CaloChallenge-ds1s are a special example of irregular geometries, since there are layers
with a single angular bin, see~\cref{fig:datasets}.
In this case, we add a minimal number of bins to reach the required patch size dimension.
For both datasets, we use a patch size of five and, therefore, add four additional bins
per radial voxel in the layers without angular sectioning.
These additional bins do not carry energy and are part of a single patch in the neural
network. Therefore, they do not introduce biases in the encoding and the overall generation
process. 
After generation, they are collapsed into a single bin with energy equal to the maximum
generated energy deposition. This approach mildly increases the number of voxels,
but we expect it not to occur for high-granular geometries where 
a common divisor is easier to find.
We train a vision transformer on the ds1-$\gamma$ and ds1-$\pi^+$ datasets.
The increase in the
total number of voxels is of a factor $\sim 1.19$ for ds1-$\gamma$ and $\sim 1.17$ for ds1-$\pi^{+}$.

\subsubsection*{CaloHadronic dataset.}
Originally generated for~\cite{Buss:2025cyw}, this dataset 
contains showers produced from an incident $\pi^{+}$.
Unlike the cylindrical geometry of the other datasets,
the detector is represented in cartesian coordinates, and it corresponds to
the high-granular calorimeter proposed for the International Linear
Collider~\cite{ILC:2019gyn}. The incoming particle is orthogonal to the 
calorimeter, and it carries incident energy sampled from
\begin{equation}
    E_\text{inc} \sim \mathcal{U}(10, 90)\,\text{GeV} \;.
\end{equation}
The main challenges coming from this dataset are the 
extremely high granularity, the separation between the electromagnetic (ECal)
and hadronic (HCal) calorimeters,
and the intrinsic complex nature of hadronic showers.
Our aim is to showcase a more realistic example where different
granularities can arise and show how a ViT can deal with them.
Therefore, we only use 100k showers to avoid
the computationally expensive training on the full dataset. 
We further simplify the generation task in two ways.
First, in~\cite{Buss:2025cyw} the original calorimeter is up-sampled
by a factor of three in each of the $x$-$y$ axes, thus increasing the generation
space by a factor of nine.
We avoid the complication of blindly increasing the number of voxels and use
the original cell size for the HCal.
Therefore, the HCal contains 43200 voxels organized as $(x, y, z)=(30, 30, 48)$,
where each cell has a lateral width of 30 mm.
Second, the ECal section is much more granular and sparse.
Instead of using the original ECal voxelization $(x, y, z) = (180, 180, 30)$,
we down-sample with a sum-pooling operation down to 2250 voxels organized
as $(x, y, z) = (15, 15, 10)$, where each cell has size $5.1\times12$ mm.
Here we do not perform the up-sampling operation which can be learned, together with the down-sampling step, with an autoencoder-like structure, especially if the showers are incredibly sparse~\cite{Ernst:2023qvn,Favaro:2024rle,Toledo-Marin:2024gqh}.
We use 80k showers for training and validation, and 20k showers for testing.

We set the total patch size to $P_\text{tot}=75$ with segmentation
$P_\text{ecal}=(5, 5, 3)$ and $P_\text{hcal}=(3, 5, 5)$ for the electromagnetic and
hadronic calorimeter, respectively. The flexibility in the patch selection allows for the straightforward transformation to patch space embeddings without adding unphysical layers which would be
needed to divide the detector into patches with a single segmentation.

\section{Evaluation metrics}
\label{sec:metrics}

\subsubsection*{High-level features.}
The simplest measure for the fidelity of the generated calorimeter showers
is the definition of high-level observables. Following~\cite{Krause:2024avx}, we calculate a large set of observables that characterize the shape of showers, 
together with energy-related observables
that instead focus on the overall energy distribution across layers.
The energy-related observables are:
\begin{itemize}
    \item the total energy deposited in the calorimeter obtained as the sum of all the energy deposits, $E_\text{tot}$, divided by the incident energy of the incoming particle, $E_\text{inc}$;
    \item $E_i$: the total energy deposition in layer $i$;
    \item shower profiles in the depth or transverse direction: we calculate the average energy deposition over the dataset at each layer in the $z$ direction,
    \begin{equation}
    \langle E_i \rangle_{x} = \frac{\sum_{j=1}^N E_{ij}}{N} \;,
    \end{equation}
    and each radius $r_i$,
    \begin{equation}
    \langle E(r_i) \rangle_{x} = \frac{\sum_{j=1}^N E(r_i)_j}{N} \;,
    \end{equation}
    where $E(r_i)$ indicates the sum of the energy depositions at radius $r_i$, the index $j$ runs over the showers and $N$ is the size of the sample.
\end{itemize}
Shower shape observables are sensitive to the distribution of energy within
one layer, hence these are useful to evaluate the large ViT shape network.
For example, our set includes:
\begin{itemize}
    \item $\langle \xi \rangle$: the shower center of energy calculated from the energy deposits $x_i$ along the axis $\xi$ in the physical space of the detector, defined as
    \begin{equation}
    \langle \xi \rangle = \frac{\sum_i\xi_i\cdot x_i}{\sum_i x_i} \;.
    \end{equation}
    \item $\sigma_{\langle \xi \rangle}$: the width of the center of energy calculated from the same quantities and defined as 
    \begin{equation}
    \sigma_{\langle \xi \rangle} = \sqrt{\frac{\sum_i\xi_i^2\cdot x_i}{\sum_i x_i} - \langle \xi \rangle^2} \;.
    \end{equation}
\end{itemize}
We calculate these observables to define either a set of features sensitive to layer-wise mismodelings or global shapes for the entire geometry. 
Additionally, we compute the sparsity of the showers defined as the active voxels with energy deposition $x_i > x_\text{th}$.

\subsubsection*{Distance-based metrics.}
Distance-based metrics like the Kernel-Physics-Distance (KPD) and Fréchet-Physics-Distance (FPD) were introduced in~\cite{Kansal:2022spb} as a contribution to the CaloChallenge \cite{Krause:2024avx} and are based on metrics used in computer science, the Fréchet Inception distance (FID)~\cite{NIPS2017_8a1d6947} and kernel Inception distance (KID)~\cite{2018arXiv180101401B}. The Fréchet distances are the Wasserstein 2 distances of the multivariate Gaussians fitted to features of generated and reference data. For images, the features are taken from the penultimate layer of a pretrained Inception-V3~\cite{2015arXiv151200567S} classifier. In the physics case, they are taken as the high-level features defined earlier. Starting from the means $\mu_i$ and covariances $C_i$ of the generated and reference data, the Wasserstein 2 distance is defined as
\begin{equation}
    \text{FPD}(\mu_1, C_1,\mu_2, C_2) = | \mu_1 - \mu_2|^2 + \mathrm{Tr}\left(C_1 + C_2 - 2 \sqrt{(C_1 C_2)}\right).    
\end{equation}
The kernel distances work on the same features (pretrained classifier activations or physical, high-level observables) and determine the distance between the generated and reference set using the kernel-based Maximum-Mean-Discrepancy (MMD). Since it tends to correlate rather strongly with the Fréchet distances~\cite{Krause:2024avx}, we report only the FPD scores in our discussion below.

Our evaluation uses the implementation of FPD as provided by the JetNet package~\cite{Kansal_JetNet_2023}, with the same hyperparameter settings as used in the CaloChallenge \cite{Krause:2024avx} evaluation.

\subsubsection*{Neural classifiers.}
Classifiers are the most powerful evaluation metrics on the
market. Given samples from two distributions, a neural classifier approximates
the likelihood ratio, hence the optimal statistic for a simple two-hypothesis
test, between the two. A simple 1D metric which can be extracted from a 
classifier is the area under the curve (AUC) score~\cite{Krause:2021ilc}.
Alternatively, a full phase space reweighting function can be estimated
from the histogram of the learned approximate likelihood ratio~\cite{Das:2023ktd}.
Although a powerful metric, the optimal training of a classifier is hard to 
achieve, especially in high-dimensional spaces. To mitigate this issue,
we train multiple classifiers as done in~\cite{Krause:2024avx}.
We include a ``high-level'' classifier trained on a set of high-level
features, a ``low-level'' classifier trained on the full set of voxels, and a ``ResNet'' convolutional architecture, which takes as inputs the same 
low-level information, but it has a stronger inductive bias towards
identifying spatial mismodelings. Our implementation of the ``ResNet'' 
classifier allows training and evaluation only on regular geometries.
All classifiers are trained to distinguish between a test \geant and a generated sample with the same number of showers. The input features, the classifier hyperparameters, and the best network selection follow
the prescription used in the CaloChallenge~\cite{Krause:2024avx}.

\subsubsection*{Generation time.}
In a fast detector simulation setting, the optimal working point for a 
generative surrogate also depends on the generation speed of the neural 
network. Since this metric depends on the architecture itself but also on the size
of the detector, i.e. the number of generated patches, it is important to measure it for each dataset.
We measure the generation time both on CPU and GPU. On GPU, we assume
that it is possible to parallelize the event generation and use a reference 
batch size of 100. The timing includes overheads from the initialization
of the CUDA kernels and the time required to move each batch from GPU back
to CPU. For all the tests, we use a single NVIDIA A100 GPU. 
Even though differentiable emulators greatly benefit from GPU parallelism,
we report the CPU generation time as well. The CPU used is an AMD EPYC Zen 3 Milan, from which we allocate a single core and thread.

\section{Results}
\label{sec:results}
We organize the results as follows. First, we set the baseline performance
of CaloDREAM++ on regular geometries and demonstrate its
capabilities on irregular geometries. We train these networks from random
initializations. Then, for the transfer learning studies, we select the networks trained on the LEMURS dataset and fine-tune on the CaloChallenge-ds2/3, and CaloHadronic. We also propose a superresolution study from the CaloChallenge-ds2 to CaloChallenge-ds3.
Details on the parameters of the neural network are described in~\labelcref{app:hyperparams}.

\subsection{Regular geometries}
\label{sec:baseline}

\subsubsection*{CaloChallenge datasets 2/3.}

Starting from the performance of CaloDREAM on the CaloChallenge datasets, already studied in~\cite{Favaro:2024rle}, 
we provide here a comparison to the CaloDREAM++ described in the
previous section. A detailed description of the datasets and their evaluation have already been done in~\cite{Krause:2024avx}. Therefore, we discuss the generation performance only in terms
of neural network classifiers, since they provide stronger discrimination performance, FPD metric, and generation time. The high-level features histograms are provided as supplementary material\footnote{Figures and generated samples are available at~\href{https://doi.org/10.5281/zenodo.18071948}{10.5281/zenodo.18071948}}.

In~\cref{tab:ds23_auc} (left), we summarize the AUC scores from the three neural network
classifiers.
As expected, we reproduce similar results to the original CaloDREAM for ds2. For ds3,
we improve the performance for all three classifiers, especially for the low-level one.
In~\cite{Favaro:2024rle}, fully training the network for ds3 exceeded the available computational
resources, limiting the size of the neural network and inhibiting complete convergence.
We instead fix the architecture of the ViT and adjust the size of one patch to reduce the
number of embedded patches in the transformer layers. We create patches with size $(z, \alpha, r) = (3, 10, 3)$ 
for a total of 450 patches for ds3, which reduces the number of patches by a factor of $\sim$3 
compared to~\cite{Favaro:2024rle}.
Although smaller patches provide better resolution, we find that reducing the training cost while
reaching a better convergence can ultimately improve the generation performance.
The up-to-date generation performance, evaluated in terms of AUC score, shows that the high-level
features are almost indistinguishable from \geant. For ds3, we observe a small deterioration which can be associated to fluctuations of the training procedure. Among the classifiers that look
at the entire shower, only the more advanced ResNet extracts mismodeled generated features.
We remark the improvement over CaloDREAM by including the AUC scores reported in~\cite{Krause:2024avx}.
We observe similar improvements in the FPD metric. In particular,
the CaloDREAM++ samples for ds2 are significantly closer to the \geant reference.

\begin{table}[t]
    \centering
    \begin{tabular}{l@{\hspace{0.5cm}}c@{\hspace{0.5cm}}c@{\hspace{0.5cm}}c@{\hspace{0.5cm}}c@{\hspace{0.5cm}}c}
    \toprule
      &  & \multicolumn{3}{c}{Classifier AUC} & FPD$\times 10^{3}$ \\
      &  & High-level & Low-level & ResNet & \\ \midrule
     \multirow{2}{*}{\geant} & ds2-e$^{-}$ & 0.499(2) & 0.500(2) & 0.500(4) & 10.7(8) \\
      & ds3-e$^{-}$ & 0.500(3) & 0.498(2) & 0.499(2) & 8.7(5) \\ \midrule
     \multirow{2}{*}{CaloDREAM} & ds2-e$^{-}$ & 0.521(2) & 0.531(3) & 0.681(15) & 25(1) \\
        & ds3-e$^{-}$ & 0.524(4) & 0.630(5) & 0.802(14) & 21(1)  \\ \midrule
     \multirow{2}{*}{CaloDREAM++} & ds2-e$^{-}$  & 0.512(1) & 0.516(1) & 0.683(9) & 16.0(5) \\
                   & ds3-e$^{-}$ & 0.538(2) & 0.524(1) & 0.799(9) & 26.3(4) \\ \bottomrule
    \end{tabular}
\caption{Summary of evaluation metrics for the baseline networks on regular geometries. The AUC score of neural classifiers, as defined in the text, and the FPD score confirm the higher fidelity of the CaloDREAM++ network. The given classifier uncertainties are the standard deviations of 5 independent classifier trainings, while we report the FPD uncertainty estimate from a single sample. We compute the \geant reference numbers by comparing the samples in the training and test datasets.}
\label{tab:ds23_auc}
\end{table}
The generation time for CFM networks depends on the number of function
evaluations used to numerically solve the integral in~\cref{eq:cfm_sampling}. We observe that the fidelity of generated
samples reaches a plateau at $\sim$20 function evaluations. In~\cref{tab:ds23_time},
we report the generation time for the full generation and for a single step of the Runge-Kutta 4 (RK4) solver.
The CPU time should be considered as a standard reference number for fully sequential generation. Generation on GPUs fully exploits the 
benefit of modern deep learning, and it shows that generation time can be easily kept below 100 ms. Single-step generation can further reduce
the sampling time with an accuracy tradeoff. We leave the study of strategies for the reduction of the function evaluations to the future.
Here, we observe that the single-step generation suffers from overheads from initialization of the CUDA kernels, as it does not show the expected $20\times$ gain.
\begin{table}[t]
    \centering
    \begin{tabular}{l@{\hspace{0.5cm}}c@{\hspace{0.5cm}}c}
    \toprule
       \multicolumn{3}{c}{ Gen. time  [ms per shower] } \\ \midrule
     CaloChallenge   & ds2-e$^{-}$  & ds3-e$^{-}$ \\ \midrule
     CPU batch 1 (1 RK4 step)        & 8.09(3)$\times 10^{2}$ & 1.39(4)$\times 10^{3}$ \\
     GPU batch 100 (1 RK4 step) & 11.0(2)     & 12.5(5) \\ \midrule
     CPU batch 1 (full gen)          & 5.43(4)$\times 10^{3}$ & 1.630(1)$\times 10^{4}$      \\
     GPU batch 100 (full gen)    & 34(1)    & 96(1) \\ \bottomrule
    \end{tabular}
    \caption{Generation time on the CaloChallenge-ds2/3 datasets on GPU, with batch size 100, and CPU, with batch size 1 and on a single-core and thread machine.}
    \label{tab:ds23_time}
\end{table}

\subsubsection*{LEMURS dataset.}
\begin{figure}[t]
    \centering
    \includegraphics[width=0.325\linewidth]{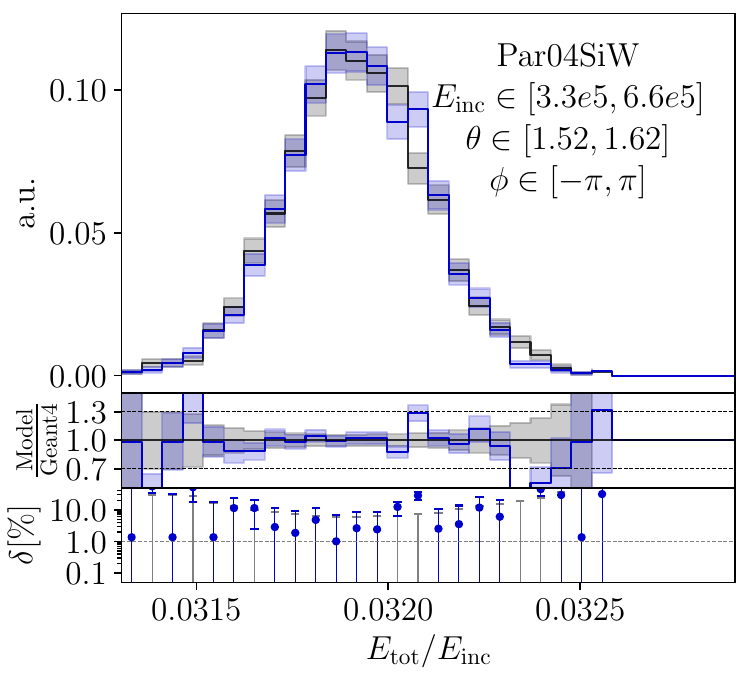}
    \includegraphics[width=0.325\linewidth]{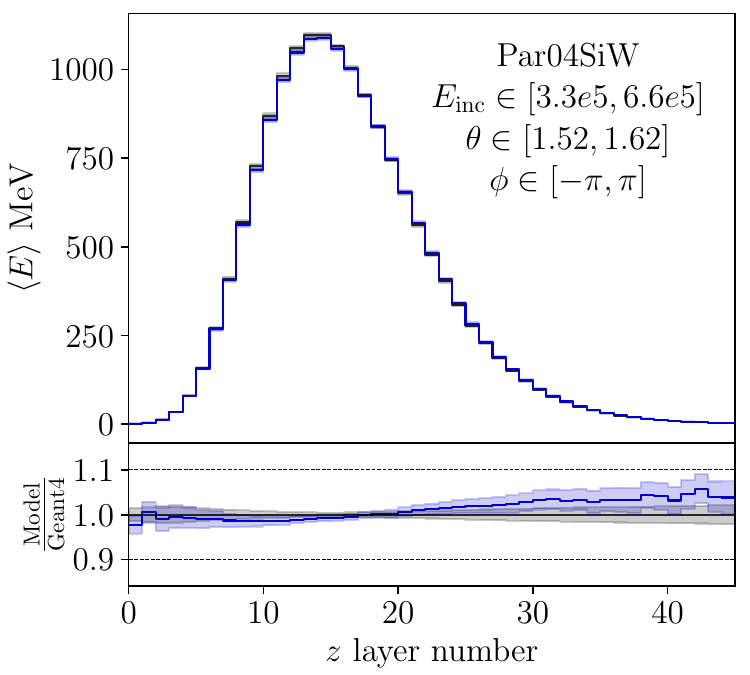}
    \includegraphics[width=0.325\linewidth]{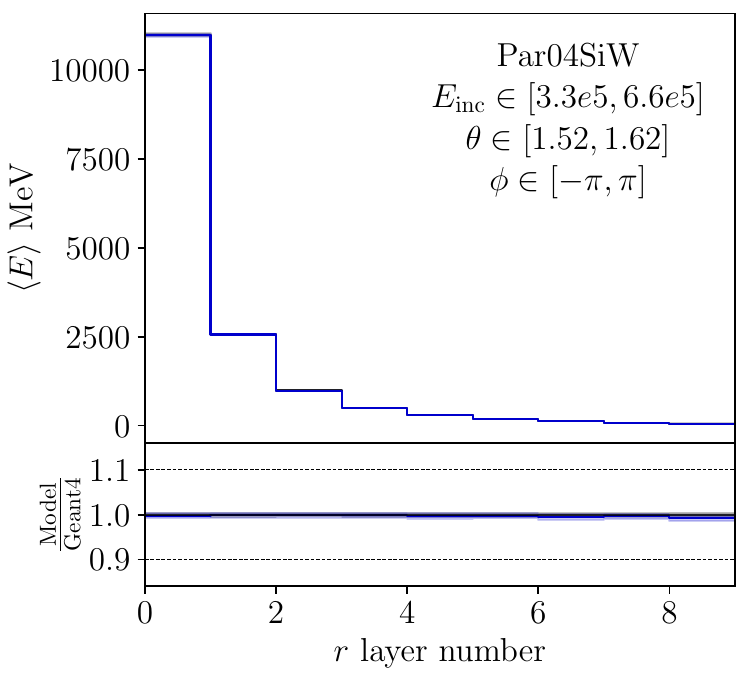} \\
    \includegraphics[width=0.325\linewidth]{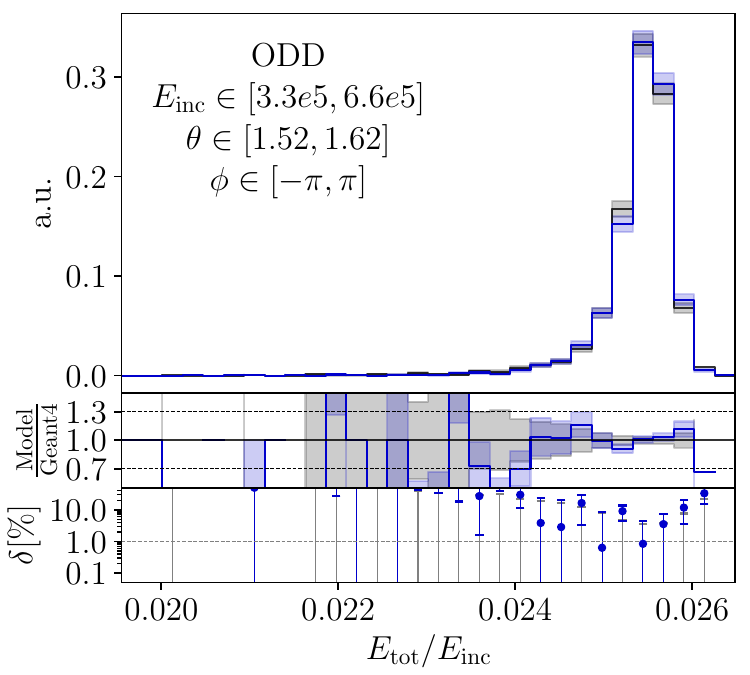}
    \includegraphics[width=0.325\linewidth]{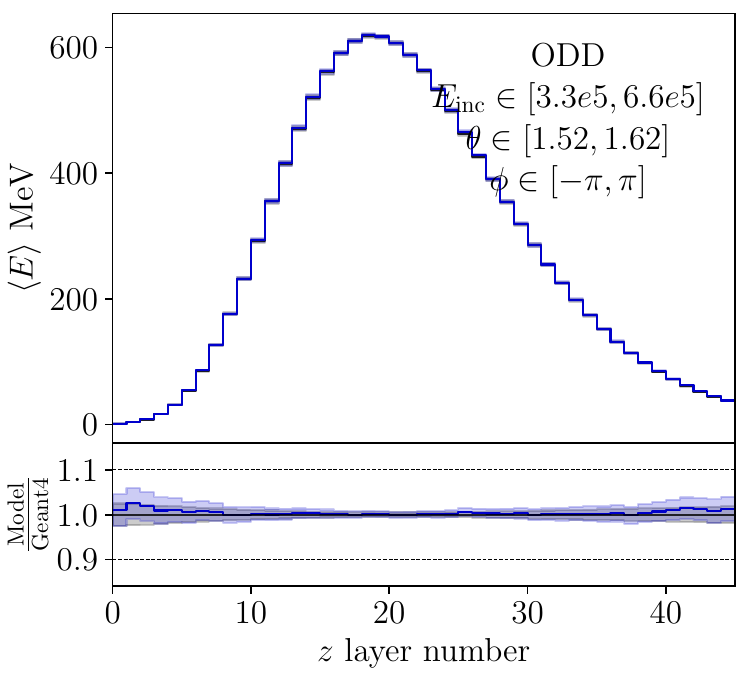}
    \includegraphics[width=0.325\linewidth]{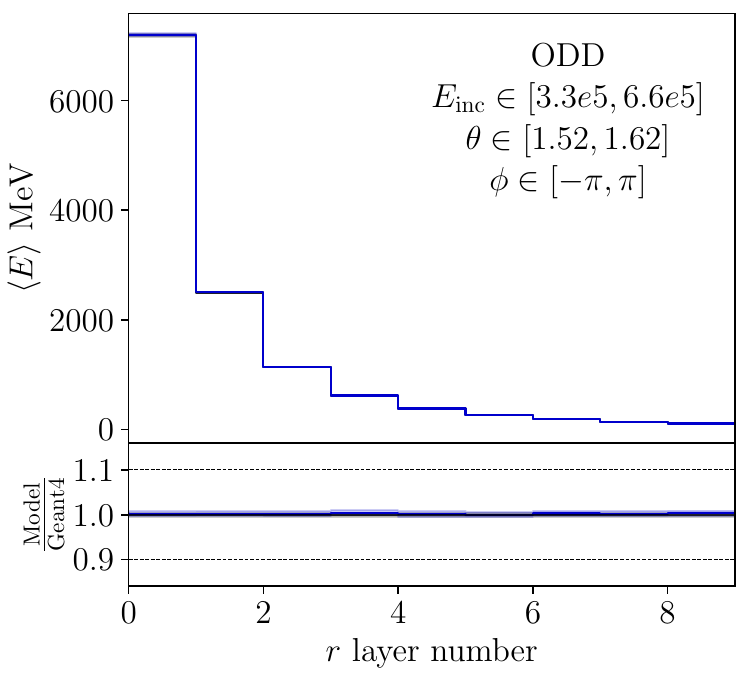} \\
    \includegraphics[width=0.325\linewidth]{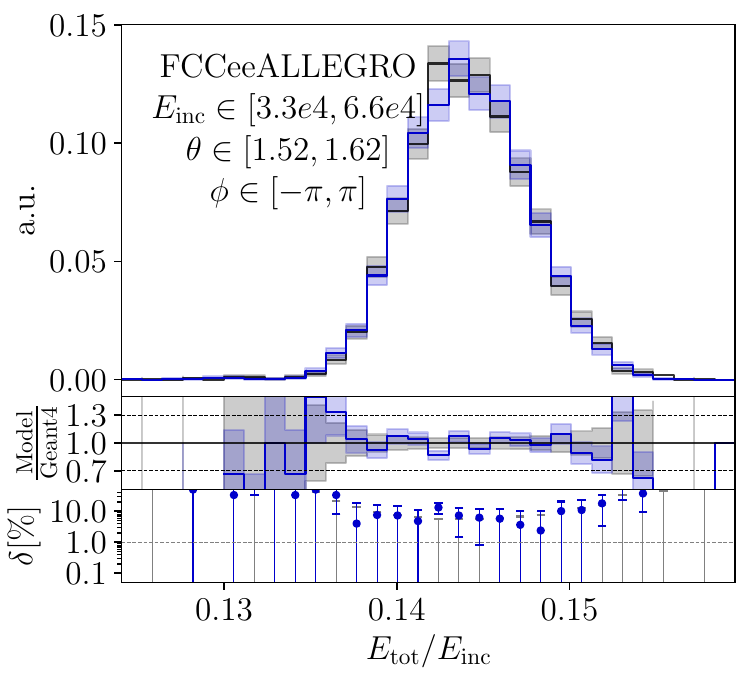}
    \includegraphics[width=0.325\linewidth]{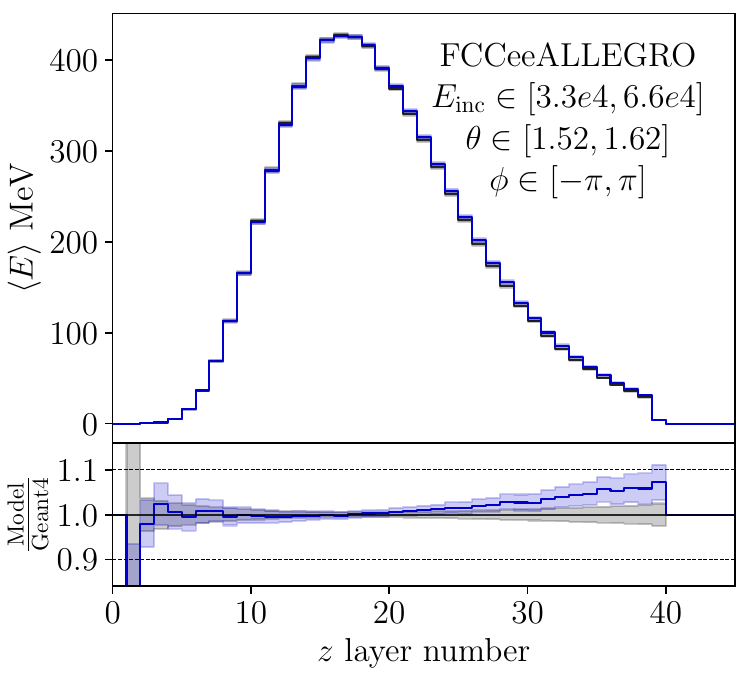}
    \includegraphics[width=0.325\linewidth]{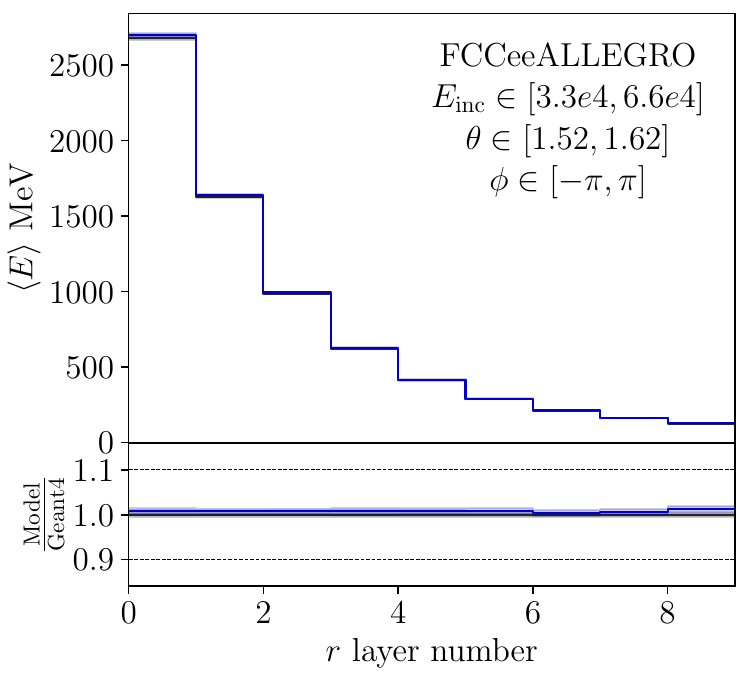} \\
    \includegraphics[width=0.60\linewidth]{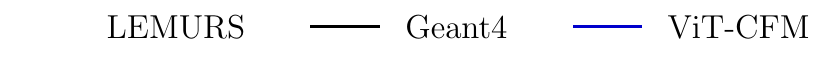}
    \caption{Summary of the sliced evaluation for $E_\text{inc}\in [0.33\cdot E_\text{i-max}, 0.66\cdot E_\text{i-max})$ GeV, $\theta\in[1.52, 1.62)$), where $E_\text{i-max}$ is 100 GeV for the FCC detectors and 1TeV for the Par04 and ODD detectors. We show the visible energy, the energy profile in the z direction, and the energy profile in the radial direction: (top) Par04SiW, (middle) ODD, and (bottom) FCCeeALLEGRO detectors. }
    \label{fig:sliced_eval_wp1}
\end{figure}
\begin{figure}[t]
    \centering
    \includegraphics[width=0.325\linewidth]{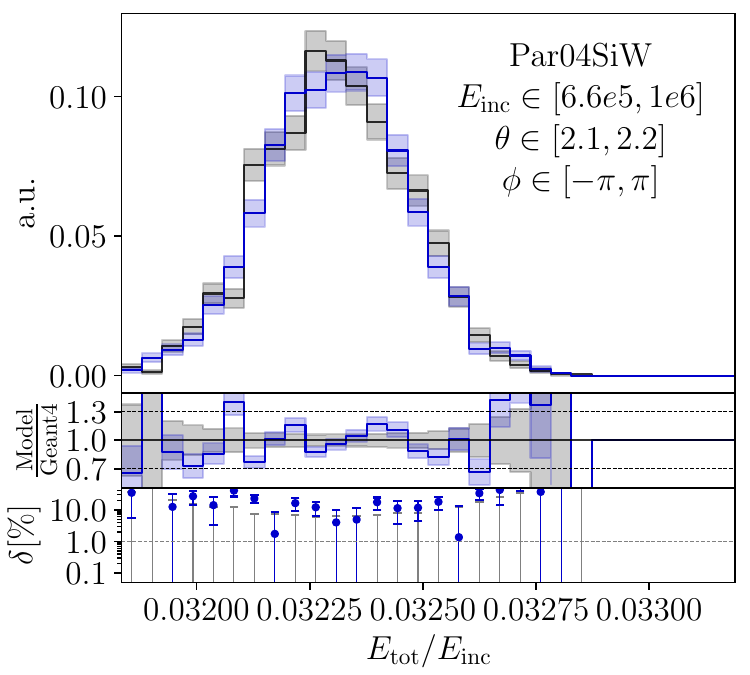}
    \includegraphics[width=0.325\linewidth]{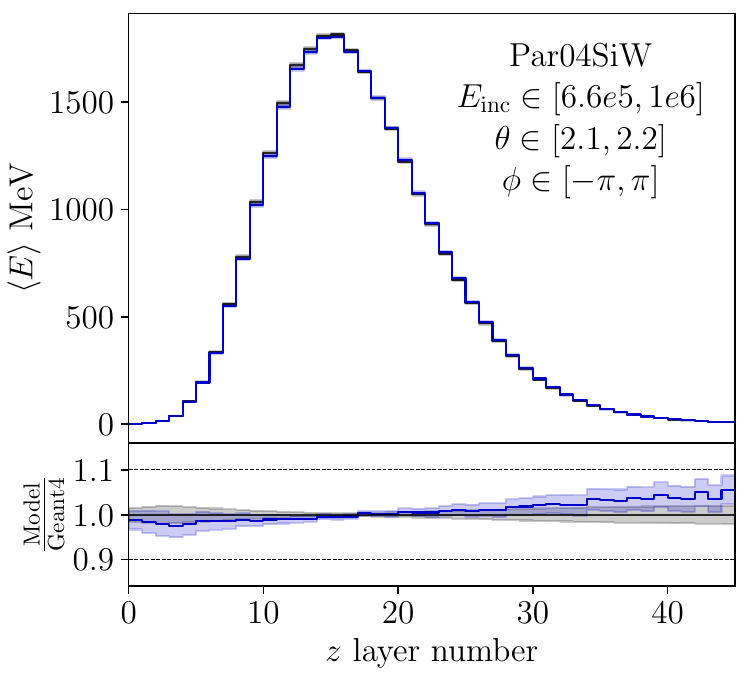}
    \includegraphics[width=0.325\linewidth]{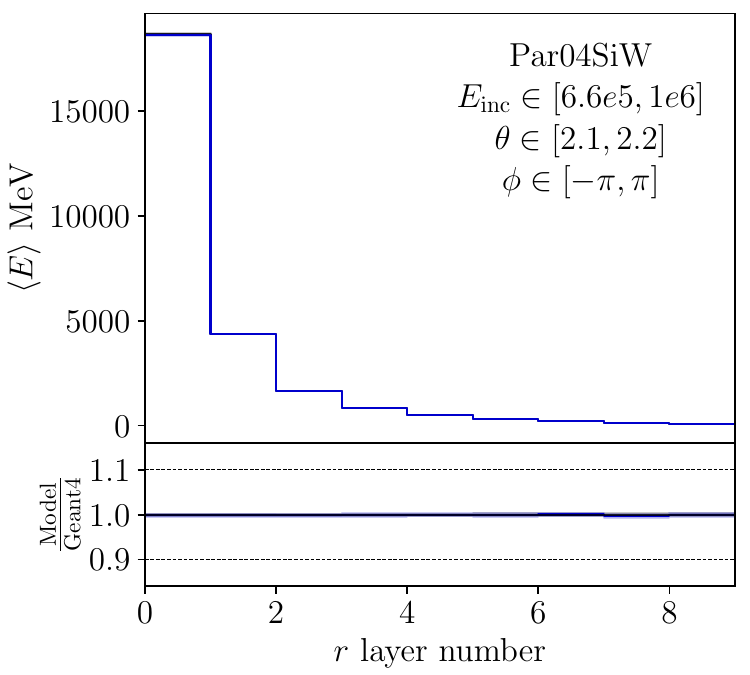} \\
    \includegraphics[width=0.325\linewidth]{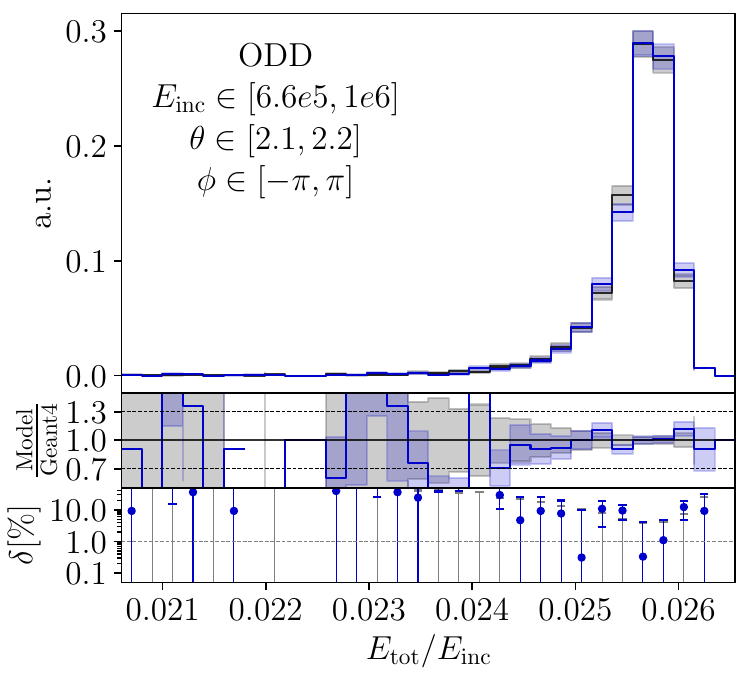}
    \includegraphics[width=0.325\linewidth]{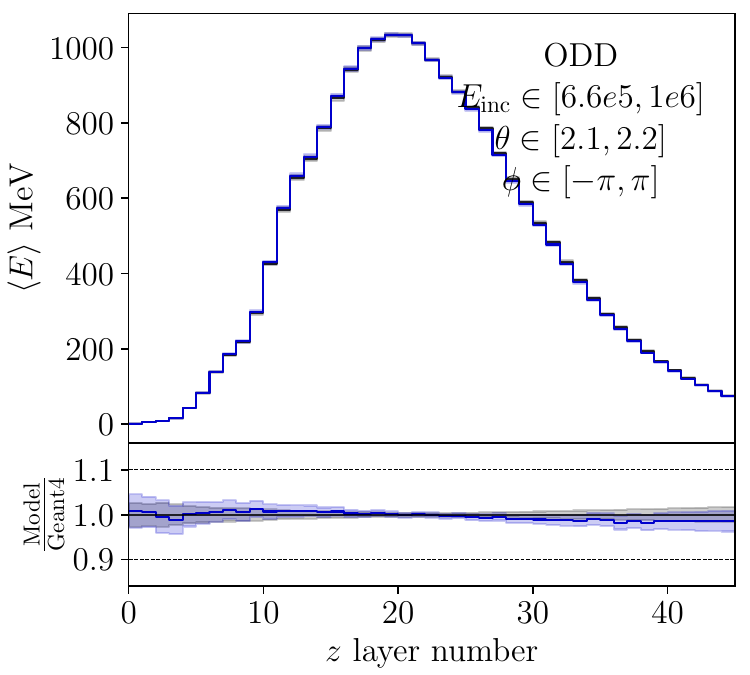}
    \includegraphics[width=0.325\linewidth]{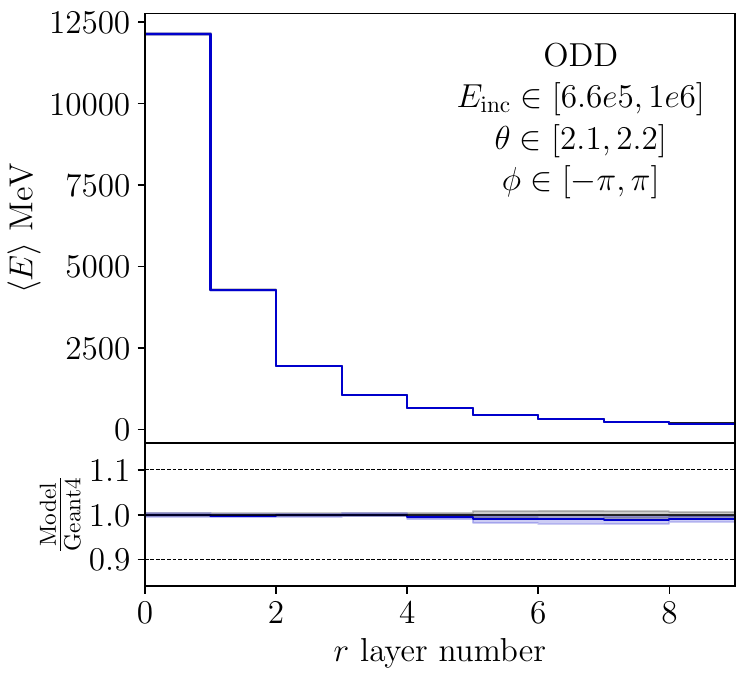} \\
    \includegraphics[width=0.325\linewidth]{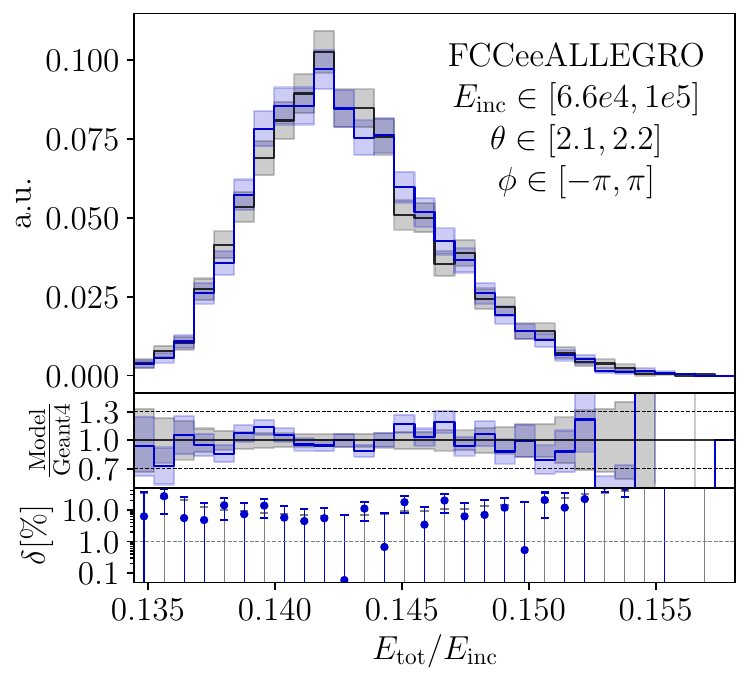}
    \includegraphics[width=0.325\linewidth]{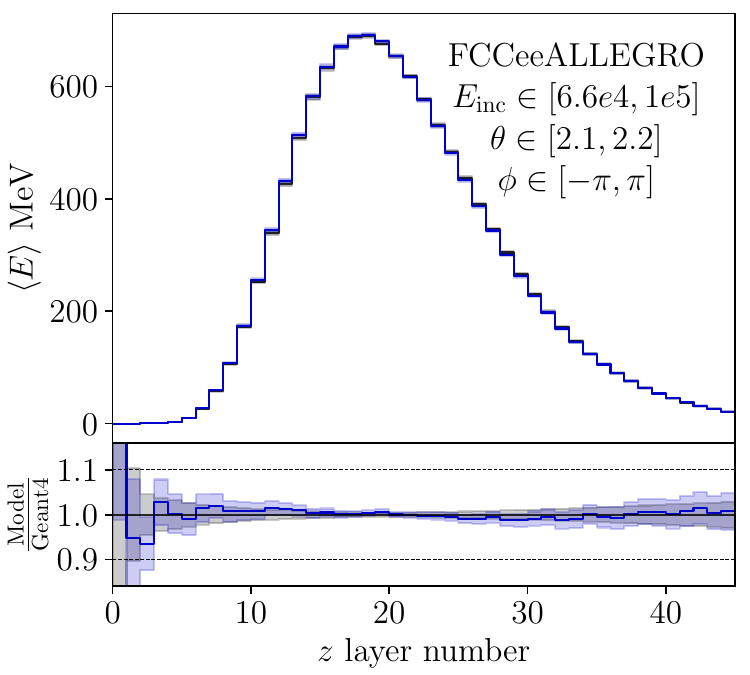}
    \includegraphics[width=0.325\linewidth]{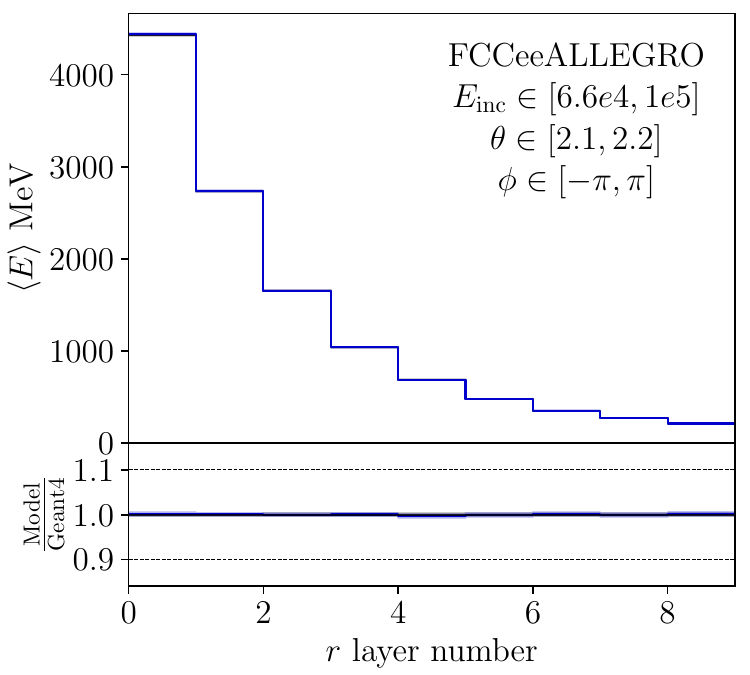} \\
    \includegraphics[width=0.60\linewidth]{figs/LEMURS/lemurs_legend.pdf}
    \caption{Summary of the sliced evaluation for $E_\text{inc}\in [0.66\cdot E_\text{i-max}, E_\text{i-max})$ TeV, $\theta\in[2.1, 2.2)$), where $E_\text{i-max}$ is 100 GeV for the FCC detectors and 1TeV for the Par04 and ODD detectors. We show the visible energy, the energy profile in the z direction, and the energy profile in the radial direction: (top) Par04SiW, (middle) ODD, and (bottom) FCCeeALLEGRO detectors. }
    \label{fig:sliced_eval_wp2}
\end{figure}

For the LEMURS dataset, we train a single vision transformer 
to model the entire range of the global conditions and the five different
detectors. Therefore, the vector of global conditions $C$ includes the 
kinematic variables of the incoming particle, $E_\text{inc}, \theta, \phi$, and a weight vector that distinguishes the detector geometry. During training, the weight vector corresponds to a one-hot encoding, e.g. the Par04SiW detector has weights $[1,0,0,0,0]$. Finally, we append the energy 
ratios, $u$, to the same vector, for a total of 53 conditions.
The energy network is also extended to depend on the full set of global kinematic variables. However, since it is easier to train and much smaller than the ViT,
we use five different energy networks, one for each detector.
We split the evaluation of the generative network into two parts:
\begin{itemize}
    \item A sliced evaluation: we compare to high-level features calculated from 2.5k samples generated in a slice in $(E_\text{inc}, \theta)$, while being inclusive in $\phi$. We define three linearly spaced bins in $E_\text{inc}$ with equal size. In $\theta$ we instead select two narrower regions,
    namely $\theta\in\{[1.52, 1.62), [2.1, 2.2)\}$.
    \item Full evaluation: we train neural classifiers on a large sample that covers the full range of conditions.
\end{itemize}
The sliced evaluation allows us to evaluate the generation performance in 
narrower phase space regions. Although the LEMURS dataset provides test samples of 1k showers
at fixed conditions, we notice that the training dataset contains many fewer events for some
of the working points. This results in large oversampling factors, 
sometimes up to a factor of 100. While exploring the amplification capability of neural
networks is an important question, in this work, we limit the evaluation to tests with 
statistical powers that do not exceed the number of samples seen during training.
Comparing a generated narrow slice to fixed test conditions is also not
feasible because of biased selection and the physics effect.
Therefore, we prefer to extract test slices from a held-out fraction of the training dataset.
Each slice contains approximately 2.5k showers.
Slicing comes at its own risks: large slices integrate physics behaviors
at different conditions, which might dilute mismodeling in particular
narrow regions.

\Cref{fig:sliced_eval_wp1} and \Cref{fig:sliced_eval_wp2} compare the generative network 
to \geant for two phase space regions: showers with incident energy in the medium 
energy bin and $\theta\in[1.52, 1.62]$ rad, and showers from the highest incident energy bin and $\theta\in[2.1, 2.2]$ rad. In this analysis, we
highlight the visible energy and the shower profiles for the three most different geometries.
Additional high-level features are provided in~\labelcref{app:lemurs_additional}. The energy network, which 
solely defines the generation quality of the visible energy and the $z$ shower profile, 
shows good agreement with \geant in both phase space regions, with relative deviations $\delta$ within the Poisson uncertainties of the slice. In particular, for the FCCeeALLEGRO
detector, the energy network correctly models the transition from central showers with a
sharp energy cut in the last five layers to lateral showers with a smooth energy dependence. We only note a small tendency to produce higher-energetic showers, which could be related to the small number of samples in the slice.
Comparing to~\cite{Raikwar:2025fky}, this highlights that a factorization into two networks can drastically improve the energy reconstruction of showers while still capturing angular dependencies. 
We observe a similar accuracy for the radial shower profile, an observable which depends on
both the energy and the shape network.
For the shower profiles, the reported error bars are the standard deviation of the mean energy deposition for 
each bin. Hence, they should not be considered as Poisson uncertainties.
For this reason, we include the relative deviation from the ground truth
only for the energy ratio $E_\text{tot}/E_\text{inc}$.

For the integrated evaluation, the large size of the sample allows us to train large 
neural network classifiers for a holistic evaluation. \cref{tab:lemurs_aucs} shows the 
AUC score of the high-level and ResNet classifiers. We exclude the low-level classifiers since they take as inputs the same information used for the ResNet classifier, but they show lower sensitivity to mismodeled showers, as already seen in the CaloChallenge studies.
The AUC and FPD scores both show good agreement with \geant across all the detectors.
The only exception is the ResNet classifier for the FCCeeALLEGRO, which shows a larger AUC score. This is indeed expected since this is the most complex detector with strong dependencies on the position of the showers, as it was already observed with the sliced evaluation. While the FPD metric scores all networks as \geant-like, the ResNet is able to identify differences between the generated and reference samples. 

The architecture and the training hyperparameters are adopted from the CaloChallenge evaluation.
We include histograms of additional high-level features for both evaluations in~\labelcref{app:lemurs_additional}.

\begin{table}[]
    \centering
    \begin{tabular}[t]{l@{\hspace{0.25cm}}c@{\hspace{0.25cm}}c@{\hspace{0.25cm}}c@{\hspace{0.25cm}}c@{\hspace{0.25cm}}}
        \toprule
         & \multicolumn{2}{c}{Classifier AUC} & \multicolumn{2}{c}{FPD} \\ \midrule
         & High-level  & ResNet & \geant & ViT-CFM \\ \midrule
        Par04SiW     & 0.502(1) & 0.530(5)  & 1.0059(3) & 1.0063(4) \\
        Par04SciPb   & 0.503(2) & 0.55(1)   & 1.0059(6) & 1.0066(4) \\
        ODD          & 0.510(5) & 0.544(4)  & 1.0077(3) & 1.0098(5)  \\
        FCCeeCLD     & 0.509(2) & 0.559(5)  & 1.0069(4) & 1.0080(3) \\
        FCCeeALLEGRO & 0.506(3) & 0.688(16) & 1.0066(4) & 1.0086(3)  \\ \bottomrule
    \end{tabular}
    \caption{Summary table of neural classifier AUC scores. Each classifier is trained to distinguish the \geant ground truth from a single detector sample from a total of 200k showers. We also report the FPD scores obtained from the JetNet library~\cite{Kansal:2023iqy,Kansal:2022spb}.}
    \label{tab:lemurs_aucs}
\end{table}
%

\subsection{Irregular geometries}
\label{sec:baseline_irreg}
Next, we show the generation performance on irregular geometries.
In the following, we only require a fixed total patch size and a segmentation of the 
voxelized space divisible by that number.

\subsubsection*{CaloChallenge dataset 1.}

\begin{figure}
    \begin{minipage}[t!]{0.75\linewidth}
    \centering
    \includegraphics[width=0.49\linewidth, page=1]{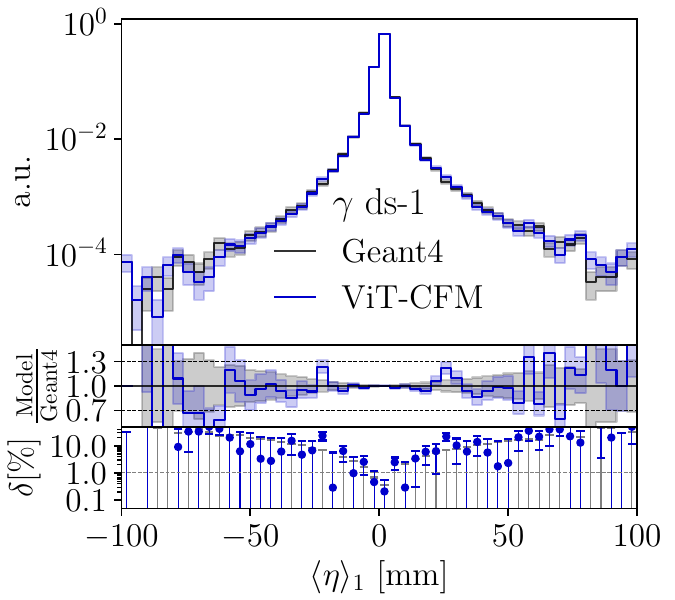}
    \includegraphics[width=0.49\linewidth, page=1]{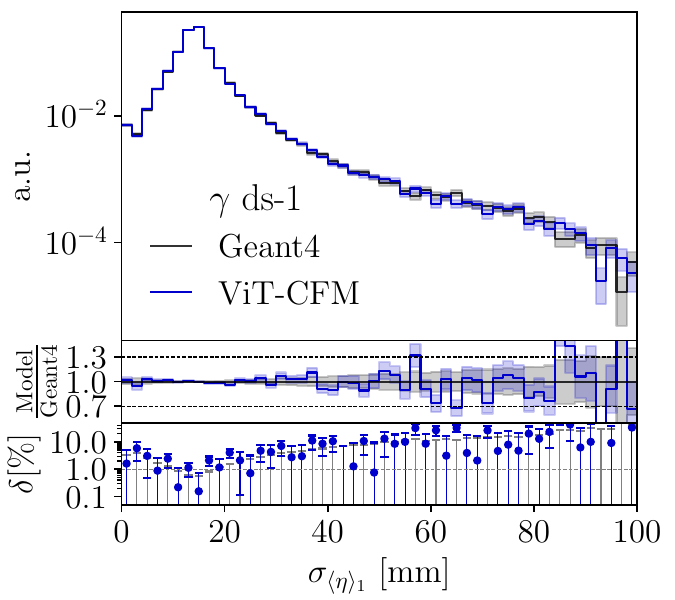}
    \end{minipage}
    \begin{minipage}{0.2\linewidth}
    \begin{footnotesize}
        \begin{tabular}{lc}
        \toprule
         Classifier &  ds1-$\gamma$  \\ \midrule
        Low-level & 0.518(1) \\
        High-level & 0.504(3) \\ \midrule
        \multicolumn{2}{c}{ \makecell{Gen. time \\ $[\text{ms per shower}]$} } \\ \midrule
        CPU batch 1   & 3.8$\times 10^{3}$  \\
        GPU batch 100 & 24.36(5) \\
        \bottomrule
    \end{tabular}
    \end{footnotesize}
    \end{minipage}
    \caption{Summary of the evaluation on the CaloChallenge-ds1-$\gamma$ dataset. We show the center of energy and the shower width in layer-1, the AUC scores of a low- and high-level neural classifier, and the generation time on CPU, with batch size 1, and GPU, with batch size 100.}
    \label{fig:summary-ds1g}
\end{figure}

\begin{figure}
    \begin{minipage}[t!]{0.75\linewidth}
    \centering
    \includegraphics[width=0.49\linewidth, page=1]{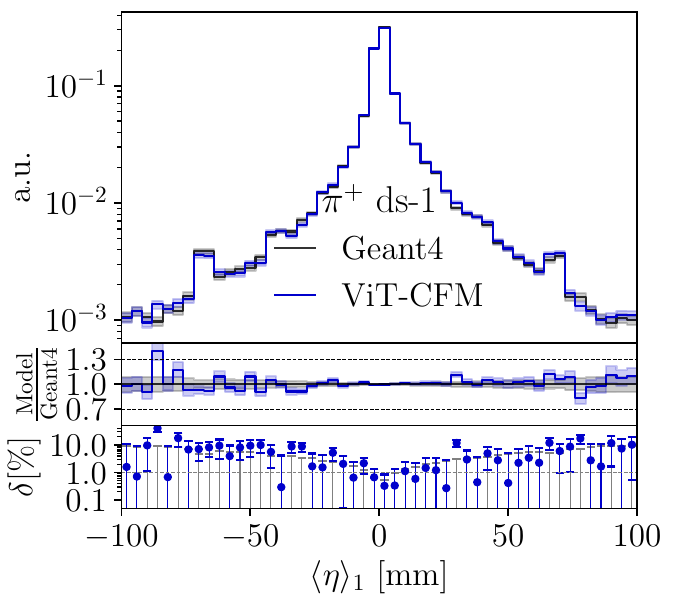}
    \includegraphics[width=0.49\linewidth, page=1]{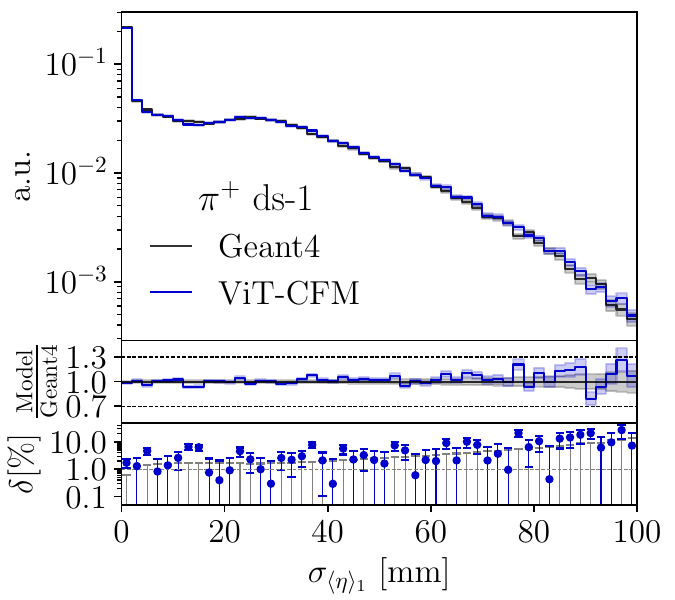}
    \end{minipage}
    \hspace{0.05cm}
    \begin{minipage}{0.2\linewidth}
    \begin{footnotesize}
        \begin{tabular}{lc}
        \toprule
         Classifier &  ds1-$\pi^{+}$  \\ \midrule
        Low-level & 0.633(3) \\
        High-level & 0.602(4) \\ \midrule
        \multicolumn{2}{c}{ \makecell{Gen. time \\ $[\text{ms per shower}]$} } \\ \midrule
        CPU batch 1   & 5.7$\times 10^{3}$ \\
        GPU batch 100 & 30.6(3) \\
        \bottomrule
    \end{tabular}
    \end{footnotesize}
    \end{minipage}
    \caption{Summary of the evaluation on the CaloChallenge-ds1-$\pi^{+}$ dataset. We show the center of energy and the shower width in layer-1, the AUC scores of a low- and high-level neural classifier, and the generation time on CPU, with batch size 1, and GPU, with batch size 100.}
    \label{fig:summary-ds1p}
\end{figure}

\Cref{fig:summary-ds1g} and~\Cref{fig:summary-ds1p} show a summary evaluation of the
generation performance. We highlight the center of energy and the width of the center of energy
for the first segmented layer. For both incident particles, the neural network reproduces the 
\geant response up to statistical uncertainties. The AUC scores of the high- and low-level classifiers 
provide a holistic view of the entire phase space. For ds1-$\gamma$, both classifiers
confirm that the network emulates \geant almost perfectly. For ds1-$\pi^{+}$, the high-level
features show a similar level of agreement, but we observe a higher level of mismodelling 
from the low-level classifier. While hadronic showers are more complex to model, this discrepancy arises from the usage of a different version of \geant in the generation of the test sample~\cite{Krause:2024avx}. The CaloChallenge found a low (high)-level AUC of  $0.609(4)$ ($0.558(2)$) when comparing \geant train and test sets~\cite{Krause:2024avx}. 
Nonetheless, these results improve over all the other submissions to the CaloChallenge.

The lower dimensionality of the dataset implies that a smaller number of patches is processed by the ViT. We observe this effect in the generation time; even though the neural network has the same number of parameters, the generation time for the full generation of a shower for ds1 is shorter than that of the regular ds2 and ds3.

\subsubsection*{CaloHadronic dataset.}

\begin{figure}
    \centering
    \includegraphics[width=0.325\linewidth, page=1]{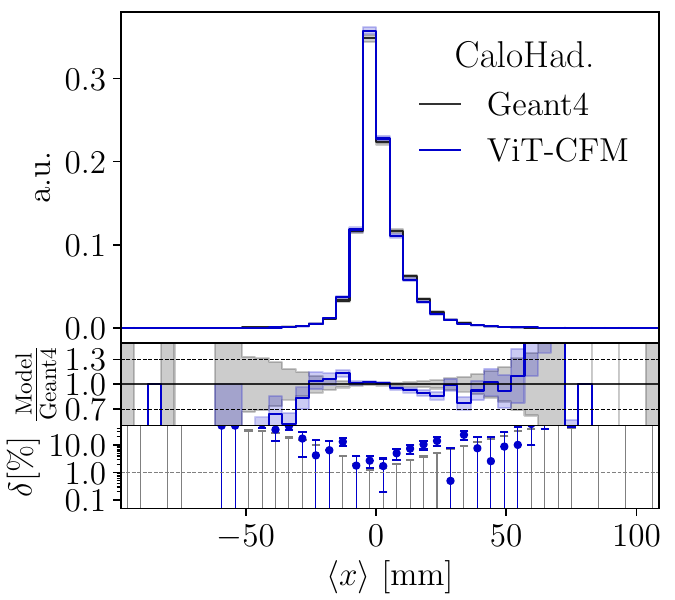}
    \includegraphics[width=0.325\linewidth, page=1]{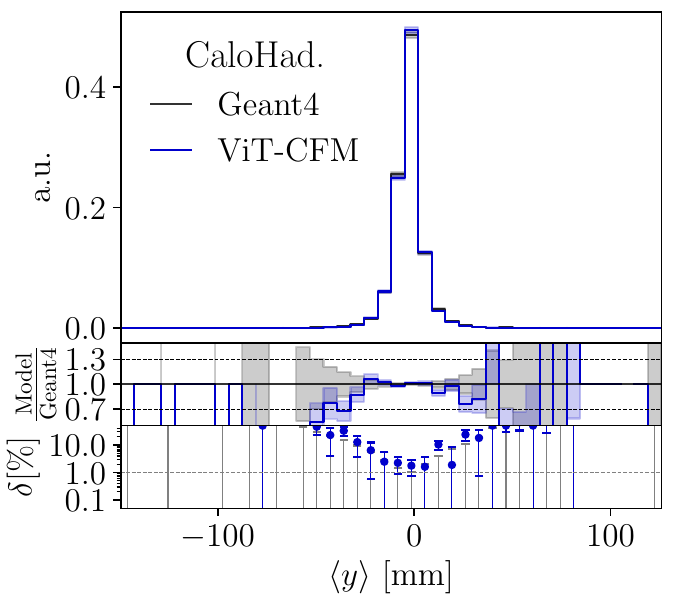}
    \includegraphics[width=0.325\linewidth, page=1]{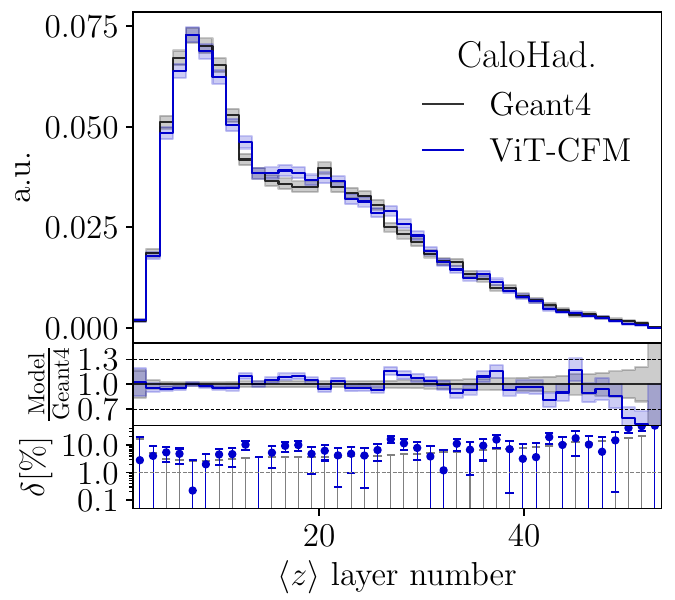} \\
    \begin{minipage}[t!]{0.75\linewidth}
    \includegraphics[width=0.437\linewidth, page=1]{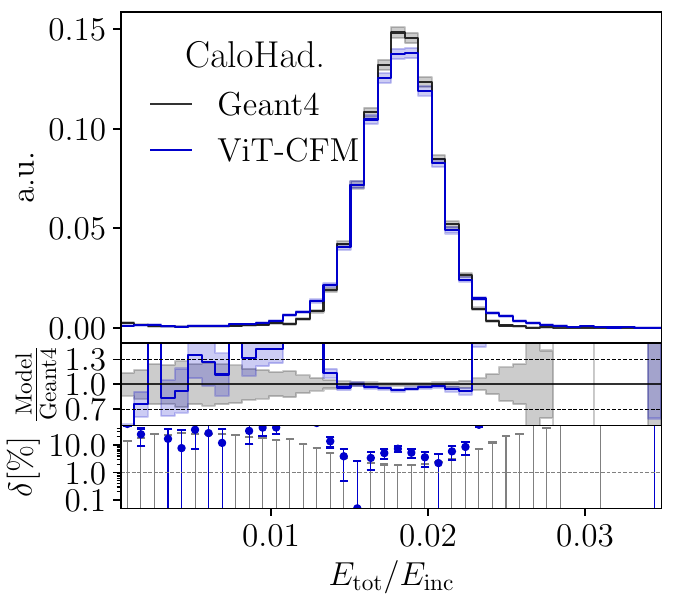}
    \includegraphics[width=0.437\linewidth, page=1]{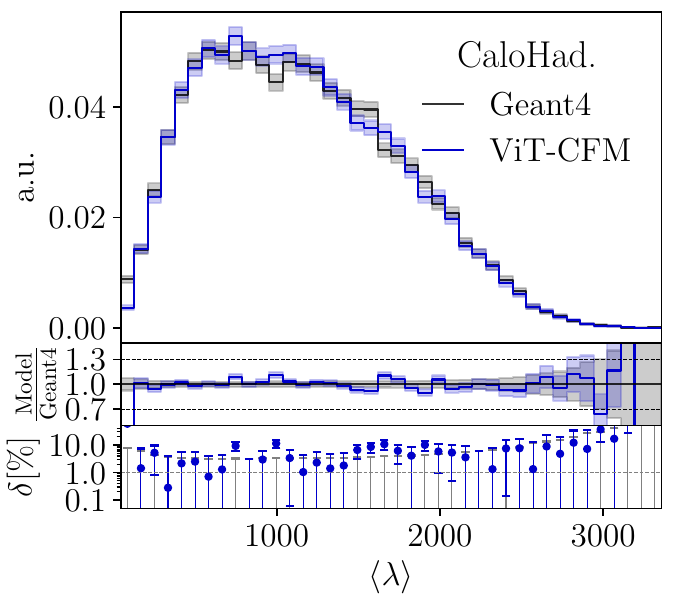}
    \end{minipage}
    \begin{minipage}{0.24\linewidth}
    \begin{footnotesize}
        \hspace{-1.2cm}
        \begin{tabular}{lc}
        \toprule
         Classifier &  CaloHadronic  \\ \midrule
        High-level & 0.614(3) \\ \midrule
        \multicolumn{2}{c}{ \makecell{Gen. time \\ $[\text{ms per shower}]$} } \\ \midrule
        CPU batch 1   & 2.25(1)$\times 10^{4}$ \\
        GPU batch 100 & 145(2) \\
        \bottomrule
    \end{tabular}
    \end{footnotesize}
    \end{minipage}
    \caption{Summary of the evaluation on the CaloHadronic dataset. We show the center of energy in the three coordinates $x$,$y$, and $z$ for the entire calorimeter, the total visible energy $E_\text{tot}$, and the average occupancy $\langle \lambda \rangle$ in the detector. In the table, we report the AUC score of a neural classifier and the generation time on CPU and GPU.}
    \label{fig:summary-calohad}
\end{figure}

To evaluate the generative network, we follow a methodology similar to~\cite{Buss:2025cyw}.
We first calculate global high-level features for the entire detector.
The top row of~\cref{fig:summary-calohad} shows the center of energy in the $x$ and $y$ directions
in units of the real detector size, and the center of energy in the $z$ direction as a function of the layer number. The bottom row shows the ratio
$E_\text{tot}/E_\text{inc}$ and the average number of hits $\langle \lambda \rangle$. Finally, we show,
in table, the AUC score of a neural classifier trained on the five observables together with the layer energies, and the 
generation time on CPU and GPU. The marginal distributions, as well as the neural classifier,
confirm that the main features of the hadronic showers are well-captured by the generative
network, with larger deviations for low-energy and low-occupancy showers, otherwise within
statistical uncertainties. 
For all the results, the \geant reference corresponds to a test sample containing 20k showers,
which is compared to a generated sample with the same statistic, and 
we apply a low-energy threshold of $x_\text{th}=1$~keV.
Given the small size of the test sample, we refrain from training a classifier
on the low-level energy deposits. The details of the classifier implementation and training
are reported in~\labelcref{app:hyperparams}. 

The generation time on both CPU and GPU increases due to the larger number of patches compared to the CaloChallenge-ds3. Nevertheless, the GPU generation time is still at $\mathcal{O}(100)$~ms.
%

\subsection{Transfer learning}
\label{sec:finetuning}

To study the transfer learning capabilities of the neural network, we start from the networks trained on the LEMURS and the CaloChallenge-ds2 datasets in the previous section.
We present two clear cases where fine-tuning can be used to reduce the 
cost of training ViTs for fast detector simulation. First, we use the large
LEMURS dataset as a pretraining objective during which the network
learns multi-detector responses and angular dependencies. The target
fine-tuning is the CaloChallenge-ds2 dataset. For the fine-tuning phase,
we set the angular conditions to $\theta=1.57$ and $\phi=0.0$.
Since the detector matches the Par04 setting, we set the weight vector equal to the Par04SiW detector. The weight vector
is an effective entry point for the introduction of prior knowledge
of the detector configuration. For instance, a third detector with
the Par04 geometry can be encoded as a weighted combination of the
two materials already seen during training.
In this case example, the fine-tuning step specializes the network
to a single angular condition and adjusts the response from a photon
initiated shower to an electron one.
The second study is a superresolution from the CaloChallenge-ds2, or LEMURS,
voxelization to detectors with higher resolutions.
For this study, we reinitialize all the components that encode and 
decode the shower information in the ViT, as described in~\cref{sec:ml}.
This choice corresponds to the ``Full fine-tuned'' strategy adopted for point clouds in~\cite{Gaede:2025shc}. Fine-tuning the entire backbone is strictly more expressive than adjusting a subset of weights. Since our networks require a manageable amount of resources, we do not explore other more parameter-efficient training strategies.

The principal figures of merit for evaluating the training efficiency of 
generative networks are the amount of data and the computational
resources used during training. We explore two scenarios: one where we 
fix the amount of available data and we vary the number of training 
iterations, and a second where we scan over the size of the training
dataset.

\subsubsection*{Training iterations}
\begin{figure}
    \centering
    \includegraphics[width=0.45\linewidth]{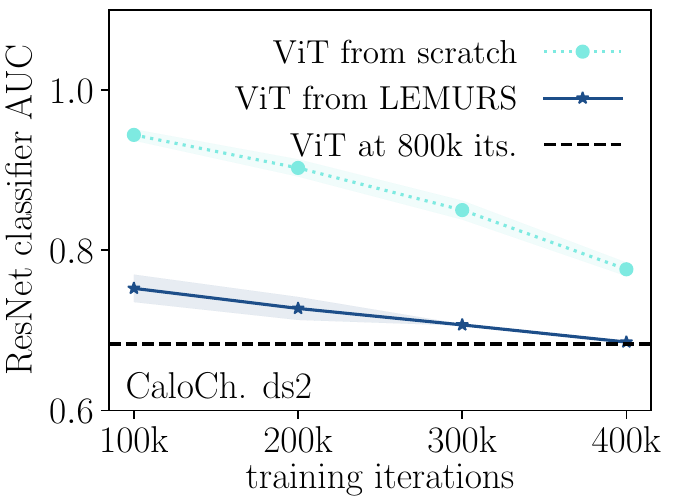}
    \includegraphics[width=0.45\linewidth]{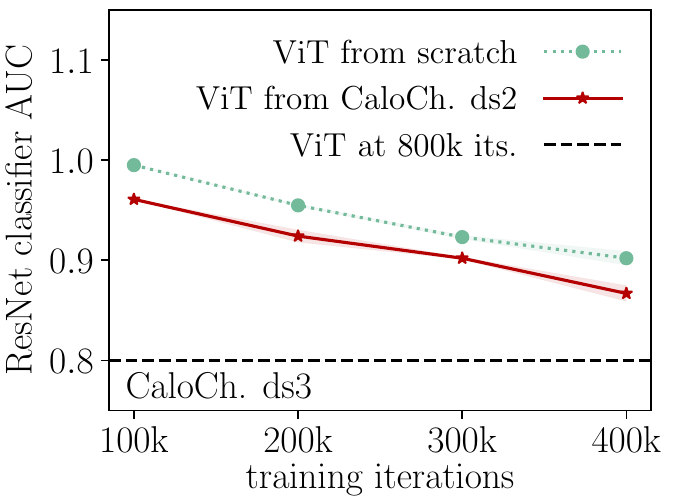} 
    \caption{Efficiency gain of fine-tuned networks for varying training iterations. The ResNet classifier AUC score of fine-tuned networks is significantly lower than that of the networks trained from scratch. (left) Pretraining on the LEMURS dataset, fine-tuned on the CaloChallenge-ds2. (right) Pretraining on the CaloChallenge-ds2, fine-tuned on the CaloChallenge-ds3.}
    \label{fig:training_its_ft}
\end{figure}

\begin{figure}
    \centering
    \includegraphics[width=0.45\linewidth]{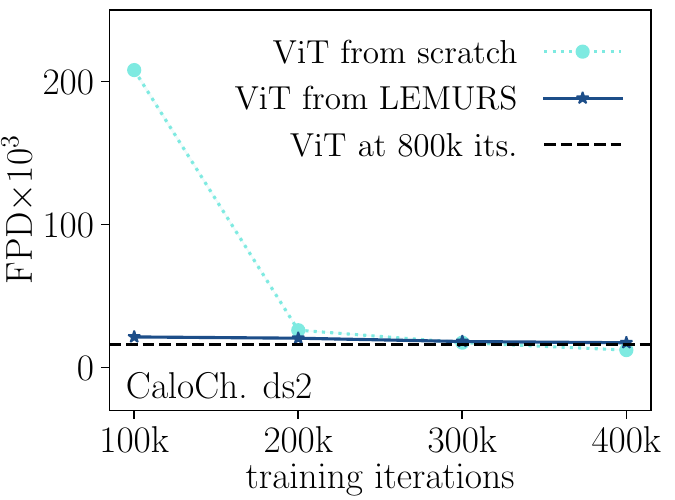}
    \includegraphics[width=0.45\linewidth]{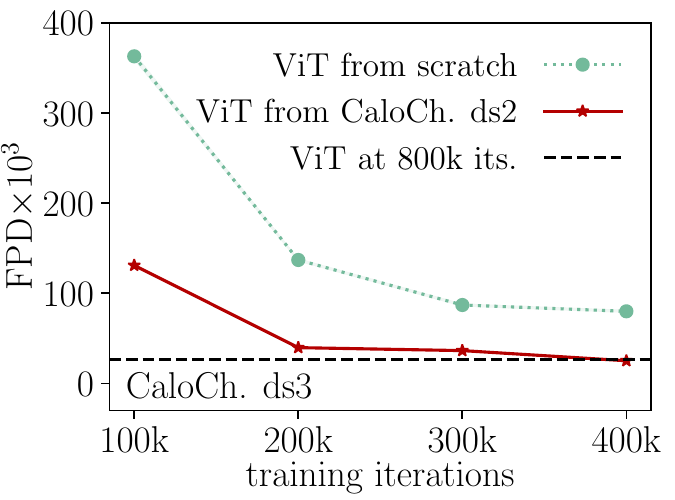} 
    \caption{Efficiency gain of fine-tuned networks for varying training iterations. The FPD score of fine-tuned networks converges to the
    best value observed from a complete training in significantly less iterations than the networks trained from scratch. Notice that the the ResNet classifier in~\cref{fig:training_its_ft} indicates smaller gains compared to the FPD. (left) Pretraining on the LEMURS dataset, fine-tuned on the CaloChallenge-ds2. (right) Pretraining on the CaloChallenge-ds2, fine-tuned on the CaloChallenge-ds3.}
    \label{fig:training_its_ft_fpd}
\end{figure}

The baseline training of~\cref{sec:baseline} converges after approximately
800k iterations. Here, we aim to find the minimum number of training iterations
needed to obtain the same performance with the fine-tuned network. 
\Cref{fig:training_its_ft} summarizes the efficiency gain from utilizing a
pretrained network. On the left, we compare a ViT trained from scratch against a
network trained on LEMURS and then fine-tuned on the CaloChallenge-ds2.
We use the ResNet classifier AUC score as our evaluation metric, as 
it demonstrated the strongest discriminative performance in our baseline
studies.
On the right, we repeat the same study for the superresolution example.
In both cases, the fine-tuned network converges more rapidly, showing
an AUC score significantly smaller than 1.0 already after 100k iterations. The performance
of the network trained from scratch is reached with roughly 400k training
iterations, therefore converging in half the training time.
Faster convergence is particularly beneficial if multiple neural networks
have to be trained because the large pretraining phase happens only
once. As an example, the calorimetric module of AtlFast3~\cite{ATLAS:2021pzo} contains 300 neural networks
trained with different particle types and at different pseudo-rapidity
angles. On our hardware, the pretraining phase becomes subdominant already
after six trainings. 
We remark that the selection of the metric can affect the evaluation
of the fine-tuning step and the following efficiency gain claims. In~\cref{fig:training_its_ft_fpd}, we show the same scan over the training iterations but we use the FPD score as the evaluation metric. Notice that generation
quality converges to the optimal value much more quickly than the
ResNet classifier. This highlights the importance of selecting the 
most discriminative metric if a holistic evaluation of the 
network is of interest.

\subsubsection*{Dataset size}
\begin{figure}
\begin{minipage}[t!]{0.55\linewidth}
    \includegraphics[width=0.80\linewidth]{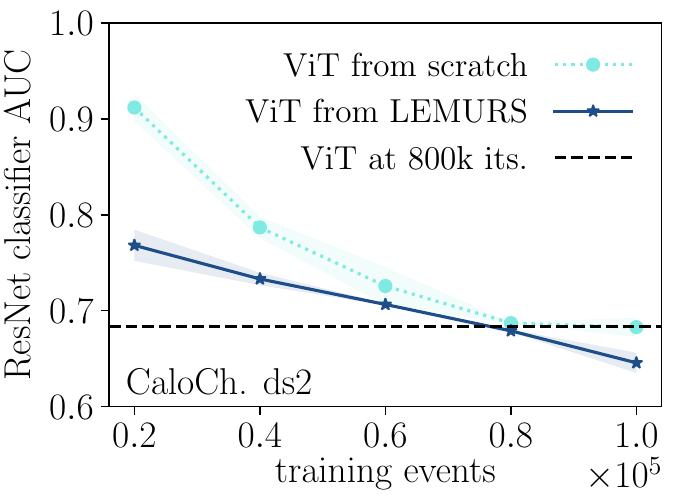}
\end{minipage}
\begin{minipage}[t!]{0.45\linewidth}
    \begin{tabular}{l @{\hspace{0.25cm}} c @{\hspace{0.25cm}} c}
    \toprule
            & from scratch & fine-tuned  \\ \midrule
            & \multicolumn{2}{c}{ResNet cls. AUC} \\ \midrule
    CaloCh. ds2 & 0.683(9) & 0.645(9) \\
    CaloCh. ds3 & 0.799(9) & 0.789(9) \\ \midrule
            & \multicolumn{2}{c}{High-level cls. AUC} \\ \midrule
    CaloHadronic & 0.614(3) & 0.600(4)   \\ \bottomrule
    \end{tabular}
\end{minipage}
    \caption{(left) Efficiency gain of fine-tuned networks for varying dataset sizes. fine-tuned networks better generalize as demonstrated by a ResNet classifier always trained with a total of 200k showers. Pretraining on the LEMURS dataset, fine-tuning on the CaloChallenge-ds2. (right) Summary table of classifier AUC scores after fine-tuning on the entire training dataset for the corresponding detectors.}
    \label{fig:training_data_ft}
\end{figure}

Generating \geant data is the most computationally intensive step of the simulation chain, and
reducing the training data necessary to reach the target accuracy and generalizability is of
large interest for a fast simulation surrogate. We explore the effect of fine-tuning on three
datasets: first on the CaloChallenge-ds2/3 as in the previous section, then on the CaloHadronic 
dataset. We perform a full performance scan over training dataset sizes only for
the CaloChallenge-ds2, since the smaller dimensionality allows for faster training. Given our
available computational resources, it was unfeasible to carry out similar studies for the higher
resolution datasets. Therefore, we look for performance improvements using the full 
training dataset.
We use as a performance metric the one that demonstrated the best discriminative power
from the previous tests; for the CaloChallenge datasets, we report the ResNet classifier AUC 
score, while for CaloHadronic, we use the neural classifier trained on the five high-level
observables.
\Cref{fig:training_data_ft} (left) shows the ResNet AUC score for the pretrained ViT on 
LEMURS and fine-tuned to the CaloChallenge-ds2. To train the classifier, we always generate
100k showers, therefore, we are also testing the
capability of the neural network to generalize beyond the number of samples seen during fine-tuning. Our results show that the fine-tuned network
provides better performance for each training dataset size, demonstrating that information captured during the pretraining phase is retained after fine-tuning.
Fine-tuned networks converge much more rapidly and we have to avoid overfitting by increasing the 
weight decay parameter as the size of the training dataset diminishes.
In particular, at 100k training showers, the fine-tuned ViT is significantly better than the
network trained from scratch, setting a new state-of-the-art for the CaloChallenge.
The high-granular ds3 shows a smaller gain. Even though the ViT converges more rapidly,
as shown in~\cref{fig:training_its_ft}, the performance plateaus at an AUC of 0.80. This
could be an indication that the neural network expressivity is the limiting factor. We observe
a similar behavior for the CaloHadronic dataset, where the high-level classifier shows a better agreement with \geant for the fine-tuned ViT.

In general, smaller differences between datasets lead to more efficient transfer learning, as we observe with the fine-tuning from the LEMURS dataset to the CaloChallenge-ds2. When the differences involve the detector geometry and the properties of the incident particle, e.g. the incident energy, as in the CaloHadronic fine-tuning, the gains from a pre-training are smaller yet observable.
%

\section{Conclusions}
\label{sec:conclusions}

Machine learning emulators are effective solutions to 
accelerate slow simulators. Fast calorimeter simulation is the
most striking and compelling example at HEP collider experiments, such as those at the LHC, ILC, or the FCC.
Once ensured that the generation speed of generative networks
is sufficient, maximizing the accuracy of the surrogate
calorimeter showers becomes the real challenge. Especially for high-granular
detectors, the computational cost of training deep learning networks becomes prohibitively expensive.

We have proposed a vision transformer architecture that can be readily applied to
any voxelized detector geometry. From the performant CaloDREAM architecture,
we have defined an efficient and flexible patching for arbitrary detector layouts,
including optimized and full detector voxelizations. This enables a reduction in both
computational and data requirements without sacrificing expressiveness.
Our results show excellent agreement with \geant on the studied datasets; multiple
evaluation metrics indicate that the generated samples are indistinguishable from \geant, while also improving over previous benchmarks.

We have further introduced a pretraining strategy for ViTs and demonstrated that pretraining, followed by a targeted fine-tuning on
the CaloChallenge~ds2/3 significantly reduces the resources needed to match the 
performance of the network trained from scratch on the most sensitive evaluation metric.
Additionally, for the large LEMURS pretraining, we have shown better data efficiency and
better generalization of the fine-tuned network.
The methodology presented in this work is specific to patch-based
transformers, which benefit from the independence of the transformer
architecture from the number of patches.
This is the first application of patch-based fast calorimeter simulation to an entire detector with both an electromagnetic and a hadronic calorimeter.
We believe our studies will open the way towards the full implementation and deployment of
transformer-based emulators in the community.

\subsubsection*{Code and data availability.}
Together with this document, we include a public release of the code at \href{https://github.com/luigifvr/vit4hep}{https://github.com/luigifvr/vit4hep}. We also publish the complete set of high-level features and samples used during evaluation at~\href{https://doi.org/10.5281/zenodo.18071948}{10.5281/zenodo.18071948}.

\section*{Acknowledgments}
We would like to thank Sofia Palacios Schweitzer and Ayodele Ore for comments on the manuscript and, together with Jonas Spinner, for many useful discussions.
We would also like to thank Martina Mozzanica for support with the CaloHadronic dataset, and Anna Zaborowska and Peter Mckeown for helpful discussions on the LEMURS dataset.

L.F. is supported by the Fonds de la Recherche Scientifique - FNRS under Grant No. 4.4503.16.

Computational resources have been provided by the supercomputing facilities of the Université catholique de Louvain (CISM/UCL) and the Consortium des Équipements de Calcul Intensif en Fédération Wallonie Bruxelles (CÉCI) funded by the Fond de la Recherche Scientifique de Belgique (F.R.S.-FNRS) under convention 2.5020.11 and by the Walloon Region.
The present research benefited from computational resources made available on Lucia, the Tier-1 supercomputer of the Walloon Region, infrastructure funded by the Walloon Region under the grant agreement n°1910247.

\clearpage

\appendix

\section{Datasets visualization}
\label{app:datasets}

\begin{figure}
    \centering
    \includegraphics[width=0.75\linewidth]{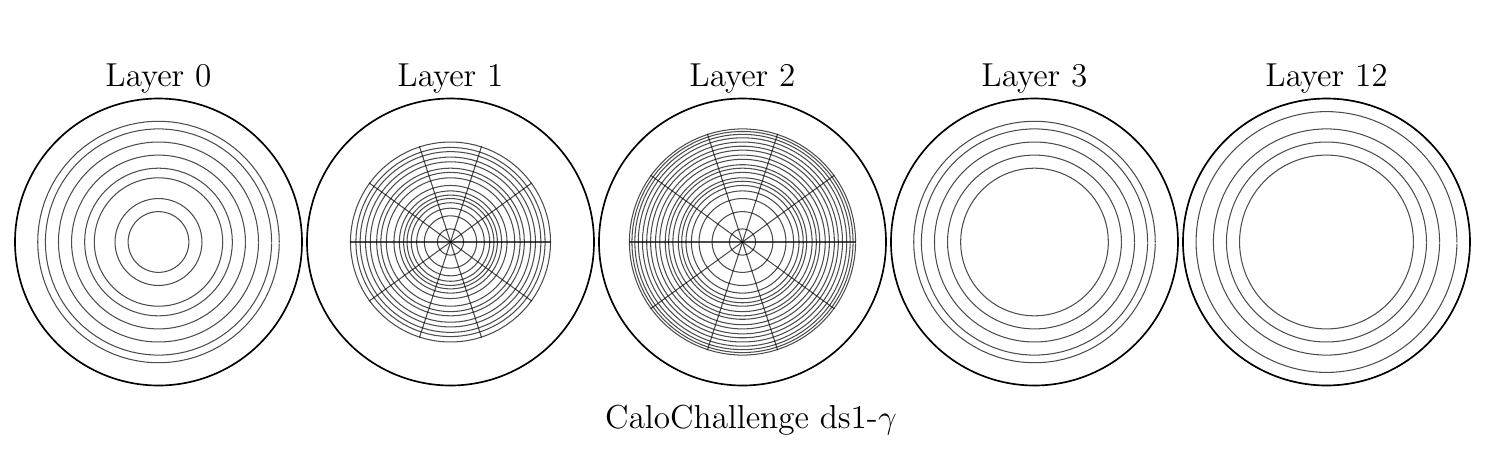}\\
    \includegraphics[width=\linewidth]{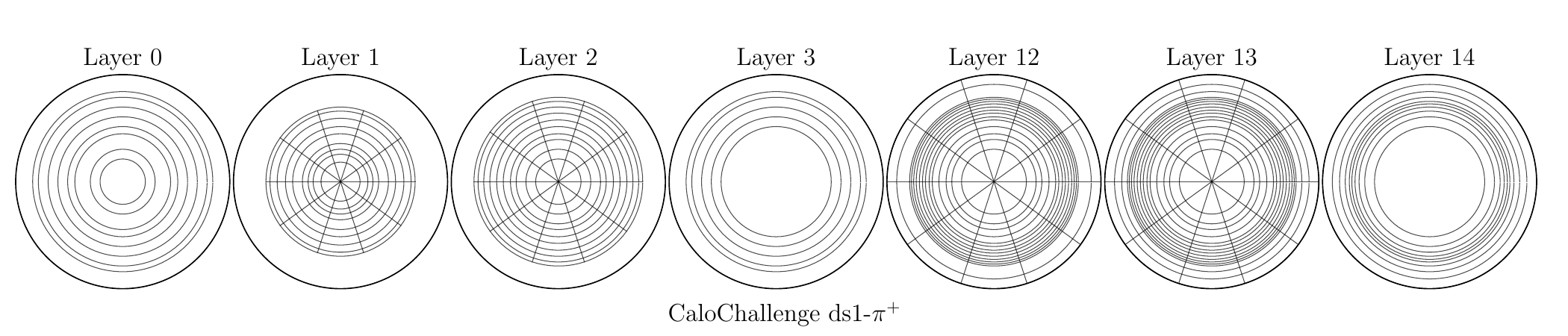} \\
    \includegraphics[width=0.25\linewidth]{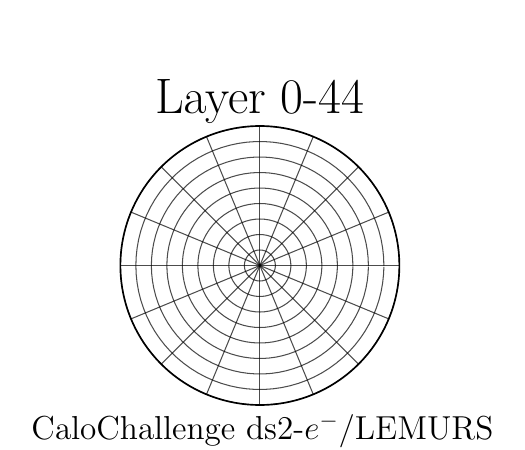}
    \includegraphics[width=0.25\linewidth]{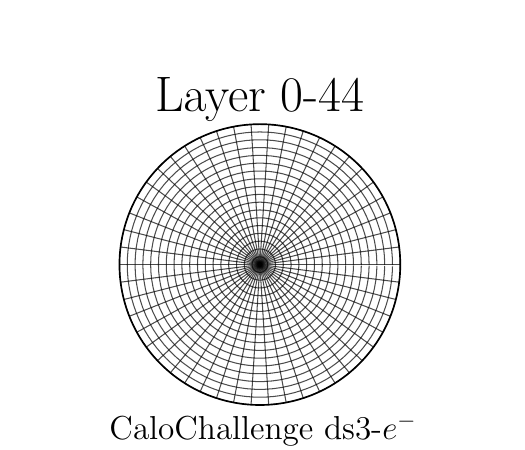}
    \includegraphics[width=0.38\linewidth]{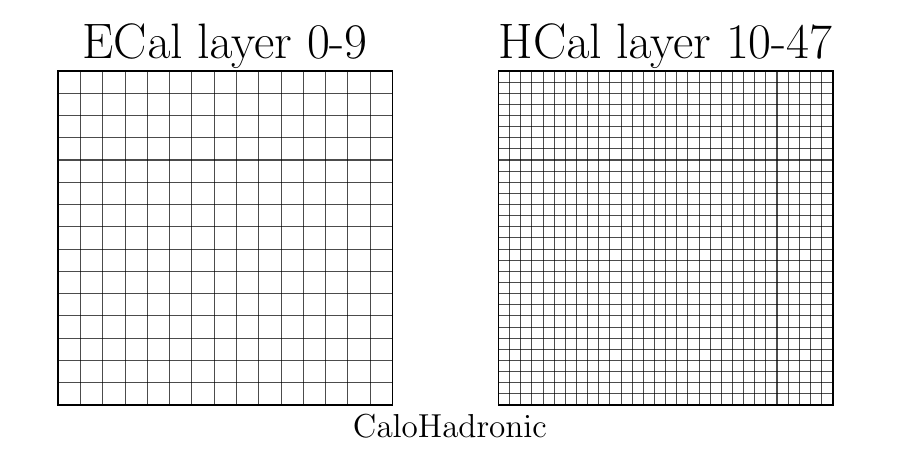}
    \caption{Visualization of the detector geometries. (top) Geometry of the five layers for the CaloChallenge-ds1-$\gamma$. (middle) Geometry of the seven layers for the CaloChallenge-ds1-$\pi^+$. (bottom) Regular geometries: voxelization of a single layer for the CaloChallenge-ds2/3, and the ECal/HCal of the CaloHadronic dataset.}
    \label{fig:datasets}
\end{figure}

In~\cref{fig:datasets}, we visualize the geometry of the datasets studied in~\cref{sec:results}.
For the irregular geometries in the CaloChallenge, we show the entire detector as layer slices. The number of radial ($n_r$) and angular ($n_\alpha$) bins in each layer for photons and pions, respectively, is
\begin{align}
  \text{photons}\; (n_r\times n_\alpha): &\qquad
  8\times1, \; 16\times10, \; 19\times10, \; 5\times1, \; 5\times1 \notag \\
  \text{pions}\; (n_r\times n_\alpha): &\qquad 
  8\times1, \; 10\times10, \; 10\times10, \; 5\times1, \; 15\times10, \; 16\times10, \; 10\times1 \;.
\label{eq:voxels_ds1}
\end{align}
The CaloChallenge-ds2 and the LEMURS datasets share the same geometry, with 45 layers organized in $16\times9$ $n_r\times n_\alpha$ bins. The CaloChallenge-ds3 increases the granularity to $50\times 18$ $n_r\times n_\alpha$ bins. Finally, for the CaloHadronic dataset, we show the voxelization in the electromagnetic calorimeter, 10 layers with $15\times15$ bins each, and the hadronic calorimeter, 48 layers with $30\times30$ bins each.

%

\section{CaloDREAM details}
\label{app:implementation}

Our architecture relies on the factorization of the generation of the total
energy deposited in a layer, from the ``energy network'', and the generation
of normalized energy deposits in each cell. The energy network takes as 
inputs energy ratios, labelled as $u$, defined in the following way
\begin{equation}
u_0=\frac{E_\text{tot}}{E_\text{inc}} \qquad \text{and} \qquad u_i = \frac{E_{i-1}}{\sum^L_{j\geq i-1} E_j}, \;\; i= 1,\ldots,L-1 \;,
\label{eq:us}
\end{equation}
where $E_\text{tot}=\sum_i E_i$, and $L$ is the total number of layers. The set of energy ratios defined in~\cref{eq:us}
is also a well-motivated preprocessing step as it ensures that the inputs
are always in the $[0,1]$ interval.
It can be shown that, given a set of energy ratios and the incident energy,
the corresponding layer energies are
\begin{align}
    &E_\text{tot} = u_0 E_\text{inc}, \notag \\ 
    &E_0 =u_0u_1E_\text{inc}, \qquad 
    E_i = u_{i+1}u_0 E_\text{inc}\prod^i_{j=1} (1-u_j), \;\;i= 1,\ldots,L-1 \;.
\end{align}
%

\section{Training hyperparameters}
\label{app:hyperparams}

\begin{table}[h]
    \centering
    \begin{small} \begin{tabular}{lc @{\hspace{10pt}} c} \toprule
        Parameter     &  \multicolumn{2}{c}{Energy network} \\ \midrule
         &  LEMURS & Others \\ \midrule
        Iterations    & 2M & 250k \\
        Batch size & 128 & 256 \\
        LR scheduler & OneCycleLR & CosineAnnealingLR \\
        Max/Initial LR & $10^{-3}$ & $10^{-4}$ \\
        Weight decay & \multicolumn{2}{c}{0.1} \\
        ODE solver  & \multicolumn{2}{c}{Runge-Kutta 4} (20 steps) \\
        \midrule
        Dim embedding  &  \multicolumn{2}{c}{64}  \\
        Intermediate dim   & \multicolumn{2}{c}{512} \\
        Num heads  & \multicolumn{2}{c}{4} \\
        Num layers   & \multicolumn{2}{c}{4} \\
        \bottomrule
    \end{tabular} \end{small}
    \caption{Training and network parameters for the energy network of~\cref{sec:ml}.}
    \label{tab:energy_params}
\end{table}

\begin{table}[h]
    \centering
    \begin{small} \begin{tabular}{lc @{\hspace{10pt}} c @{\hspace{10pt}} c} \toprule
        Parameter     & \multicolumn{3}{c}{Universal ViT block} \\ \midrule
                    & CaloChallenge & LEMURS & CaloHadronic \\ \midrule
        Iterations    & 800k & 2M & 1.2M \\
        Batch size    & 64 & 64 & 32 \\
        Weight decay  & 0.1 & $10^{-5}$ & $10^{-5}$ \\
        LR scheduler     & \multicolumn{3}{c}{CosineAnnealingLR} \\
        Initial LR    & \multicolumn{3}{c}{$10^{-4}$} \\
        ODE solver & \multicolumn{3}{c}{Runge-Kutta 4 (20 steps)} \\ \midrule
        Embedding dimension   & \multicolumn{3}{c}{480}  \\
        Attention heads       &  \multicolumn{3}{c}{6}   \\
        MLP hidden dimension  & \multicolumn{3}{c}{1920} \\
        Blocks                &  \multicolumn{3}{c}{6}   \\
        \bottomrule
    \end{tabular} \end{small}
    \caption{Training and network parameters for the universal ViT block presented in~\cref{sec:ml}.}
    \label{tab:vit_params}
\end{table}

\begin{table}[h]
    \centering
    \begin{small} \begin{tabular}{lc} \toprule
    Dataset     & Patch sizes \\  \midrule
                & $(z, r, \alpha)$ \\ \midrule
    CaloChallenge-ds1$\gamma$    & (1, 1, 5) \\
    CaloChallenge-ds1$\pi^{+}$   & (1, 1, 5)  \\
    CaloChallenge-ds2$e^{-}$     & (3, 16, 1) \\
    CaloChallenge-ds3$e^{-}$     & (3, 10, 3) \\
    LEMURS                       & (3, 16, 1) \\ \midrule
    CaloHadronic                 & $(z, x, y)$ \\ \midrule
    ECal                         & (5, 5, 3) \\ 
    HCal                         & (3, 5, 5) \\ \bottomrule
    \end{tabular} \end{small}
    \caption{Patch sizes used for the CaloChallenge, LEMURS, and CaloHadronic datasets.}
    \label{tab:patch_sizes}
\end{table}

\begin{table}[h]
    \centering
    \begin{small} \begin{tabular}[t]{lc}
    \toprule
    Parameter & Value \\ \midrule
    \multicolumn{2}{c}{high-level and low-level} \\ \midrule
    Optimizer & Adam \\
    Epochs & 100 \\
    Number of layers & 3 \\
    Hidden nodes & 2048 \\ \midrule
    \multicolumn{2}{c}{ResNet} \\ \midrule
    Optimizer & AdamW \\
    Epochs & 50 \\
    Number of layers & 18 \\ \midrule
    \multicolumn{2}{c}{Common training parameters} \\ \midrule
    Learning rate & $2\cdot10^{-4}$ \\
    Batch size & 1000 \\
    Training samples & 60k \\
    Validation samples & 20k \\
    Testing samples & 20k \\
    Activation function & leaky ReLU \\
    \bottomrule
    \end{tabular} \end{small}
    \caption{Parameters for the high-level, low-level, and ResNet classifiers network.}
    \label{tab:cls_params}
\end{table}

\clearpage
\section{Additional LEMURS high-level features}
\label{app:lemurs_additional}

\begin{figure}[b]
    \centering
    \includegraphics[width=0.49\linewidth]{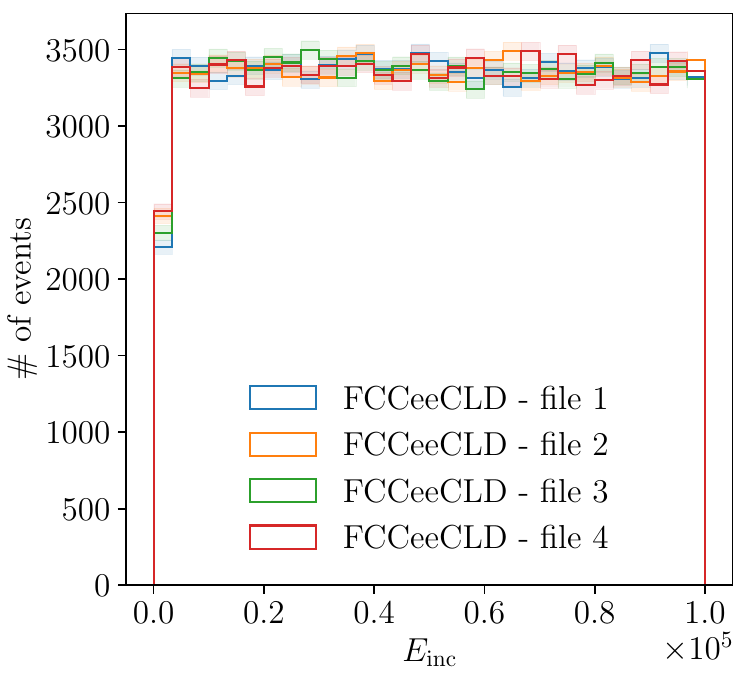}
    \includegraphics[width=0.49\linewidth]{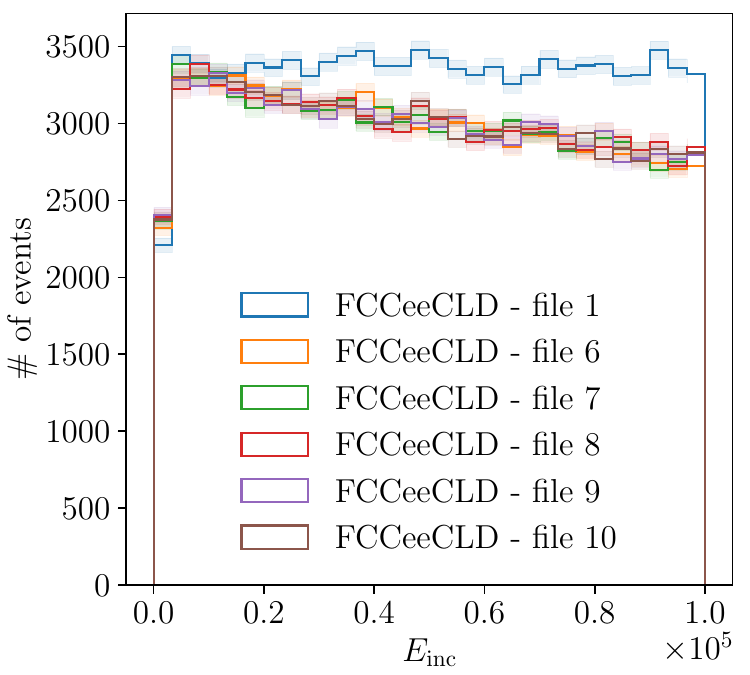}
    \caption{Distribution of the incident energies for the FCCeeCLD detector. The training files 1-4 (left) have the expected uniform distribution while the files 6-10 (right) have a smoothly falling behavior towards larger energies.}
    \label{fig:einc_fcceecld}
\end{figure}

\begin{figure}
    \centering
    \includegraphics[width=0.325\linewidth]{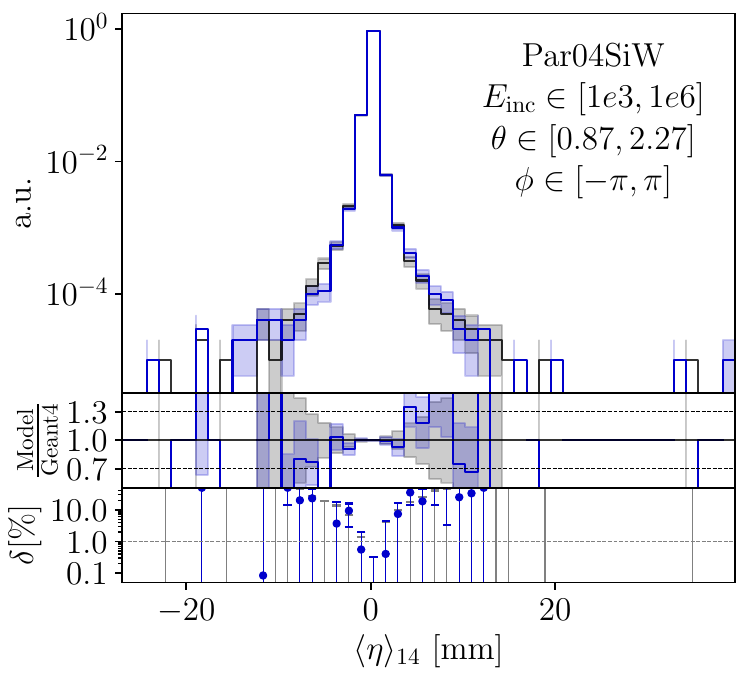}
    \includegraphics[width=0.325\linewidth]{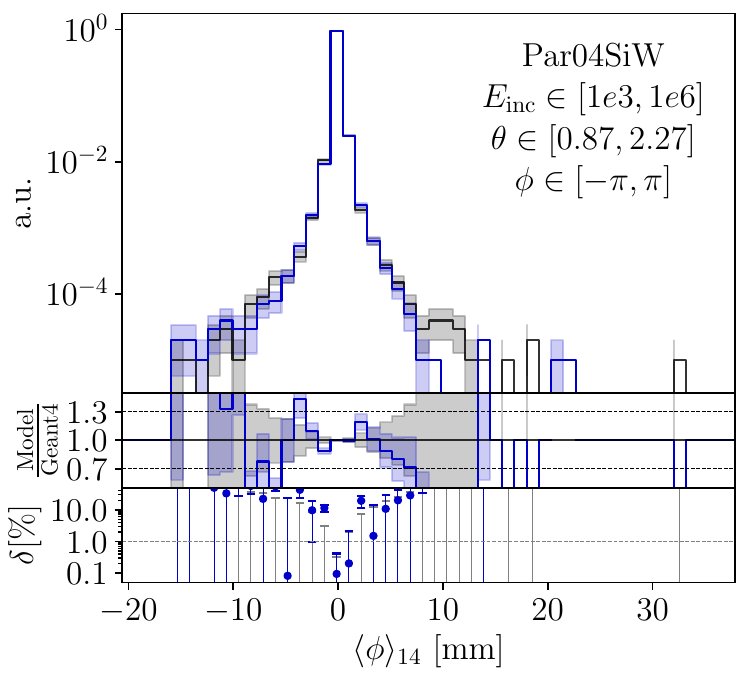}
    \includegraphics[width=0.325\linewidth]{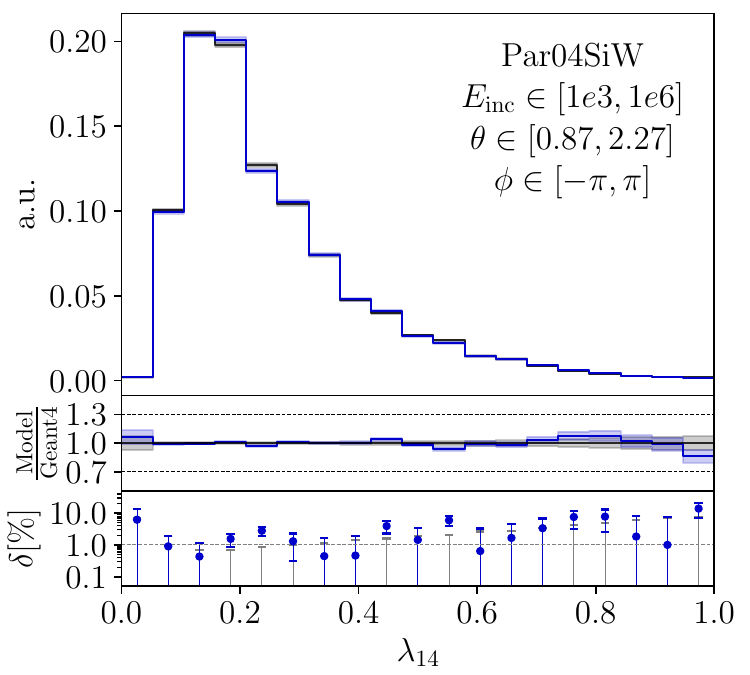} \\
    \includegraphics[width=0.325\linewidth]{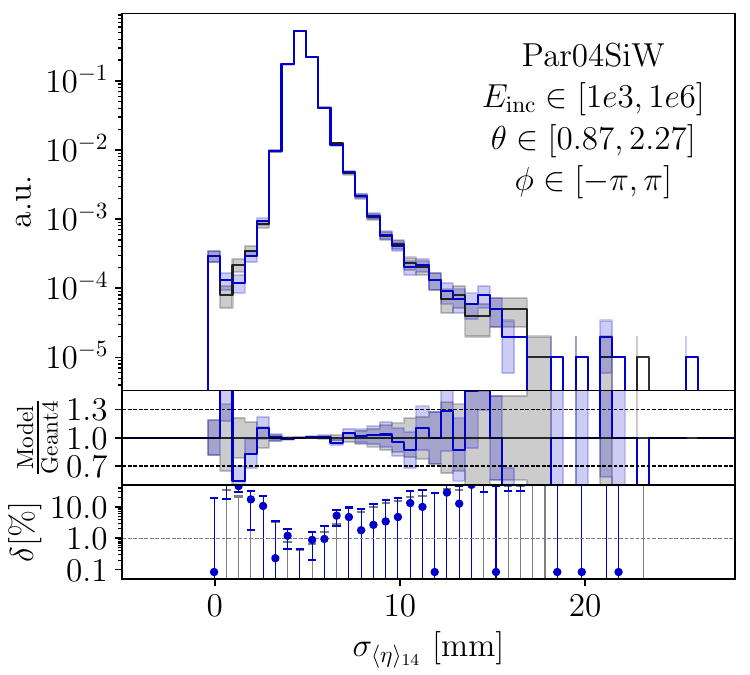}
    \includegraphics[width=0.325\linewidth]{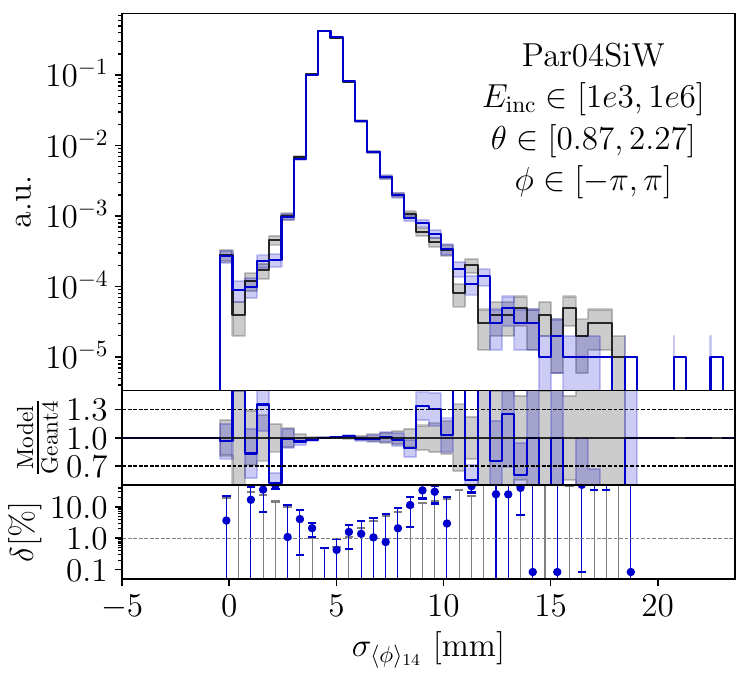}
    \includegraphics[width=0.325\linewidth]{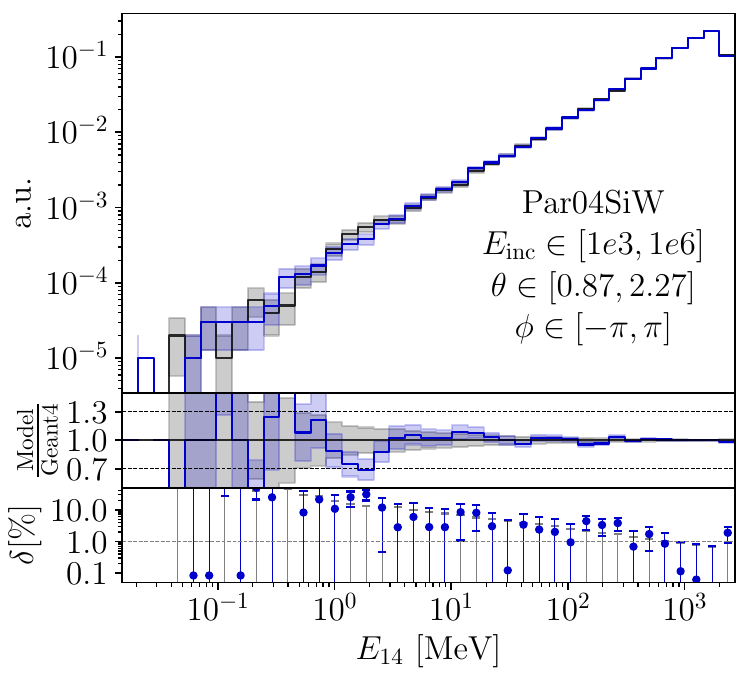} \\
    \includegraphics[width=0.325\linewidth]{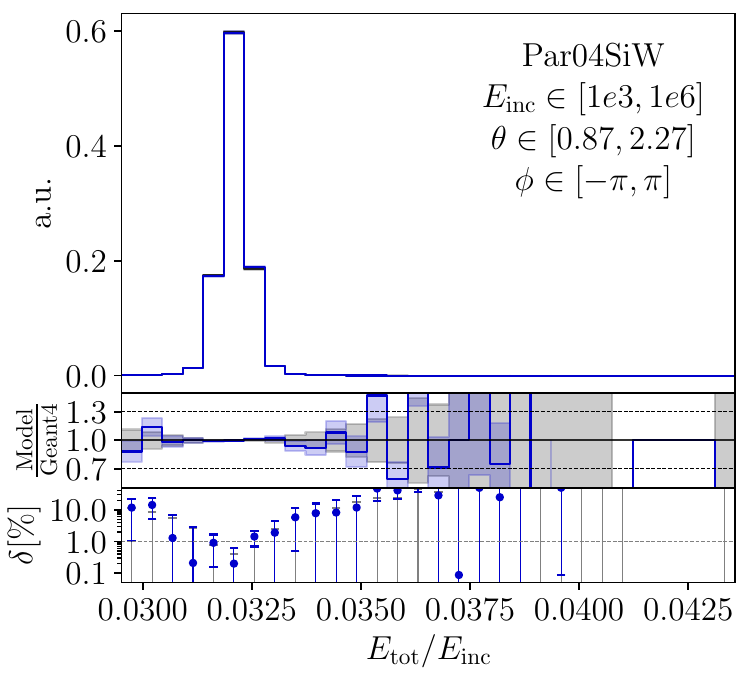}
    \includegraphics[width=0.325\linewidth]{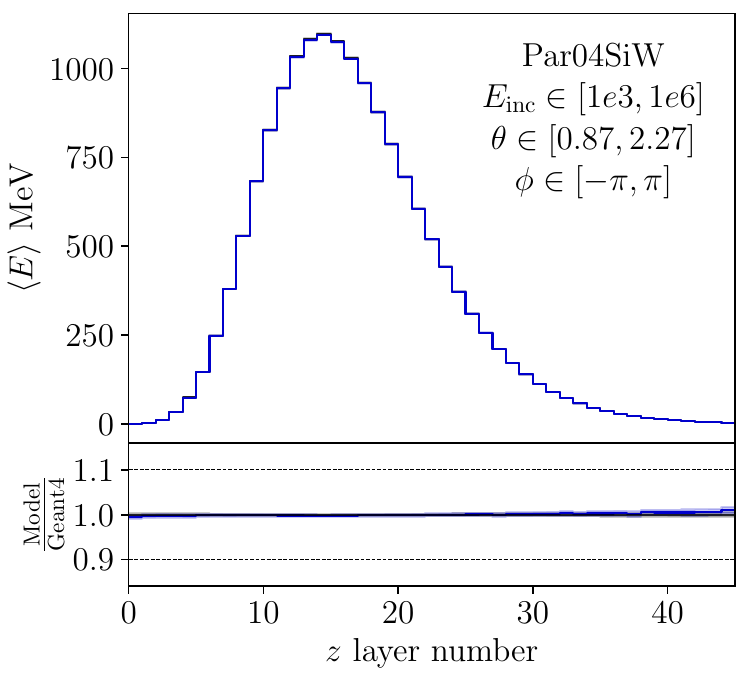}
    \includegraphics[width=0.325\linewidth]{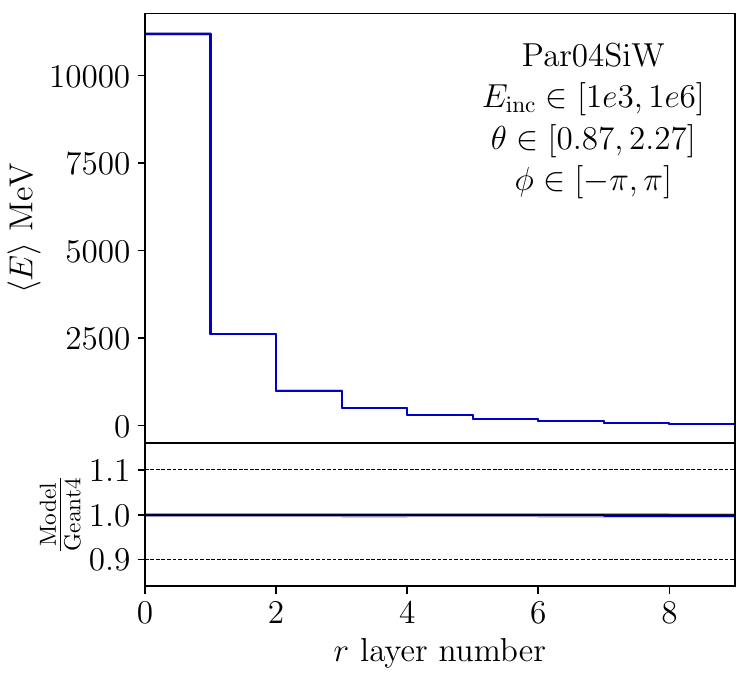} \\
    \includegraphics[width=0.60\linewidth]{figs/LEMURS/lemurs_legend.pdf}
    \caption{Summary of high-level observables for the Par04SiW detector. From left to right, top to bottom: center of energy in the $\eta$ and $\phi$ directions, sparsity, width of the center of energy in the $\eta$ and $\phi$ directions, layer energy depositions, energy ratio $E_\text{inc}/E_\text{tot}$, energy profiles in the $z$ and $r$ directions.}
    \label{fig:par04siw_hlfs}
\end{figure}

\begin{figure}
    \centering
    \includegraphics[width=0.325\linewidth]{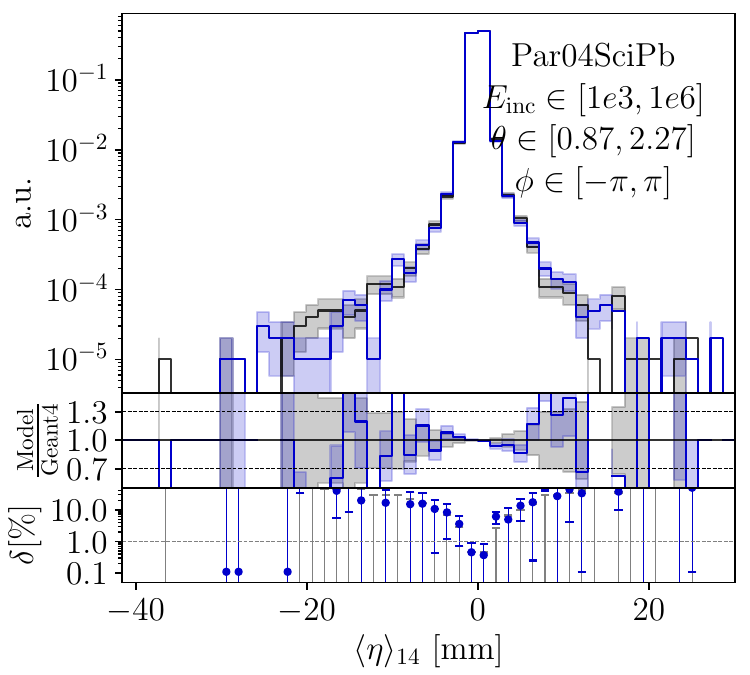}
    \includegraphics[width=0.325\linewidth]{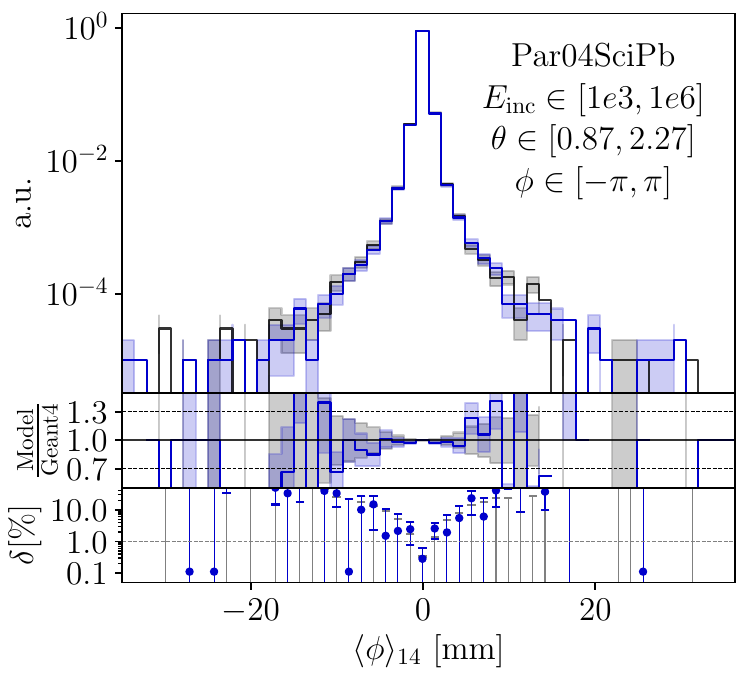}
    \includegraphics[width=0.325\linewidth]{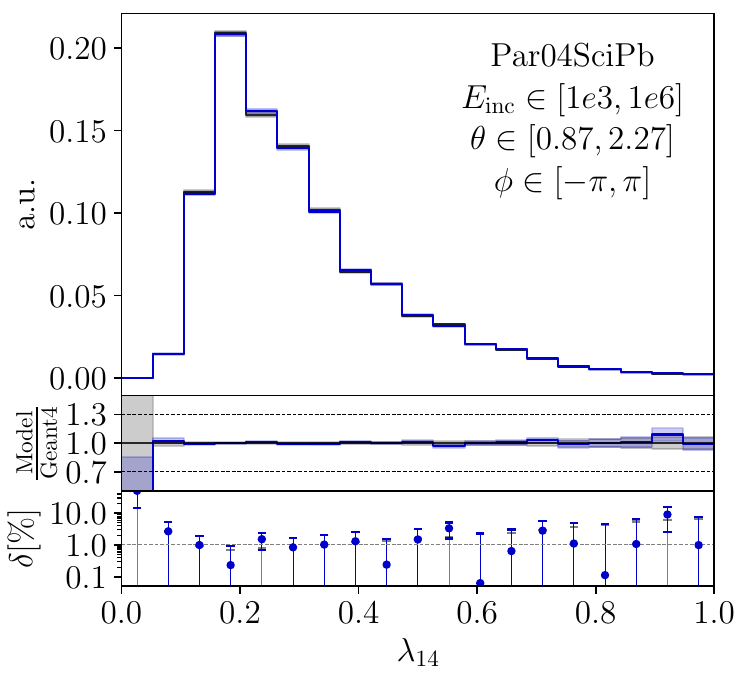} \\
    \includegraphics[width=0.325\linewidth]{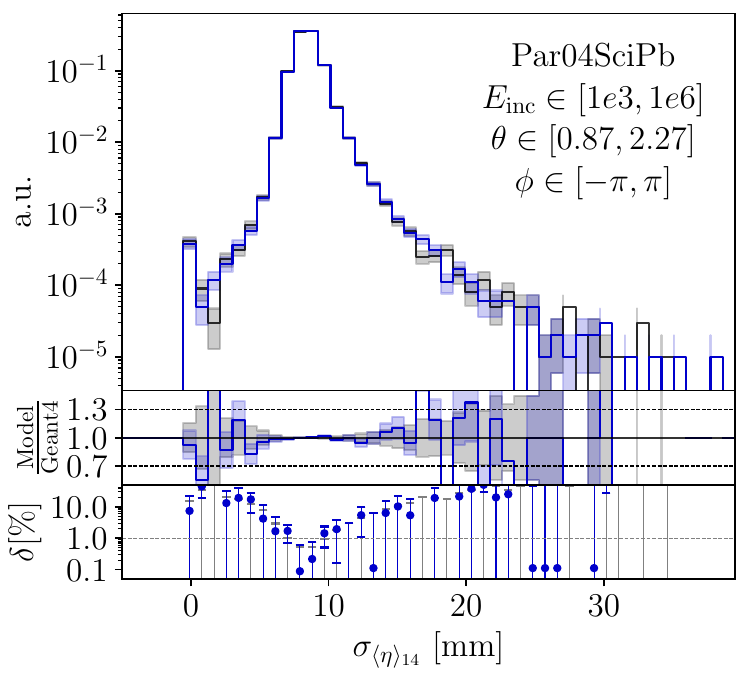}
    \includegraphics[width=0.325\linewidth]{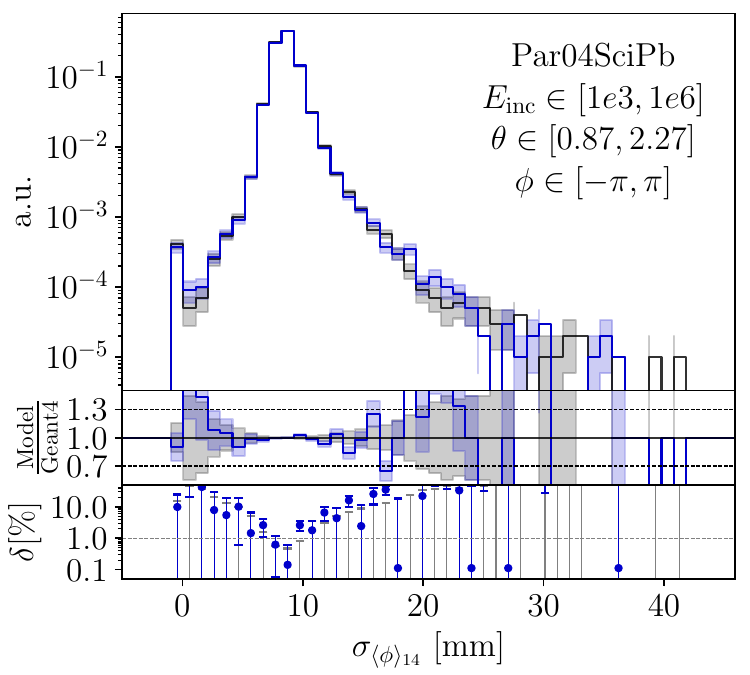}
    \includegraphics[width=0.325\linewidth]{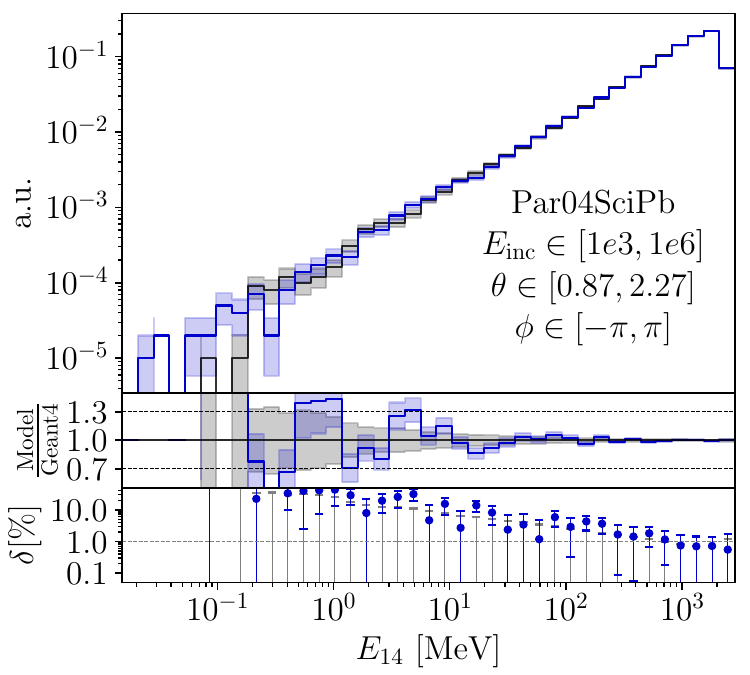} \\
    \includegraphics[width=0.325\linewidth]{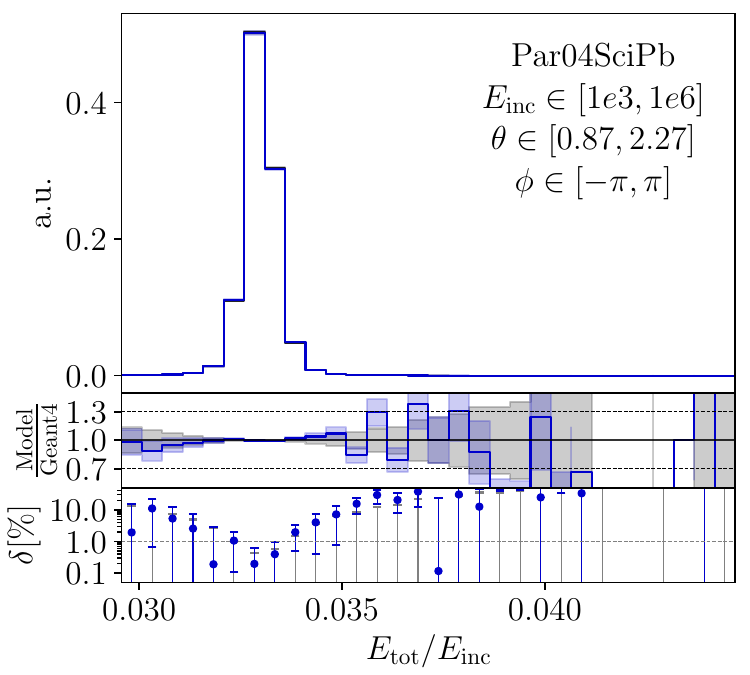}
    \includegraphics[width=0.325\linewidth]{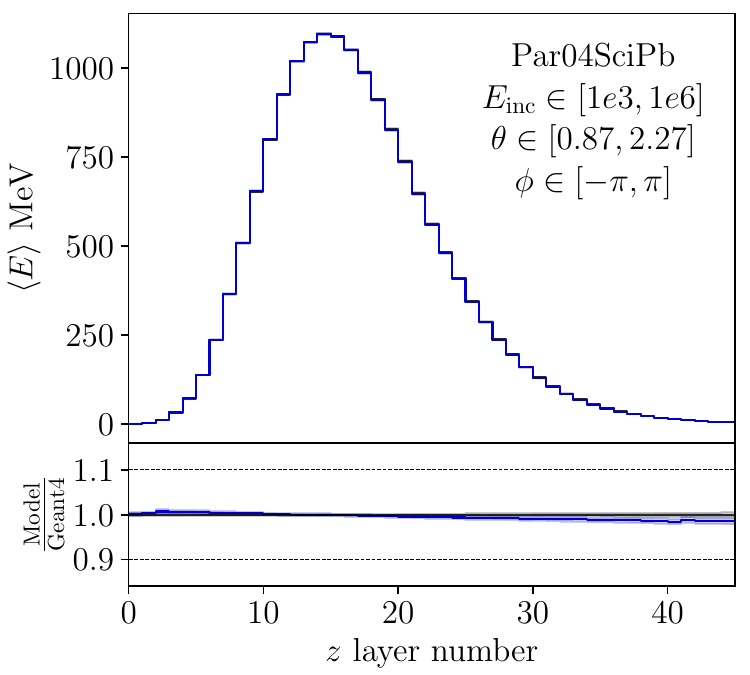}
    \includegraphics[width=0.325\linewidth]{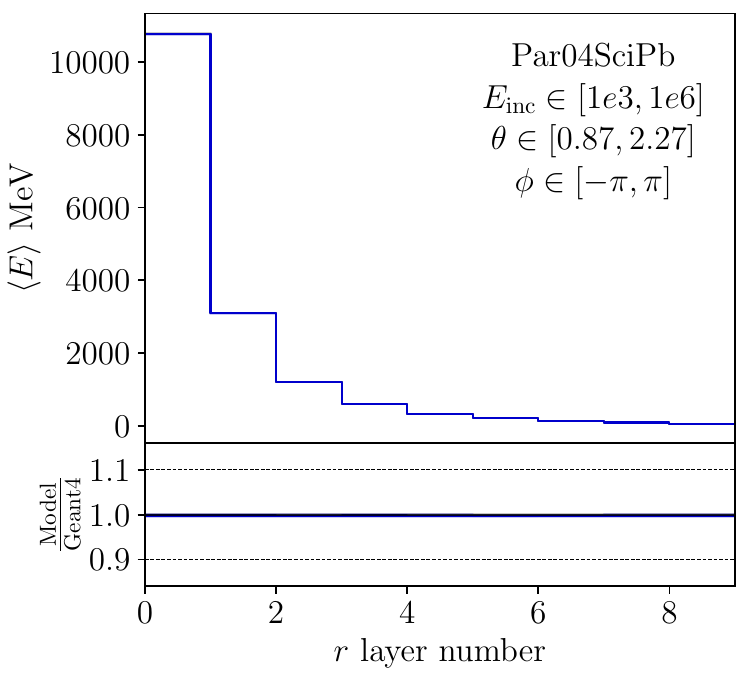} \\
    \includegraphics[width=0.60\linewidth]{figs/LEMURS/lemurs_legend.pdf}
    \caption{Summary of high-level observables for the Par04SciPb detector. From left to right, top to bottom: center of energy in the $\eta$ and $\phi$ directions, sparsity, width of the center of energy in the $\eta$ and $\phi$ directions, layer energy depositions, energy ratio $E_\text{inc}/E_\text{tot}$, energy profiles in the $z$ and $r$ directions.}
    \label{fig:par04scipb_hlfs}
\end{figure}

\begin{figure}
    \centering
    \includegraphics[width=0.325\linewidth]{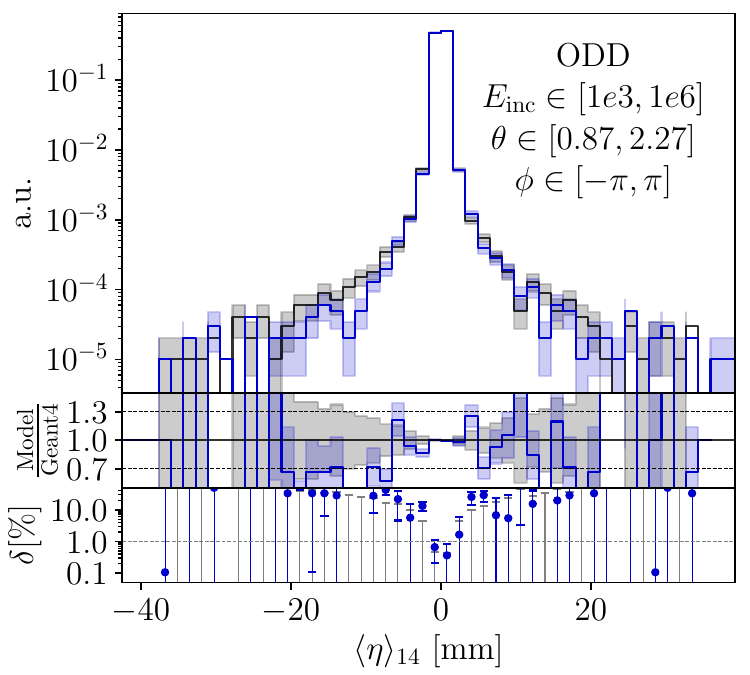}
    \includegraphics[width=0.325\linewidth]{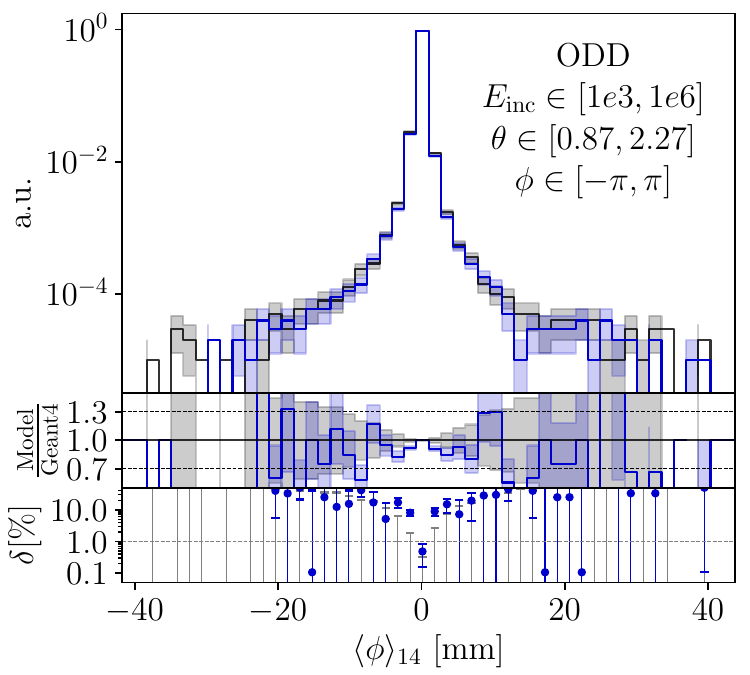}
    \includegraphics[width=0.325\linewidth]{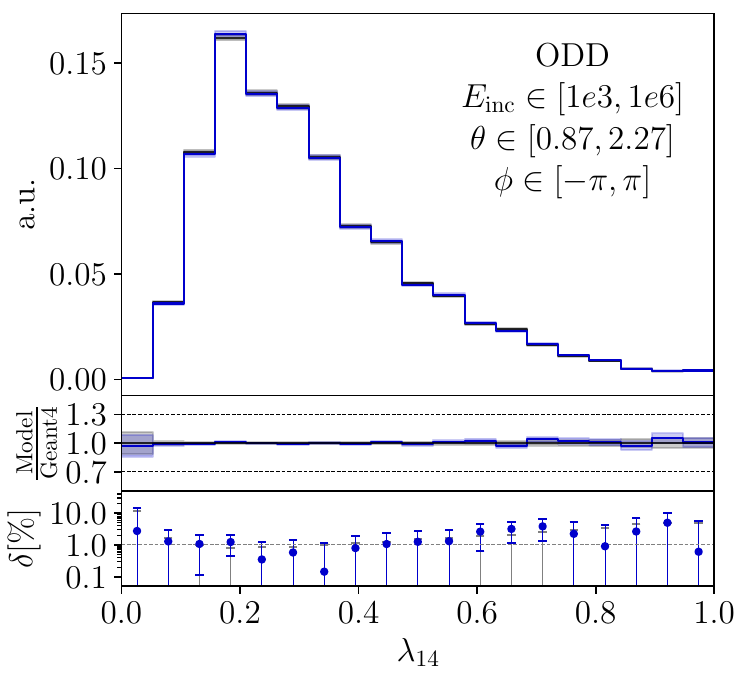} \\
    \includegraphics[width=0.325\linewidth]{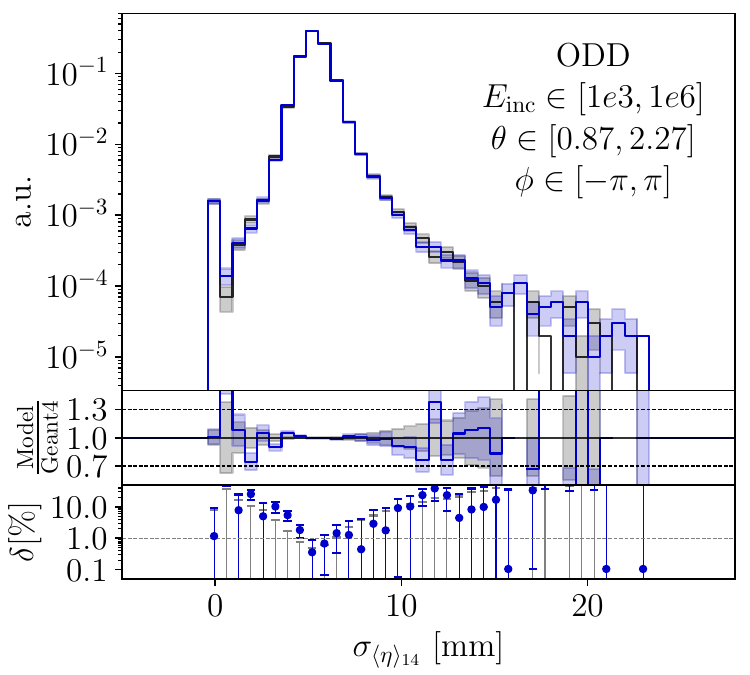}
    \includegraphics[width=0.325\linewidth]{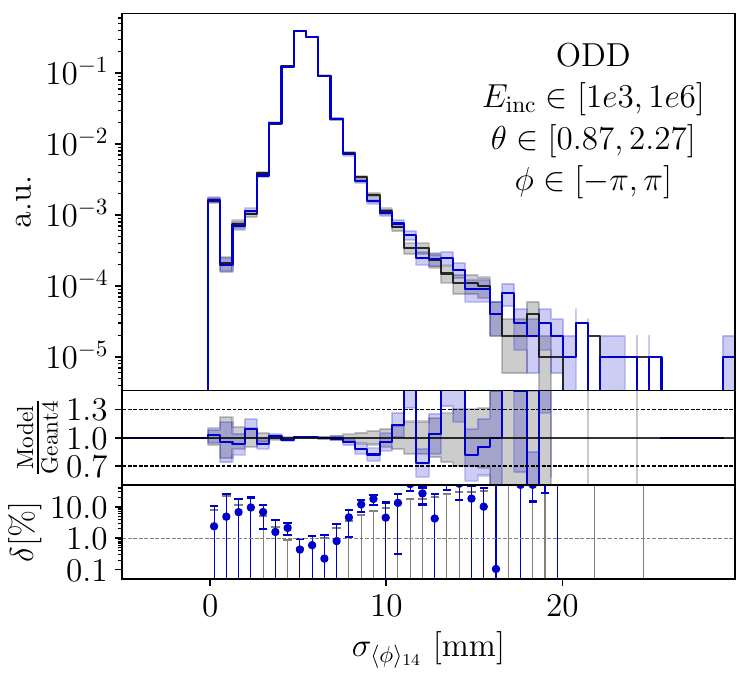}
    \includegraphics[width=0.325\linewidth]{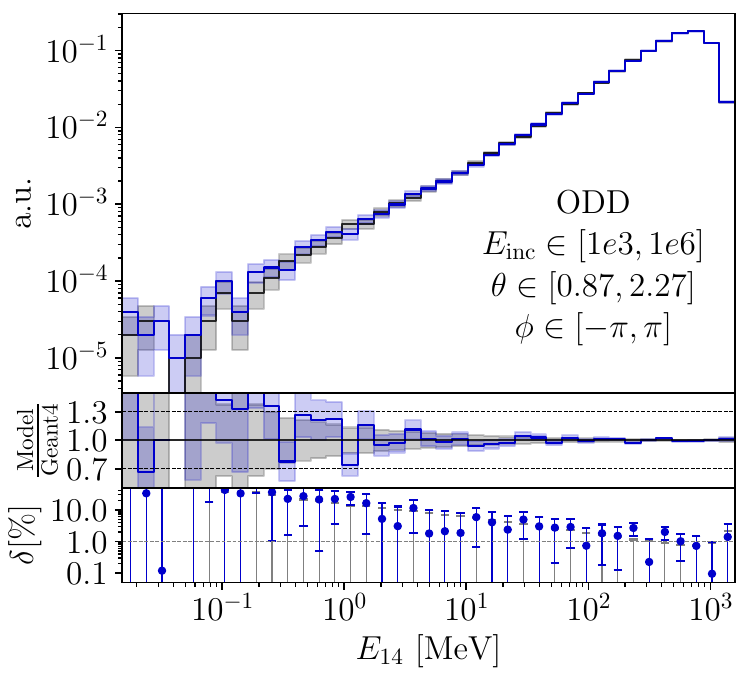} \\
    \includegraphics[width=0.325\linewidth]{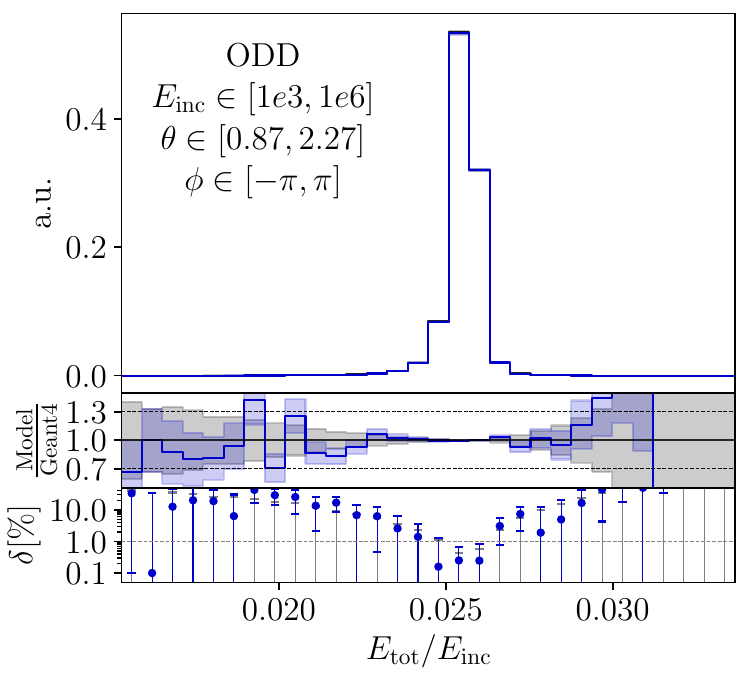}
    \includegraphics[width=0.325\linewidth]{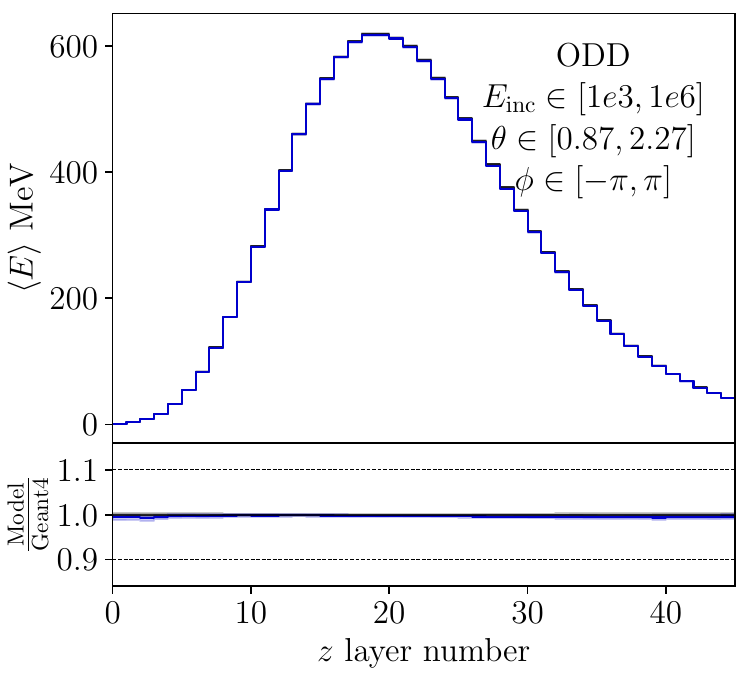}
    \includegraphics[width=0.325\linewidth]{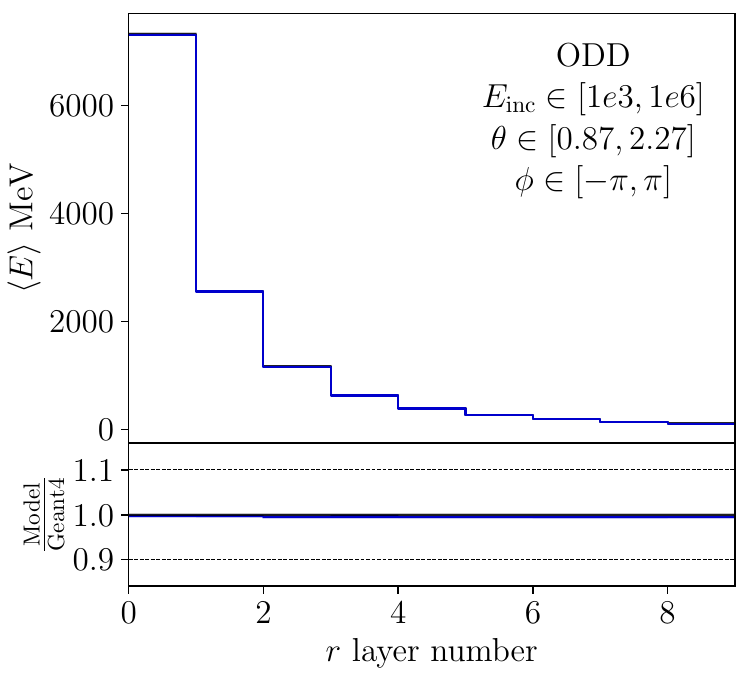} \\
    \includegraphics[width=0.60\linewidth]{figs/LEMURS/lemurs_legend.pdf}
    \caption{Summary of high-level observables for the ODD detector. From left to right, top to bottom: center of energy in the $\eta$ and $\phi$ directions, sparsity, width of the center of energy in the $\eta$ and $\phi$ directions, layer energy depositions, energy ratio $E_\text{inc}/E_\text{tot}$, energy profiles in the $z$ and $r$ directions.}
    \label{fig:odd_hlfs}
\end{figure}

\begin{figure}
    \centering
    \includegraphics[width=0.325\linewidth]{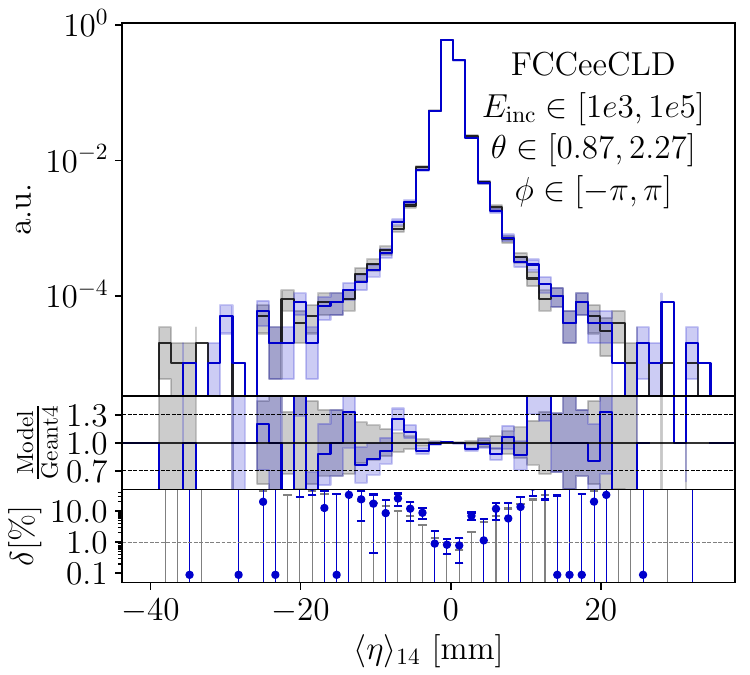}
    \includegraphics[width=0.325\linewidth]{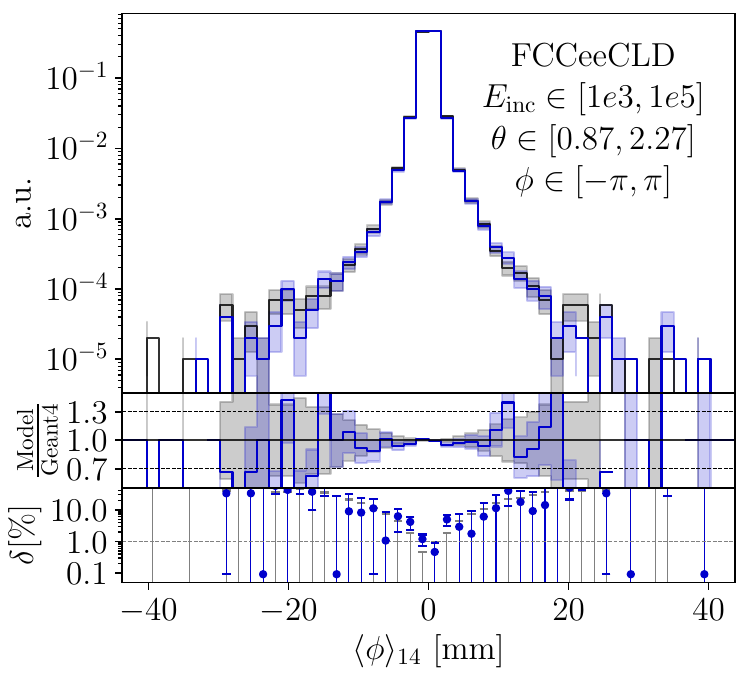}
    \includegraphics[width=0.325\linewidth]{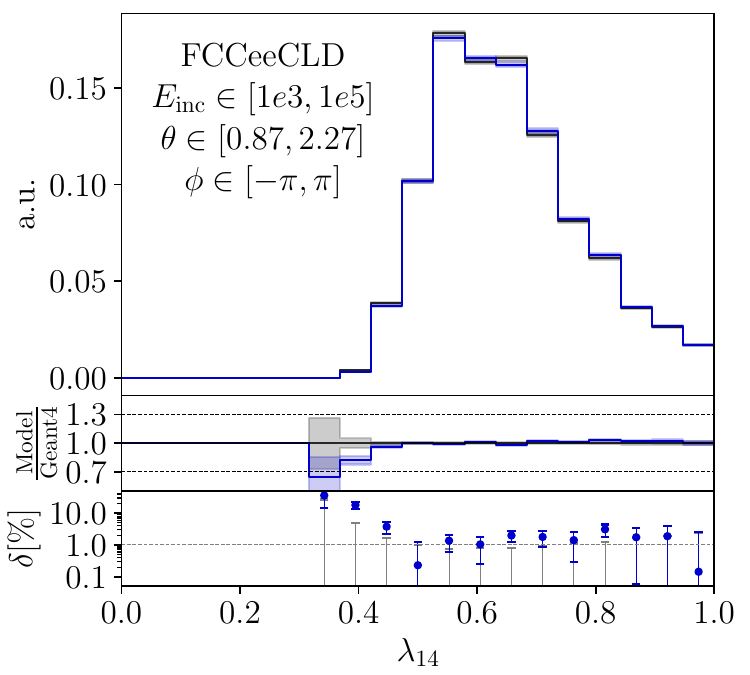} \\
    \includegraphics[width=0.325\linewidth]{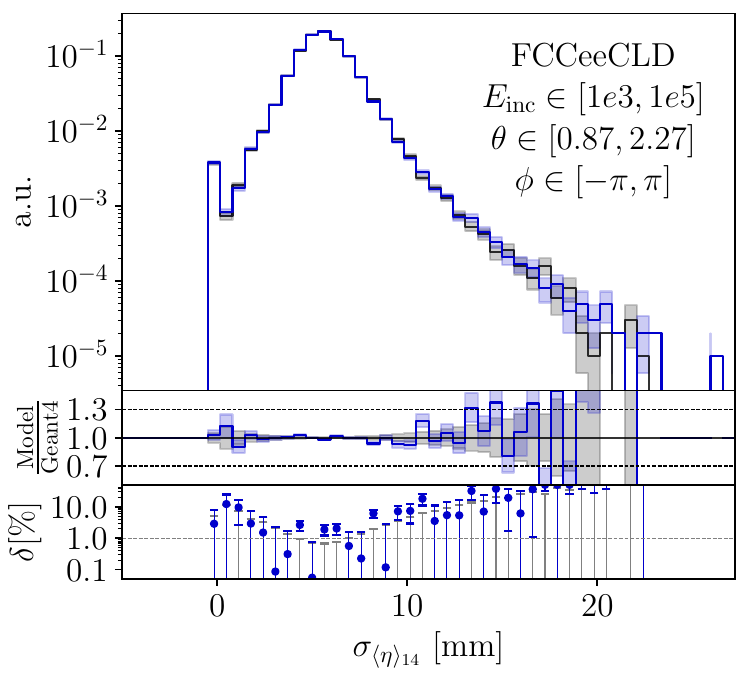}
    \includegraphics[width=0.325\linewidth]{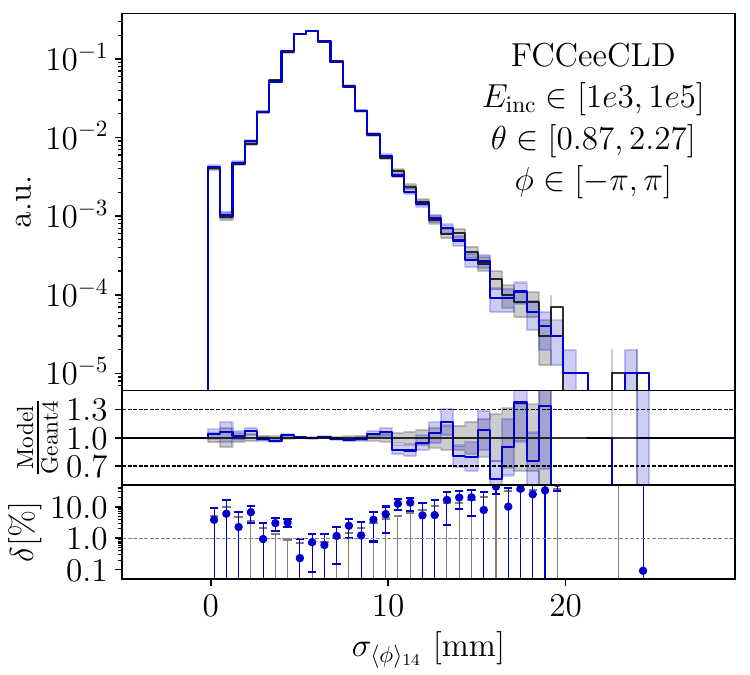}
    \includegraphics[width=0.325\linewidth]{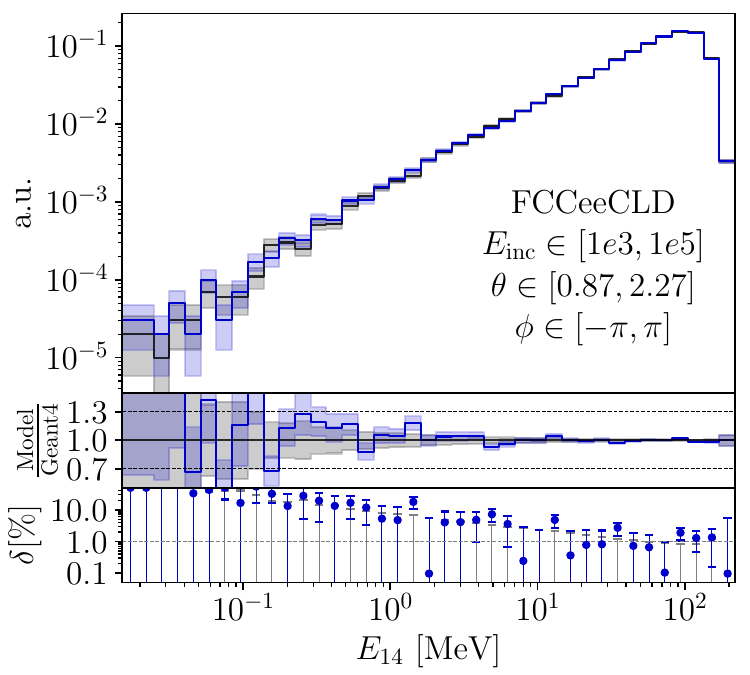} \\
    \includegraphics[width=0.325\linewidth]{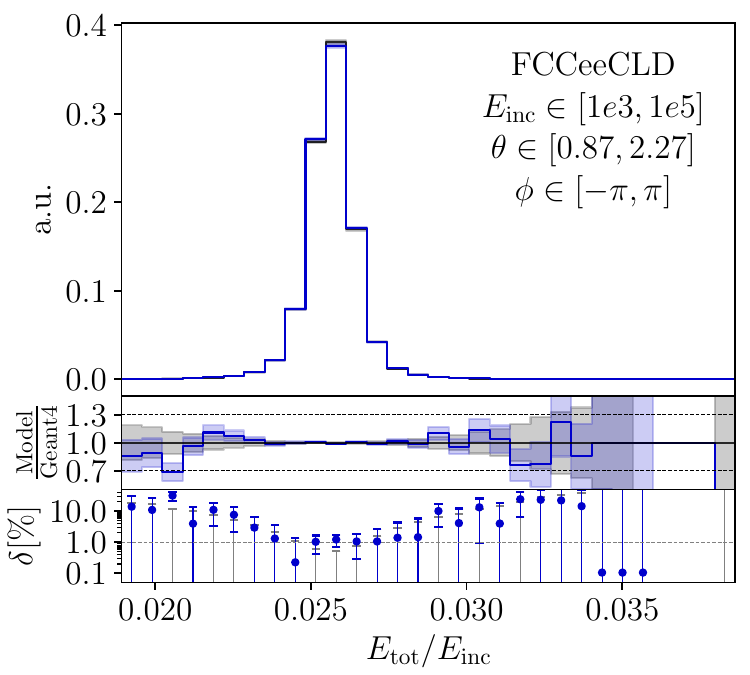}
    \includegraphics[width=0.325\linewidth]{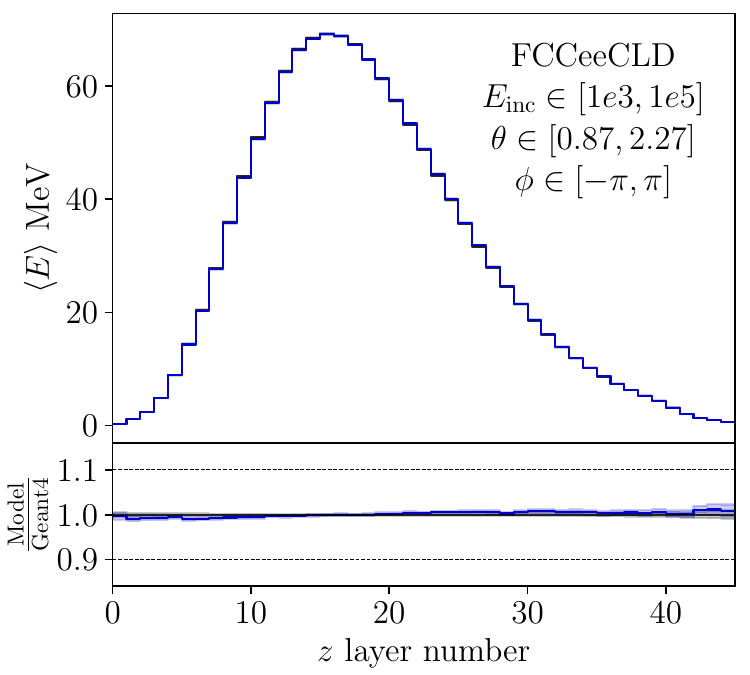}
    \includegraphics[width=0.325\linewidth]{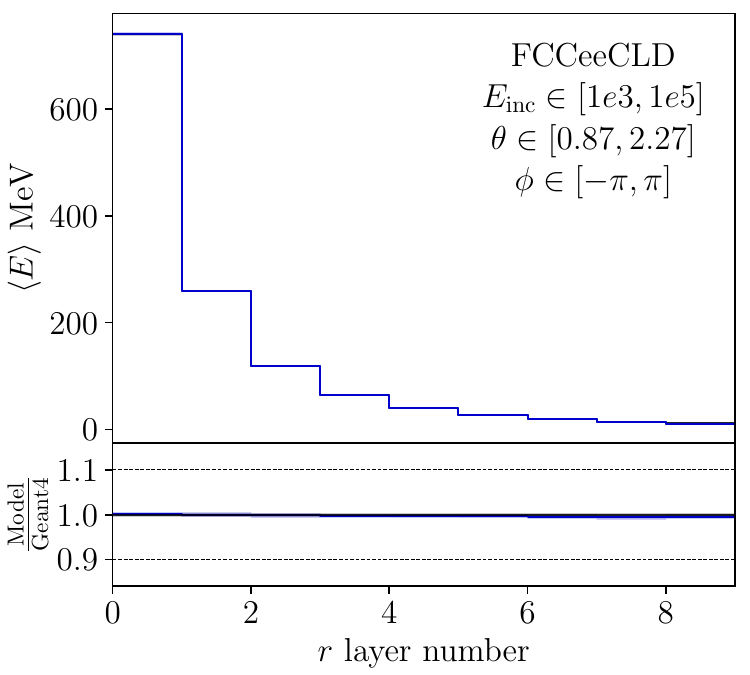} \\
    \includegraphics[width=0.60\linewidth]{figs/LEMURS/lemurs_legend.pdf}
    \caption{Summary of high-level observables for the FCCeeCLD detector. From left to right, top to bottom: center of energy in the $\eta$ and $\phi$ directions, sparsity, width of the center of energy in the $\eta$ and $\phi$ directions, layer energy depositions, energy ratio $E_\text{inc}/E_\text{tot}$, energy profiles in the $z$ and $r$ directions.}
    \label{fig:fcceecld_hlfs}
\end{figure}

\begin{figure}
    \centering
    \includegraphics[width=0.325\linewidth]{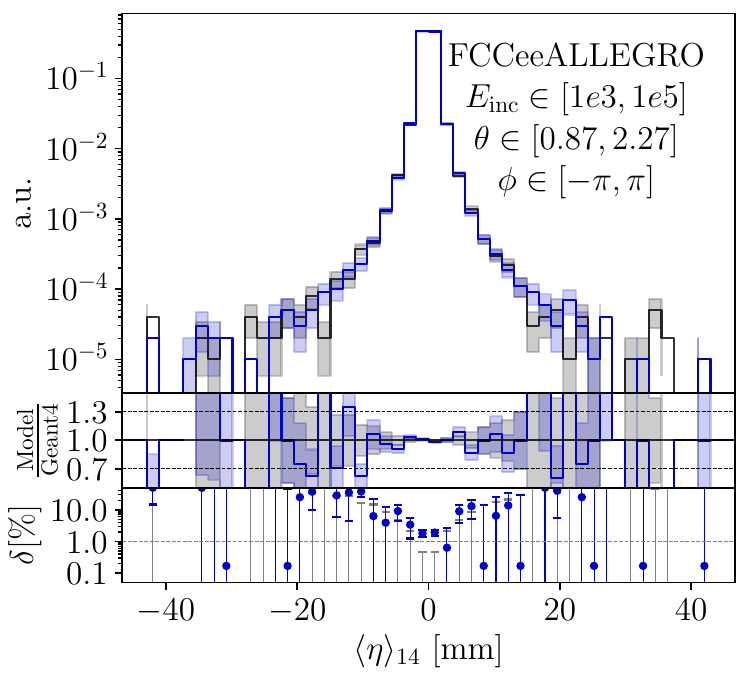}
    \includegraphics[width=0.325\linewidth]{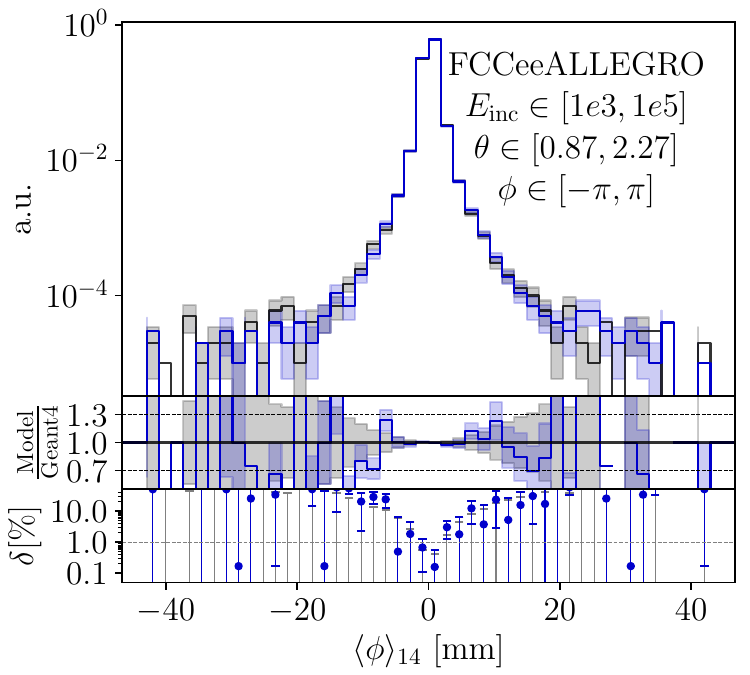}
    \includegraphics[width=0.325\linewidth]{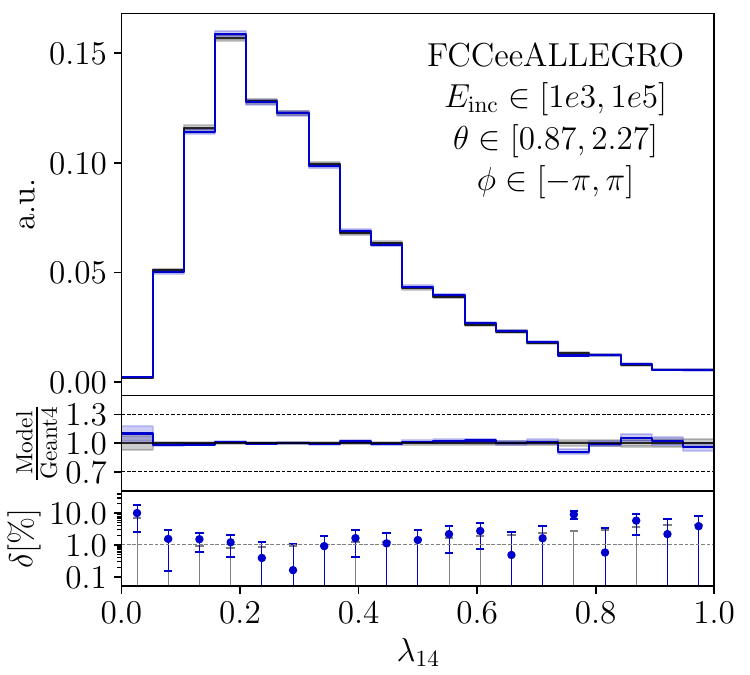} \\
    \includegraphics[width=0.325\linewidth]{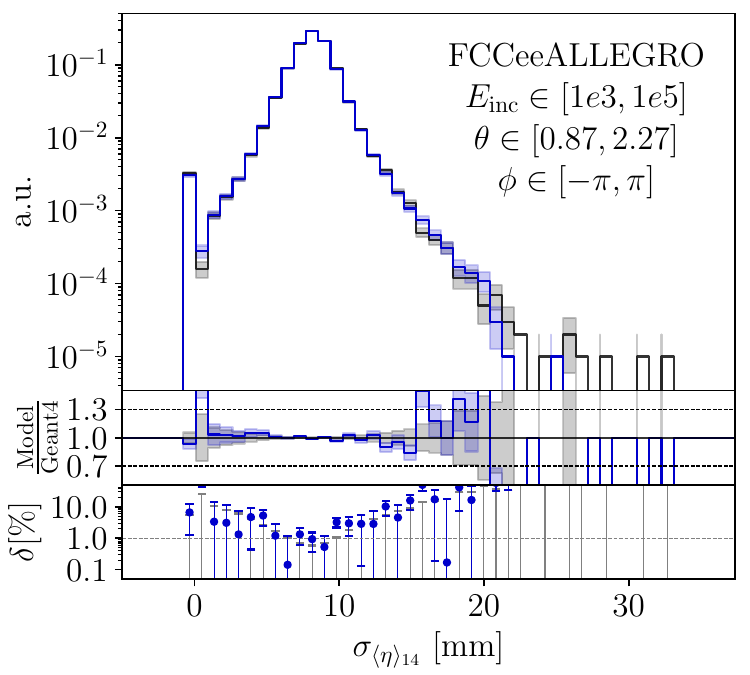}
    \includegraphics[width=0.325\linewidth]{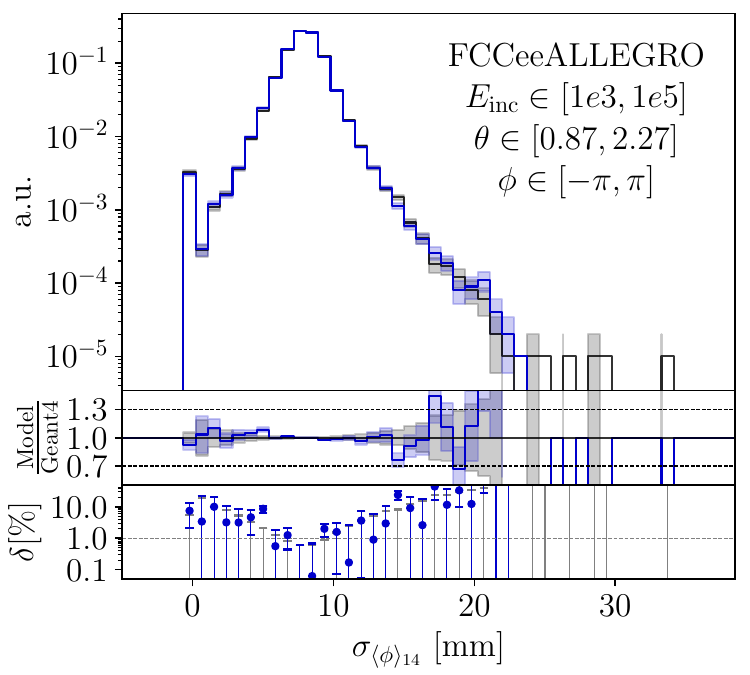}
    \includegraphics[width=0.325\linewidth]{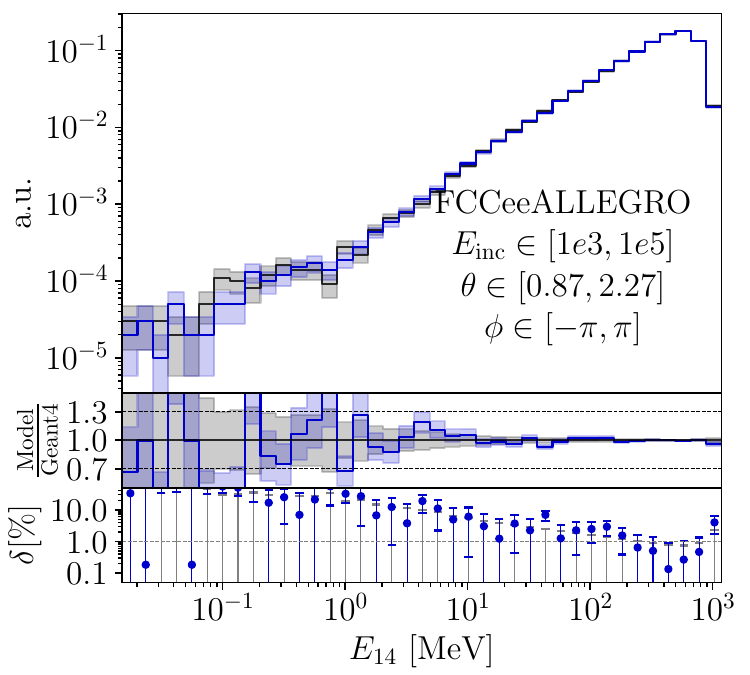} \\
    \includegraphics[width=0.325\linewidth]{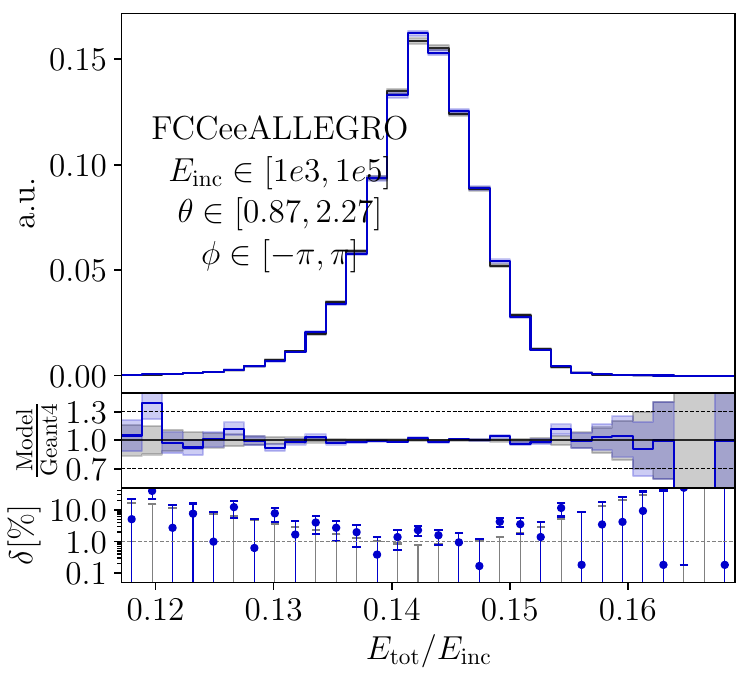}
    \includegraphics[width=0.325\linewidth]{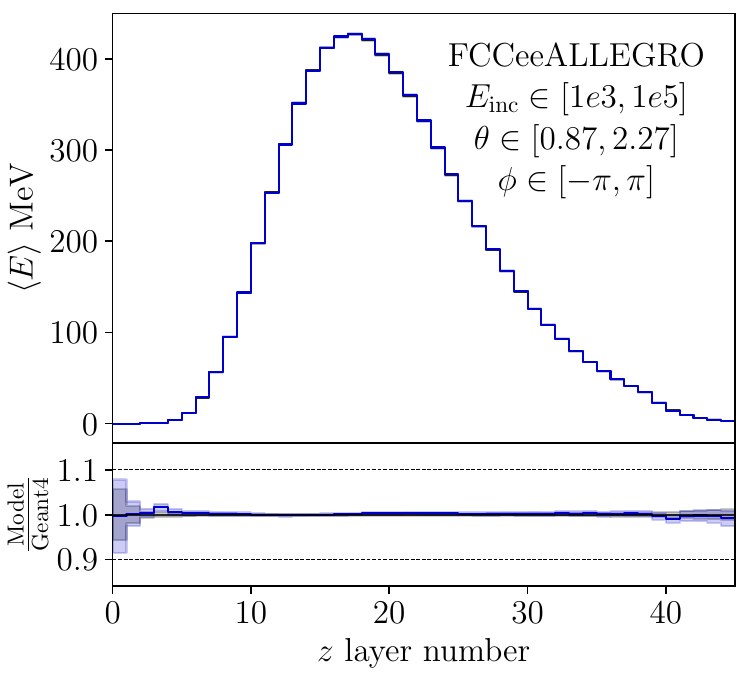}
    \includegraphics[width=0.325\linewidth]{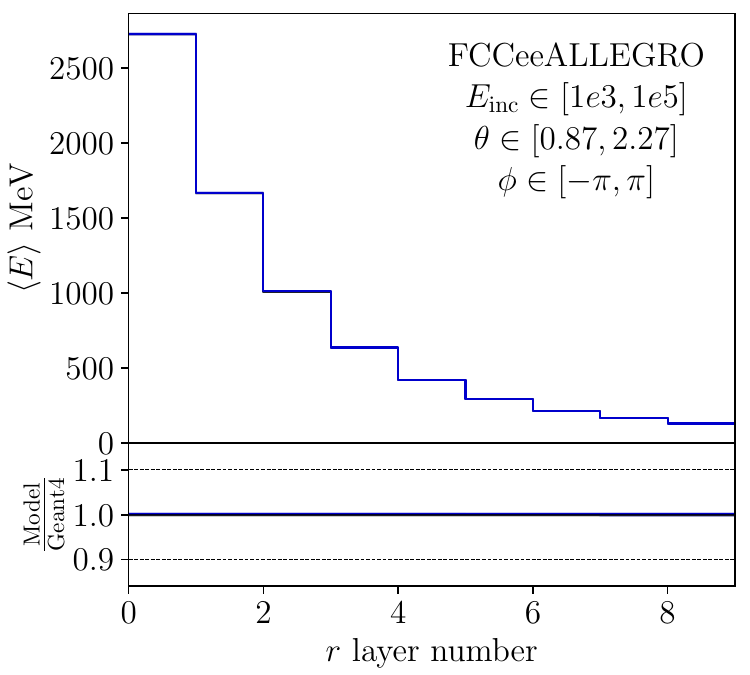} \\
    \includegraphics[width=0.60\linewidth]{figs/LEMURS/lemurs_legend.pdf}
    \caption{Summary of high-level observables for the FCCeeALLEGRO detector. From left to right, top to bottom: center of energy in the $\eta$ and $\phi$ directions, sparsity, width of the center of energy in the $\eta$ and $\phi$ directions, layer energy depositions, energy ratio $E_\text{inc}/E_\text{tot}$, energy profiles in the $z$ and $r$ directions.}
    \label{fig:fcceeallegro_hlfs}
\end{figure}

We provide, in Figs.~\labelcref{fig:par04siw_hlfs},~\labelcref{fig:par04scipb_hlfs},~\labelcref{fig:odd_hlfs},~\labelcref{fig:fcceecld_hlfs}, and~\labelcref{fig:fcceeallegro_hlfs}, histograms of the high-level features for the layer
with the largest energy deposition. These features, together with
all the other layers, have been used to train the high-level classifier
for the full LEMURS evaluation.
The held-out dataset, used for this evaluation, corresponds to the last file of the LEMURS\footnote{Our studies utilize the 1.0.0 version of the LEMURS dataset.} training dataset. The FCCeeCLD evaluation is the only exception: we noticed that the distribution of the incident energies for the files with less than 100k events does not agree with a uniform distribution. We show the discrepancy between complete and incomplete files in~\cref{fig:einc_fcceecld} and, to avoid biased comparison, we use the first file as the \geant reference.

\clearpage
\bibliographystyle{iopart-num} 
\bibliography{literature} 
\end{document}